\documentclass[11pt]{article}

\pdfoutput=1
\usepackage[parsep]{collref}

\usepackage{simpler-wick}
\usepackage{datetime2}
\usepackage{comment}
\usepackage{xcolor}

\usepackage{graphicx}
\usepackage{amsmath}
\usepackage{amsthm}
\usepackage{mathtools}
\usepackage{tabularx}
\usepackage{multirow}
\usepackage{multicol}
\usepackage{enumitem}
\usepackage[utf8]{inputenc}
\usepackage{comment}
\usepackage{amsmath}
\usepackage{amsfonts}
\usepackage{braket}
\usepackage{cancel}
\usepackage{upgreek}

\usepackage[compat=1.1.0]{tikz-feynman}

\theoremstyle{definition}

\theoremstyle{remark}

\title{Non-perturbative membrane corrections to the supersymmetric free energy in $\mathrm{AdS}_{4}/\mathrm{CFT}_{3}$ holography}
\date{\today}

\usepackage{amsmath, amssymb,amsthm}
\usepackage{stackrel}
\numberwithin{equation}{section}
\usepackage{bm,environ,mathrsfs,array,arydshln}
\usepackage{booktabs,float,slashed}
\usepackage{appendix}
\usepackage{tensor}
\usepackage[vcentermath]{youngtab}

\usepackage{simplewick}

\newcommand{\tr}{\mathrm{tr}}

\newcommand{\id}{\mathbb{I}}

\newcommand{\ads}{\mathrm{AdS}_{4}}
\newcommand{\ys}{\mathrm{Y}_{7}}
\newcommand{\Le}{\mathrm{L}}
\newcommand{\vol}{\mathrm{vol}}
\newcommand{\ses}{\mathrm{SE}_{7}}
\newcommand{\cyf}{\mathrm{CY}_{4}}
\newcommand{\hodge}{\,{\ast}}
\newcommand{\eun}{\mathrm{e}}
\newcommand{\nn}{\mathrm{n}}
\newcommand{\psin}{\psi_{-}}
\newcommand{\adj}{\mathcal{C}_{\alpha}}
\newcommand{\adjb}{\mathcal{C}_{\bar{\alpha}}}
\newcommand{\Xr}{\mathrm{X}}
\newcommand{\Hr}{\mathrm{H}}
\newcommand{\ind}{\mathrm{ind}}

\usepackage{xcolor}

\colorlet{linkred}{red!50!black}

\usepackage[
  colorlinks=true,
  linkcolor=black,
  citecolor=linkred,
  urlcolor=linkred,
  linktoc=none
]{hyperref}

\makeatletter
\renewcommand{\eqref}[1]{%
  \textup{(}%
  \hyperref[#1]{\textcolor{linkred}{\ref*{#1}}}%
  \textup{)}%
}
\makeatother

\begin{document}

\vspace{-3cm}
\thispagestyle{empty}

\begin{comment}
\begin{flushright}
Imperial-TP-2026-XX\\
\today
\end{flushright}
\end{comment}

\begin{center}
\vspace{1.2cm}

{\Large\bf Membrane instantons
and
\\
non-perturbative effects in $\mathrm{AdS}_{4}/\mathrm{CFT}_{3}$

}

\vspace{0.8cm}

Stefan A. Kurlyand

\vskip 0.5cm

{\em
Abdus Salam Centre for Theoretical Physics, Imperial College London, SW7 2AZ, UK
}

\vskip 0.35cm

E-mail: \texttt{s.kurlyand23@imperial.ac.uk}

\end{center}

\vskip 1.2cm

\begin{abstract}
We study Euclidean M2-brane instantons in Freund-Rubin backgrounds $\ads\times \ys$. For a seven-dimensional weak $G_{2}$ manifold $\ys$, we show that the BPS condition for an M2-brane wrapping a three-cycle $\Sigma\subset \ys$ is equivalent to the associativity condition with respect to the nearly parallel $G_{2}$-structure. When $\ys$ is Sasaki-Einstein, we identify a special class of BPS M2-branes that preserve both real internal Killing spinors and correspond to invariant three-dimensional submanifolds inheriting a Sasakian structure. We analyse the quadratic fluctuations around BPS M2-brane instantons in these backgrounds. For the special class of M2-branes in Sasaki-Einstein manifolds, the fluctuation problem reduces to transversely elliptic complexes, and the one-loop partition function can be expressed in terms of the corresponding equivariant indices. We then apply the index formula for the one-loop partition function to invariant M2-branes in $S^{7}/\mathbb Z_{k}$, recovering the known result for the $S^3/\mathbb Z_k$ instantons and discussing more general invariant BPS cycles. As a further application, we consider M2-brane instantons with $S^3$-quotient worldvolumes in the $(p,q)$-model geometry.

\end{abstract}

\newpage
\tableofcontents

\section{Introduction and summary of results}

 The study of non-perturbative corrections in M-theory is a long-standing topic. In the context of Calabi-Yau compactification to five dimensions of eleven-dimensional supergravity, supersymmetric M2-brane configurations wrapping three-cycles in a Calabi-Yau threefold $\mathrm{CY}_{3}$ correspond to BPS instantons correcting the moduli-space geometry of hypermultiplets of 5d $\mathcal{N}=2$ supergravity \cite{Becker:1995kb}. Since $\mathrm{CY}_{3}$ admits a nowhere-vanishing closed holomorphic $(3,0)$ form, the BPS condition can be expressed in the language of the general theory of calibrations \cite{Harvey:1982xk}. Calibration conditions are phrased in terms of closed forms, and the corresponding supersymmetric cycles minimise volume in a given homology class. 
 
 Similarly, in compactifications of eleven-dimensional supergravity on manifolds of $G_{2}$ holonomy, supersymmetric three-cycles $\Sigma$ are calibrated by the parallel $G_{2}$-structure \cite{Becker:1996ay}:
\begin{align}\label{rst1}
d\varphi_{0} = 0 \ , \qquad \mathrm{Vol}(\Sigma) = \int_{\Sigma} \varphi_{0} \ ,
\end{align}
where the covariantly constant closed three-form $\varphi_{0}$ can be constructed as a spinor bilinear from a parallel spinor on the $G_{2}$ manifold.

In \cite{Harvey:1999as}, this analysis was developed further by computing the one-loop partition function around such $G_{2}$-calibrated cycles. Due to the preserved worldvolume supersymmetry, the quadratic modes split into two sectors, corresponding to modes in the flat four-dimensional space $ \mathbb{R}^4$ and modes transverse to $\Sigma$ in the $G_{2}$ manifold. The former sector is described by a linearised Rozansky-Witten model \cite{Rozansky:1996bq} and determines the absolute value of the one-loop partition function
\begin{align}
|Z_{1}'|=\left|\frac{\det' L_-}{(\det'\Delta)^2}\right| \ .
\end{align}
Thus, for a closed BPS M2-brane $\Sigma$, the absolute value of the one-loop partition function is a topological invariant of the three-manifold, namely the analytic Ray-Singer torsion, while the transverse sector contributes only through the phase encoded by an $\eta$-invariant of a twisted Dirac operator.
\begin{comment}
In \cite{Harvey:1999as}, this analysis was developed further by computing the one-loop partition function around such $G_{2}$-calibrated cycles. Due to the preserved worldvolume supersymmetry, the quadratic fluctuations can be organised into two sectors, corresponding to fluctuations in the flat four-dimensional space $ \mathbb{R}^4$ and fluctuations transverse to $\Sigma$ in the $G_{2}$ manifold. The quadratic action for the modes in $\mathbb{R}^4$ is described by a linearised Rozansky-Witten model \cite{Rozansky:1996bq}. The absolute value of the one-loop partition function for M2-brane $\Sigma$ is controlled only by the fluctuations in $\mathbb{R}^4$ and gives the analytic Ray-Singer torsion
\begin{align}
|Z_{1}'|=\left|\frac{\det' L_-}{(\det'\Delta)^2}\right| \ .
\end{align}
Thus, for a closed BPS M2-brane $\Sigma$, the absolute value of the one-loop partition function is a topological invariant of the three-manifold, while the fluctuations transverse to $\Sigma$ in a $G_2$ manifold contribute only through the phase encoded by an $\eta$-invariant of a twisted Dirac operator.
\end{comment}

This raises the question whether a similar one-loop analysis can be carried out for BPS M2-brane instantons in more general M-theory backgrounds. From the holographic perspective, particularly relevant examples are backgrounds of the form $\ads\times \ses$, where $\ses$ is a seven-dimensional Sasaki-Einstein manifold. Such backgrounds are expected to be dual to $\mathcal{N}=2$ three-dimensional Chern-Simons matter theories whose $S^3$ partition functions can be computed by supersymmetric localisation, leading to matrix-model integrals \cite{Kapustin:2009kz}.

A well-studied example is the duality between M-theory on $\ads \times S^7/\mathbb{Z}_k$ and three-dimensional ABJM theory with gauge group $U(N)_{k}\times U(N)_{-k}$ \cite{Aharony:2008ug}. In \cite{Drukker:2010nc,Drukker:2011zy}, the large-$N$ solution of the ABJM matrix model was obtained in the 't Hooft limit $N\to\infty$, with $\lambda=N/k$ fixed, using its relation to the lens-space Chern-Simons matrix model. The resulting ribbon-graph $1/N$ expansion is identified with the topological-string genus expansion on local $\mathbb{P}^{1}\times\mathbb{P}^{1}$. 
%The planar solution is encoded in a two-cut resolvent and determines the genus-zero topological string free energy. 
At strong coupling, the fixed-genus terms contain worldsheet-instanton corrections, while the large-order behaviour of the genus expansion is controlled by matrix-model instantons whose leading action has the expected D2-brane scaling. As noted in \cite{Herzog:2010hf}, the 't Hooft expansion does not directly provide the M-theory expansion, where $k$ is kept fixed while $N$ is taken large. To analyse the fixed $k$ regime, in \cite{Marino:2011eh} the ABJM matrix model was reformulated as the canonical partition function of an ideal one-dimensional Fermi gas. After passing to the grand-canonical ensemble, and combining the WKB expansion of the Fermi gas with the relation to topological strings and the pole-cancellation mechanism, this led to the proposal for the non-perturbative part of the modified grand potential \cite{Hatsuda:2012dt, Calvo:2012du, Hatsuda:2013gj,Hatsuda:2013oxa}. Its structure may be indicated schematically as
\begin{align}\label{ferminp}
J^{\mathrm{np}}(\mu,k)=
\sum_{m, l}
f_{m,l}(k,\mu)
\exp\left[-\left(\frac{4m}{k}+2l\right)\mu\right] \ .
\end{align}
Here $m$ and $l$ label, respectively, the worldsheet and D2-brane instanton sectors, while terms with \mbox{$m,l\neq0$} correspond to their bound states. The functions $f_{m,l}(k,\mu)$ can be organised as polynomials in the chemical potential $\mu$ with $k$-dependent coefficients, while the purely worldsheet terms $f_{m,0}(k)$ are $\mu$-independent. At leading order in the saddle-point approximation to the inverse Laplace transform, one has \mbox{$N\simeq 2\mu^{2}/\pi^{2}k$}.

From the M-theory perspective, the non-perturbative corrections to the matrix-model free energy are expected to arise from M2-brane instantons with non-vanishing volume in $\ses$. In the M-theory background $\ads\times S^{7}/\mathbb{Z}_{k}$ dual to ABJM theory, the leading non-perturbative contribution in \eqref{ferminp} in the large $N$, fixed $k>2$ limit corresponds to the $1/2$-BPS M2-brane instanton wrapping $S^{3}/\mathbb{Z}_{k}\subset S^{7}/\mathbb{Z}_{k}$ \cite{Hatsuda:2013oxa, Park:2020hgt, Beccaria:2023ujc}. In the large $k$ regime, this M2-brane configuration reduces to the $\mathbb{CP}^{1}$ instanton in type IIA string theory on $\ads\times\mathbb{CP}^{3}$ \cite{Cagnazzo:2009zh}. The IIA superstring computation of the leading large $k$ term in the instanton prefactor was presented in \cite{Gautason:2023igo}, including related computations for mass-deformed and orbifolded ABJ(M) theories. In \cite{Beccaria:2023ujc}, the finite $k$ one-loop partition function was computed directly from quantum M2-brane fluctuations around the $S^{3}/\mathbb{Z}_{k}$ saddle, for $k>2$. The on-shell value of the corresponding M2-brane action is
\begin{align}
T_{2}\mathrm{Vol}(S^3/\mathbb{Z}_{k})
=
\frac{\mathrm{L}^3}{(2\pi)^2\ell_{p}^3}
\frac{2\pi^2}{k}
=
2\pi \sqrt{\frac{2N}{k}}\ ,
\end{align}
where $\mathrm{L}$ denotes the radius of $S^7/\mathbb{Z}_k$. After non-trivial cancellations between bosonic and fermionic fluctuation determinants, the one-loop contribution of the non-zero modes gives
\begin{align}\label{bgt1}
Z_{1} = \frac{1}{4\sin^2(\frac{2\pi}{k})}\ .
\end{align}
The simplicity of the final result suggests that the observed cancellations between bosonic and fermionic determinants should be governed by an underlying supersymmetric structure of the BPS M2-brane fluctuation spectrum.

Motivated by the quantum M2-brane computation of \cite{Beccaria:2023ujc}, our aim is to identify the geometric structure underlying such one-loop cancellations for a broader class of Euclidean M2-brane instantons in Freund-Rubin backgrounds of the form $\ads\times \ys$. We first consider the case where $\ys$ is a weak $G_{2}$ manifold with at least one real Killing spinor. The corresponding spinor bilinear defines a nearly parallel $G_{2}$-structure $\varphi$. In contrast to the case of $G_{2}$ holonomy manifolds in \eqref{rst1}, the three-form $\varphi$ is not closed and hence is not a calibration in the usual sense. Nevertheless, the BPS condition for an M2-brane wrapping a three-dimensional submanifold $\Sigma\subset \ys$ can be written locally as the associativity condition
\begin{align}
\mathrm{vol}(\Sigma)=\varphi|_{\Sigma} \ .
\end{align}
Equivalently, such BPS M2-branes can be described as links $\Sigma=C(\Sigma)\cap \ys$ of Cayley-calibrated cones $C(\Sigma)$ in the Spin(7) cone $C(\ys)$.

We then specialise to seven-dimensional Sasaki-Einstein manifolds $\ses$, for which the cone $C(\ses)$ is a Calabi-Yau fourfold. In this case there is a distinguished class of BPS M2-branes, which we call invariant M2-branes, satisfying
\begin{align}
\mathrm{vol}(\Sigma)=\frac{1}{2}\eta\wedge d\eta\big|_{\Sigma} \ .
\end{align}
These M2-branes preserve the two real internal Killing spinors associated with the Sasaki-Einstein structure. As a result, the links $\Sigma$ inherit a Sasakian structure and are invariant under the $U(1)_r$ action generated by the Reeb vector field on $\ses$. Equivalently, $C(\Sigma)$ is a holomorphic two-cone in the Calabi-Yau fourfold $\cyf=C(\ses)$.

The special properties of invariant M2-branes extend beyond their classical volumes. For a generic associative M2-brane in a weak $G_{2}$ manifold, the nearly parallel $G_{2}$-structure produces additional mass shifts in the fluctuation operators, and the one-loop partition function is no longer expected to reduce to a topological invariant such as the Ray-Singer torsion. For invariant M2-branes $\Sigma\subset \ses$, however, the quadratic actions are controlled by Dolbeault-type operators commuting with the $U(1)_r$ action induced on $\Sigma$ by the Reeb vector field. Therefore, the fluctuation fields can be decomposed into $U(1)_r$ Fourier modes, and, for regular Reeb flow, the transversely elliptic operators on $\Sigma$ reduce mode by mode to elliptic operators on the quotient $\Sigma/U(1)_r$.

For a single invariant M2-brane with worldvolume $\widehat\Sigma=\Sigma/G\subset \ses/G$, where $G$ is a finite group commuting with the Reeb action, we find for the one-loop partition function:
\begin{align}\label{indexf}
Z_1 = \prod_{n\in \mathbb{Z}}
    \Big(\frac{2\pi n}{\ell_{\xi}}+2\Big)^{-2\widehat{\chi}^{G}_{n}}
    \prod_{n\in \mathbb{Z}}
     \Big(\frac{2\pi n}{\ell_{\xi}}\Big)^{-\widehat\chi^{G}_{n}(N\Sigma)} \ ,
\end{align}
where $\ell_{\xi}$ is the length of the Reeb circle fibre on the covering link $\Sigma$, so that a mode with Fourier label $n$ has $-i\mathcal L_\xi$ eigenvalue $2\pi n/\ell_\xi$. The two products correspond to the two multiplets appearing in the fluctuation spectrum. The multiplicities $\widehat{\chi}^{G}_{n}$ and $\widehat{\chi}^{G}_{n}(N\Sigma)$ denote the $G$-invariant contributions obtained after expanding the multiplet fields into $U(1)_r$ Fourier modes. For an isolated BPS saddle, when the relevant zero modes are absent, the product formula \eqref{indexf} gives the Riemann $\zeta$-function regularised one-loop partition function.
%around the corresponding M2-brane instanton.

The one-loop partition function \eqref{indexf}, however, requires a separate treatment when such zero modes are present, namely for the $U(1)_r$ Fourier modes satisfying $2\pi n/\ell_{\xi}+2=0$ in the first multiplet and $n=0$ in the second multiplet. In the present setup, bosonic zero modes correspond to tangent directions to the BPS moduli space, while fermionic zero modes lie in the obstruction space for integrating these infinitesimal deformations to finite BPS deformations. The difference between the numbers of bosonic and fermionic zero modes is controlled by the indices $\widehat\chi^G_n$ at $2\pi n/\ell_\xi+2=0$ and $\widehat{\chi}_{0}^{G}(N\Sigma)$. If either index is negative, so that unsaturated fermionic zero modes are present, the corresponding M2-brane saddle is not expected to contribute to the partition function without additional insertions or interaction terms that absorb these modes.

As an illustration of the general index formula \eqref{indexf}, we first apply it to invariant M2-branes in $S^{7}/\mathbb Z_{k}$. The simplest case corresponds to the $S^{3}/\mathbb Z_k\subset S^7/\mathbb{Z}_k$ instanton considered in \cite{Beccaria:2023ujc}. In this case, with $\ell_{\xi}=2\pi$ for the covering $S^{3}\subset S^7$, \eqref{indexf} gives \eqref{bgt1}. A closely related isolated saddle appears in the M-theory dual of the $(p,q)$-model \cite{Imamura:2008nn,Imamura:2008dt}. Applying the same index formula to the analogous M2-brane in $(S^7/(\mathbb{Z}_{p}\times \mathbb{Z}_{q}))/\mathbb{Z}_k$, we find that, for $pk>2$ and $qk>2$:
\begin{align}
    Z_{1}
    =
    \frac{1}{4\sin(\frac{2\pi}{pk})\sin(\frac{2\pi}{qk})} \ .
\end{align}
Up to an overall normalisation factor $4pq$, this reproduces the characteristic sine dependence conjectured from the matrix model for the leading worldsheet-instanton coefficient in the $(p,q)$-model \cite{Hatsuda:2015lpa}. Returning to the ordinary $S^7/\mathbb{Z}_k$ quotient, we also consider more general invariant M2-brane cycles. In these cases we find fermionic zero modes in the obstruction space, suggesting that the corresponding M2-brane instantons do not contribute to the partition function unless these fermionic zero modes are saturated in the path integral.

The rest of the paper is organised as follows. In Section \ref{sec2} we derive the BPS condition for M2-brane instantons in weak $G_{2}$ backgrounds and relate associative links to Cayley cones, with a special focus on invariant M2-branes in Sasaki-Einstein manifolds. We then describe invariant M2-branes in $S^{7}/\mathbb Z_{k}$ in terms of holomorphic cones in $\mathbb{C}^4/\mathbb{Z}_k$. In Section \ref{sec3} we derive the quadratic fluctuation operators around invariant M2-branes and identify the worldvolume supersymmetry organising the fluctuations into multiplets. In Section \ref{sec4} we analyse the structure of the zero modes that appear and compute the one-loop partition function in terms of equivariant indices. We apply the result to the invariant M2-branes in $S^{7}/\mathbb Z_k$ and $(S^7/(\mathbb{Z}_{p}\times \mathbb{Z}_{q}))/\mathbb{Z}_k$ corresponding to the leading matrix-model worldsheet-instanton, and then discuss more general cycles in $S^7/\mathbb{Z}_k$ for which fermionic obstruction zero modes are present. Section \ref{sec5} contains concluding remarks and comments on the holographic interpretation and open problems. Appendix \ref{A} summarises our Clifford-algebra, Killing-spinor and $G_{2}$/Sasaki-Einstein conventions. Appendix \ref{B} gives the quadratic fluctuation operators for a general associative M2-brane in a weak $G_{2}$ manifold. Appendix~\ref{bosonicinv} collects technical details on the Dolbeault-type normal-bundle operators and the bundle maps used in the fluctuation analysis around invariant M2-branes.

\section{BPS M2-branes, associative links and holomorphic cones} \label{sec2}

The simplest class of supersymmetric $\ads$ M-theory backgrounds is of Freund-Rubin type \cite{Freund:1980xh}. The eleven-dimensional geometry is a direct product of $\ads$ and a weak $G_2$ manifold $\ys$, with the four-form flux proportional to the volume form on $\ads$. The Euclidean background can be written as
\begin{align} \label{2.1aa}
ds^2_{11} = \Le^2\big(\frac{1}{4}ds^2_{\ads}+ds^2_{\ys}\big)\ ,
\qquad
F_{4} = \frac{3i}{8}\Le^3 \mathrm{vol}(\ads)\ ,
\end{align}
where $\Le$ denotes the radius of $\ys$, measured in eleven-dimensional Planck units. Since $\Le$ is the only length scale in the problem, we set $\Le/\ell_{p}=1$ in what follows and restore its dependence at the end by dimensional analysis. The superisometry algebra of the M-theory background is $\mathfrak{osp}(\mathcal{N}|4)$, corresponding to four Killing spinors on $\ads$ and $\mathcal{N}\geq 1$ %preserved
real Killing spinors on $\ys$.
Given a real Killing spinor $\psi$ on $\ys$, one can construct a nearly parallel $G_2$-structure $\varphi$, satisfying
\begin{align}
d\varphi = 4 \hodge \varphi\ , \qquad d\hodge\varphi = 0 \ ,
\end{align}
where $\ast$ is the Hodge star operator on $\ys$.  In fact, the existence of a spin structure admitting a Killing spinor \(\psi\) on a seven-dimensional manifold is equivalent to the existence of a nearly parallel $G_{2}$-structure \cite{friedrich1997nearly}.  It is also important to note that $\varphi$ does not define a calibration on $\ys$ in the sense of \cite{Harvey:1982xk}, since $\varphi$ is not closed. Nevertheless, as we discuss in this section, for an M2-brane wrapping an oriented three-dimensional submanifold $\Sigma$, the condition for preserving the Killing spinor $\psi$ can be written locally as the associativity condition
\begin{align}
\mathrm{vol}(\Sigma) = \varphi\big|_{\Sigma}\ . \label{2.2c}
\end{align}

\subsection{BPS condition for M2-brane instanton}\label{sec2.1}

A probe M2-brane in an on-shell eleven-dimensional supergravity background is described by the BST action \cite{Bergshoeff:1987cm, Bergshoeff:1987qx}, which is invariant under target-space supersymmetry and local $\kappa$-symmetry. For an M2-brane instanton in $\mathrm{AdS}_{4}\times \ys$, the worldvolume wraps a three-dimensional submanifold $\Sigma \subset \mathrm{Y}_{7}$.
%, which simplifies the BPS analysis.

For a generic Euclidean eleven-dimensional background, the Killing spinors $\epsilon$ preserved by the M2-brane are determined by the $\kappa$-symmetry projection
\begin{align} \label{2.1}
   (1-\Gamma)\epsilon = 0 \ , \qquad 
   \Gamma = \frac{i}{3!\sqrt{g}}\epsilon^{\alpha\beta\gamma}
   E^{A}(\partial_{\alpha})
   E^{B}(\partial_{\beta})
   E^{C}(\partial_{\gamma})
   \Gamma_{ABC} \ , \qquad \alpha =1,2, 3 \ ,
\end{align}
where $\Gamma_{A}$ are the eleven-dimensional Dirac matrices, $g_{\alpha\beta}$ is the induced worldvolume metric, and $E^{A}(\partial_{\alpha})$ denotes the pullback of the target-space vielbein. In the $\mathrm{AdS}_{4}\times \ys$ background, the BPS condition \eqref{2.1} for M2-branes entirely embedded in $\ys$ is conveniently expressed using the $4+7$ decomposition of the eleven-dimensional Clifford algebra
\begin{align} \label{1.12}
    &\{\Gamma_{A}, \Gamma_{B}\} = 2\delta_{AB} \ , \qquad
    \Gamma_{A}=(\Gamma_{\hat{a}}, \Gamma_{i}) \ , \qquad A = 1, \dots , 11 \ , \\
    &\Gamma_{\hat{a}} = \gamma_{\hat{a}}\otimes \id \ , 
    \qquad \hat{a} = 1, \dots, 4 \ , \\
    &\Gamma_{i} = \gamma_{\hat{5}}\otimes \gamma_{i} \ , \qquad
    \gamma_{\hat{5}} = \gamma_{\hat{1}}\gamma_{\hat{2}}\gamma_{\hat{3}}\gamma_{\hat{4}} \ , 
    \qquad i = 1, \dots, 7 \ .
\end{align}
Accordingly, an eleven-dimensional Killing spinor may be taken in factorised form\footnote{The inner product for eleven-dimensional spinors decomposes as $
    \langle \epsilon , \epsilon'\rangle  = \langle \chi, \chi'\rangle \cdot \langle \psi, \psi' \rangle$. The inner product for seven-dimensional spinors is specified in Appendix \ref{A}. For the inner product of four-dimensional spinors, we take the antisymmetric pairing $\langle \chi, \chi'\rangle = \chi^{T}C\chi', \  C^{T}= -C, \ C\gamma_{\hat{a}}C^{-1}= \gamma_{\hat{a}}^{T}$. Furthermore, at a given point a four-dimensional spinor can be decomposed with respect to $\gamma_{\hat{5}}$ chirality as $\chi = \chi_{+}\oplus \chi_{-} , \  \gamma_{\hat{5}}\chi_{\pm}=\pm \chi_{\pm} $, so that $\langle\chi, \chi'\rangle = \langle \chi_{+}, \chi_{+}'\rangle+\langle \chi_{-}, \chi_{-}'\rangle$.}
\begin{align}
    \epsilon = \chi \otimes \psi \ , \label{2.5}
\end{align}
where $\chi$ is a four-dimensional spinor on $\ads$ and $\psi$ is a seven-dimensional real spinor on $\ys$. For the background \eqref{2.1aa}, the Killing spinor equation splits into its $\mathrm{AdS}_{4}$ and $\ys$ components
\begin{align}
    &\mathcal{D} \epsilon  = \hat{D}\chi \otimes \psi + \chi \otimes D\psi=0 \ ,  \label{2.9b} \\
    &\hat{D}\chi = \hat{\nabla}\chi - i  e^{\hat{a}}\gamma_{\hat{a}}\gamma_{\hat{5}}\chi=0 \ , 
    \qquad 
    D\psi = \nabla\psi + \frac{i}{2}e^{i}\gamma_{i}\psi=0 \ , \label{2.9}
\end{align}
where $e^{\hat{a}}$ and $e^{i}$ are orthonormal coframes on $\mathrm{AdS}_{4}$ and $\ys$ respectively. For an M2-brane localised at a point in $\mathrm{AdS}_{4}$ and wrapping a three-dimensional submanifold of $\ys$, the condition \eqref{2.1} reduces to
\begin{align}
    & \chi \otimes \psi
    =
    \gamma_{\hat{5}}\chi \otimes \gamma \psi \ , \qquad
    \gamma =
    \frac{i}{3!\sqrt{g}}
    \epsilon^{\alpha\beta\gamma}
    e^{i}(\partial_{\alpha})
    e^{j}(\partial_{\beta})
    e^{k}(\partial_{\gamma})
    \gamma_{ijk} \ , \label{2.9a} \\
    & g^{\alpha\beta}e^{i}(\partial_{\beta})\gamma_{i}\gamma
    =
    \frac{i}{2!\sqrt{g}}
    \epsilon^{\alpha\beta\gamma}
    e^{i}(\partial_{\beta})e^{j}(\partial_{\gamma})\gamma_{ij} \ , 
    \qquad \gamma^{2}=\id \ . \label{2.9c}
\end{align}
%We assume that the background \(\mathrm{AdS}_{4}\times \mathrm{SE}_{7}\) preserves at least two real supercharges. 
Since the $\mathrm{AdS}_{4}$ Killing spinor derivative $\hat{D}$ does not commute with the chirality operator $\gamma_{\hat{5}}$, the projection
\begin{align}\label{2.13chiral}
(1\pm\gamma_{\hat{5}})\chi = 0 \ ,
\end{align}
cannot be imposed globally on $\mathrm{AdS}_{4}$. As noted in \cite{Gautason:2025per}, based on \cite{BenettiGenolini:2019jdz} (see also \cite{BenettiGenolini:2024lbj}), one may restrict to the fixed locus $U_{0}\subset \ads$ of the supersymmetric Killing vector, on which the chirality projection is satisfied. For more general four-dimensional solutions of 4d $\mathcal{N}=2$ gauged supergravity admitting Killing spinors, this subset may consist of isolated fixed points or two-dimensional fixed loci of the Killing vector defined by the bilinear $i\langle \chi^{c}, \gamma_{\hat{a}}\gamma_{\hat{5}}\chi\rangle$, where $\chi^c$ denotes the charge-conjugate spinor. In the case of interest, $U_{0}\subset \ads$ consists of a single point, which can be identified with the centre of $\mathrm{AdS}_{4}$, and one may choose a basis of four independent Killing spinors such that each spinor has definite $\gamma_{\hat{5}}$ chirality at this point.
\begin{comment}
\footnote{For instance, in geodesic polar coordinates $(\rho, \beta_{\alpha})$ on $\mathrm{AdS}_{4}$, with radius $1/2$, a generic Killing spinor $\chi$ satisfying \eqref{2.9} can be written as
$$
\chi = \exp(\tfrac{i}{2}\rho\gamma_{\hat{\rho}}\gamma_{\hat{5}})\big[\prod_{\alpha=1}^{3}\exp(-\tfrac{1}{2}\beta_{\alpha}\gamma_{\hat{\alpha}\,\hat{\alpha}+1})\big]\chi_{0} \ , 
$$
where $\chi_{0}$ is a constant spinor. Since $[\gamma_{\hat{5}}, \gamma_{\hat{\alpha}\,\hat{\alpha}+1}]=0$, the angular factor preserves $\gamma_{\hat{5}}$ chirality. Hence, in the limit $\rho \rightarrow 0$, the chirality of $\chi$ is determined by the chirality of the constant spinor $\chi_{0}$.}
\end{comment}

We consider the case where the M2-brane preserves an eleven-dimensional Killing spinor $\epsilon$ whose four-dimensional component $\chi$ has positive chirality at the point $U_{0}$. The case of negative chirality can be treated analogously. Once the M2-brane is localised at $U_{0}$, equation \eqref{2.9a} implies
\begin{align}
    \gamma \psi = \psi \ ,
\end{align}
so that the BPS condition is written purely in terms of the internal spinor $\psi$. With the spinor conventions summarised in Appendix \ref{A}, the unit real Killing spinor $\psi$ defines a $G_{2}$-structure
\begin{align} \label{2.11}
    \varphi = 
    \frac{i}{3!}\langle \psi, \gamma_{ijk}\psi\rangle \,
    e^{i}\wedge e^{j}\wedge e^{k} \ , \qquad ||\psi||^2 = 1 \ .
\end{align}
It follows from the Killing spinor equation on $\ys$ \eqref{2.9} that the $G_{2}$-structure $\varphi$ is nearly parallel
\begin{align} 
    &d\varphi
    =
    -\frac{1}{3!}
    \langle \psi , \gamma_{ijkl}\psi\rangle \, 
    e^{i}\wedge e^{j}\wedge e^{k}\wedge e^{l}
    = 4 \hodge\varphi \ , \label{1.24} \\
    & \nabla_{X} \varphi = \iota_{X}(\ast \varphi) \ , \qquad \forall X \in T\ys \ . \label{2.15a}
\end{align} 
By the definition of $\gamma$ in \eqref{2.9a}, the condition that the M2-brane wraps a $G_{2}$-associative three-dimensional submanifold
$\Sigma \subset \ys $ is equivalent to the BPS condition
\begin{align}\label{1.22a}
    \varphi\big|_{\Sigma}=\mathrm{vol}(\Sigma)
    \quad \Longleftrightarrow \quad
    \psi=\gamma\psi \ ,
\end{align}
where $\mathrm{vol}(\Sigma)$ denotes the induced volume form on $\Sigma$. Moreover, three-dimensional submanifolds satisfying \eqref{1.22a} saturate the local BPS bound, as follows from the standard argument \cite{Becker:1995kb, Becker:1996ay}:
\begin{align} \label{1.23}
    &\int d^{3}\sigma\sqrt{g}\, \langle (\id-\gamma)\psi, (\id-\gamma)\psi\rangle  \geq 0 
    \ \ \Longrightarrow\ \ 
    \varphi\big|_{\Sigma}\leq \mathrm{vol}(\Sigma) \ .
\end{align}

It is also possible to rewrite the BPS condition \eqref{1.22a} as a local first-order equation. To this end, it is convenient to introduce a local orthonormal frame on the spinor bundle $\mathrm{S}$ over $\ys$, adapted to the Killing spinor $\psi$. The choice of a unit Killing spinor $\psi$ defines a $G_{2}$-structure, since its pointwise stabiliser in $\mathrm{Spin}(7)$ is $G_{2}\subset \mathrm{Spin}(7)$, under which the spinor representation decomposes as $\mathbf{8}_{s}=\mathbf{1}\oplus\mathbf{7}$.
 Concretely, we consider an orthonormal frame for the spinor bundle
\begin{align}
    (\psi, \ e_{i}^{\psi}) \ , \qquad e_{i}^{\psi} = -i \gamma_{i}\psi \ ,
\end{align}
so that an arbitrary section of the spinor bundle $\theta \in \Gamma(\mathrm{S}, \ys)$ can be locally decomposed with respect to this frame. In particular, by using the identities of Appendix \ref{A}, for the spinor $\gamma \psi$, defined on an arbitrary M2-brane $\Sigma\subset \ys$, we obtain a local decomposition
\begin{align}
    &  \sqrt{g} \, \gamma \psi =\frac{1}{3!}\epsilon^{\alpha\beta\gamma}\varphi(\partial_{\alpha}, \partial_{\beta}, \partial_{\gamma})\psi+ \mathrm{A}_{i}e_{i}^{\psi}   \ , \quad   \gamma \psi \in \Gamma(\mathrm{S}, \Sigma)  \ ,  \\
    &\mathrm{A}_{i}
    = \langle \sqrt{g}\, \gamma \psi, e_{i}^{\psi}\rangle 
    = \frac{1}{3!}\epsilon^{\alpha\beta\gamma}\hodge\varphi(e_{i}, \partial_{\alpha}, \partial_{\beta}, \partial_{\gamma}) \ .
\end{align}
Since the norm of $\sqrt{g} \, \gamma \psi$ determines the volume form on $\Sigma$, one can write the volume functional of $\Sigma\subset \ys$ in the BPS form
\begin{align} \label{2.23}
    \mathrm{Vol}(\Sigma)
    = \int d^{3}\sigma\, \sqrt{g}
    = \int d^{3}\sigma\, \sqrt{ |\mathrm{A}|^{2}
    + \varphi(\partial_{1}, \partial_{2}, \partial_{3})^{2} }
    \geq \int_{\Sigma}\varphi \ ,
\end{align}
where in the last inequality we have chosen the orientation on $\Sigma$ so that, on M2-branes satisfying \eqref{1.22a}, $\varphi(\partial_{1}, \partial_{2}, \partial_{3})>0$. It follows that the BPS bound is saturated precisely on those three-manifolds for which $\mathrm{A}_{i}$ vanishes
\begin{align}
    \mathrm{A}_{i}
    = \hodge\varphi (e_{i}, \partial_{1}, \partial_{2}, \partial_{3}) = 0 \ , \label{2.24a}
\end{align}
which gives the desired BPS equation.\footnote{This equation is analogous to the Dirac-type equation introduced in \cite{Harvey:1982xk} for the calibration condition of three-manifolds associated with the \(G_{2}\)-structure in $\mathbb{R}^{7}$.}

\subsection{Cone construction} \label{sec22}

Although the BPS conditions \eqref{1.22a}, \eqref{2.24a} can be verified case by case, they are not convenient for classifying BPS solutions in a given background. At the same time, the geometry of a weak $G_{2}$ manifold $\ys$ is captured by its Riemannian cone $C(\ys)$ with holonomy $\mathrm{Hol}(C(\ys)) \subseteq \mathrm{Spin}(7)$. From the cone perspective, the spinor bundle $\mathrm{S}$ on $\ys$ is naturally identified with the restriction of the positive-chirality spinor bundle $\tilde{\mathrm{S}}^{+}$ of $C(\ys)$ to a fixed radial slice. In particular, Bär’s correspondence \cite{Bar:1993} establishes a one-to-one relation between Killing spinors $\psi$ on $\ys$ and parallel spinors $\Psi$ on $C(\ys)$. On the cone, a unit positive-chirality parallel spinor $\Psi \in \tilde{\mathrm S}^{+}$ has stabiliser $\mathrm{Spin}(7)\subset \mathrm{Spin}(8)$ and therefore defines a torsion-free $\mathrm{Spin}(7)$-structure on $C(\ys)$ or, in other words, a parallel Cayley four-form $\Phi$. Correspondingly, the fibres of the bundle $\tilde{\mathrm S}^{+}$ decompose under $\mathrm{Spin}(7)$ as $\tilde{\mathrm S}^{+} = \mathbf{1}\oplus \mathbf{7}$. 

The corresponding $G_2$-structure $\varphi$ on $\ys$ is induced from the Cayley four-form $\Phi$ on $C(\ys)$ constructed from $\Psi$. As a result, associative submanifolds of $\ys$ are realised as links of Cayley-calibrated cones in $C(\ys)$. When the cone $C(\ys)$ is Euclidean space, this relation follows from Theorem~5.6 of \cite{Harvey:1982xk} and was studied in detail for $\ys=S^{7}$ in \cite{Lotay2012Associative}.

We define the Cayley four-form on $C(\ys)$ by taking a unit positive-chirality parallel spinor $\Psi$ on $C(\ys)$ and constructing a spinor bilinear
\begin{align}\label{1.25}
    &\Phi = -\frac{1}{4!}\langle \Psi, \tilde{\gamma}_{mnpq}\Psi \rangle \, \tilde{e}^{m}\wedge\tilde{e}^{n}\wedge\tilde{e}^{p}\wedge\tilde{e}^{q} \ ,   \quad d\Phi = 0 \ ,
\end{align}
where $\tilde{e}^{m}$ is an orthonormal coframe on $C(\ys)$ and $\tilde{\gamma}_{m}$ satisfy the eight-dimensional Clifford algebra. The overall minus sign is fixed by our orientation conventions. The metric on the cone $C(\ys)$ is then
\begin{align}
    d\tilde{s}_{C(\ys)}^2 = dr^2+r^2ds^2_{\ys} \ , \quad r\in \mathbb{R}_{+} \ .
\end{align}
The Euler, or radial, vector field $\zeta$ on $C(\ys)$ in the cone coordinates takes the form
\begin{align}
    \zeta = r\partial_{r} \ , \qquad |\zeta|^2=r^2 \ .
\end{align}
Using the identification of Clifford modules discussed in Appendix \ref{A}, the spinor bundle on $\ys$ is identified with the restriction of the positive-chirality spinor bundle $\tilde{\mathrm{S}}^{+}$ of $C(\ys)$ to the slice $r=1$. Under this identification
\begin{align}
    \psi = \Psi\big|_{r=1}\ , \qquad \gamma_i = i \tilde\gamma_r \tilde\gamma_i \ ,
\end{align}
where $\gamma_i$ are the seven-dimensional gamma matrices acting on $\psi$. Using further the expression \eqref{2.11} for the $G_{2}$-structure on $\ys$, one finds
\begin{align}\label{1.28}
    \iota_{\zeta}\Phi\big|_{r=1}
    = -\frac{r^4}{3!}\langle \Psi,\tilde\gamma_r\tilde\gamma_{ijk}\Psi\rangle\, e^i\wedge e^j\wedge e^k\big|_{r=1}
    = \varphi \ .
\end{align}
The Cayley form $\Phi$ is homogeneous of degree four with respect to the cone scaling, and it is self-dual because $\Psi$ has positive chirality. It follows that on the cone $\Phi$ takes the standard form
\begin{align}
    \Phi = r^{3}\,dr\wedge \varphi + r^{4}\hodge\varphi \ . \label{2.32}
\end{align}

The relation \eqref{2.32} between $\Phi$ and $\varphi$ makes the correspondence between calibrated cones in $C(\ys)$ and BPS M2-branes on $\ys$ explicit. We call a fourfold $C(\Sigma) \subset C(\ys)$ a cone if it is invariant under the dilations $f_{t}$ generated by the radial vector field $\zeta$:
\begin{align}
    f_{t}\big(C(\Sigma)\big)=C(\Sigma)\ , \quad t\in \mathbb{R}\ .
\end{align}
Assuming that $C(\Sigma)$ is smooth away from the apex, %and intersects $\ys$ transversely
its link
$\Sigma=C(\Sigma)\cap \ys$ is a compact smooth three-manifold. The induced cone metric and volume form on $C(\Sigma)$ are
\begin{align}
    &ds^2_{C(\Sigma)} = dr^2+r^2ds_{\Sigma}^2 \ , \qquad \mathrm{vol}(C(\Sigma)) = r^3\, dr\wedge \mathrm{vol}(\Sigma) \ . \label{2.32aa}
\end{align}
From \eqref{2.32} and \eqref{2.32aa} it follows that the cone $C(\Sigma)$ is Cayley if and only if its link $\Sigma$ is associative with respect to $\varphi$.

Among seven-dimensional weak $G_2$ manifolds, a special subclass is given by Sasaki-Einstein manifolds $\ses$, whose metric cone $C(\ses)$ is a Calabi-Yau fourfold $\cyf$ (see e.g. \cite{Sparks:2010sn}). Equivalently, the holonomy of the Riemannian cone satisfies $\mathrm{Hol}(\cyf)\subseteq SU(4)\subset \mathrm{Spin}(8)$.
After complexification, the positive-chirality spinor representation decomposes under the subgroup $SU(4)\times U(1)_{r}\subset \mathrm{Spin}(8)$ as
\begin{comment}
\footnote{This realises the standard isomorphism of positive-chirality spinors with antiholomorphic forms
$$
    \tilde{\mathrm{S}}^{+}{\otimes_{\mathbb{R}}\mathbb{C}} \cong \Lambda^{0,\mathrm{even}}T^{\ast}\cyf 
    = \Lambda^{0,0}T^{\ast}\cyf \oplus \Lambda^{0,2}T^{\ast}\cyf \oplus \Lambda^{0,4}T^{\ast}\cyf \ ,
$$
where $\Lambda^{0,0}T^{\ast}\cyf$ and $\Lambda^{0,4}T^{\ast}\cyf$ correspond to the singlets $\mathbf{1}_{\pm 2}$.}
\end{comment}
\begin{align}
    \mathbf{8}_{s}^{\mathbb{C}}=\mathbf{1}_{+2} \oplus \mathbf{6}_{0}\oplus \mathbf{1}_{-2} \ ,
\end{align}
where the subscripts denote the $U(1)_{r}$ charges. This implies that there are two $SU(4)$-invariant %complex
parallel spinors on $\cyf$. To describe them explicitly, one can consider the action $\mathcal{J}$ of the complex structure $J$ on the complexified spinor module.\footnote{Any element $A \in \mathfrak{so}(8)$ can be lifted to an element $\mathcal{A}\in \mathrm{Cl}_{8}$ via $\mathcal{A} = \frac{1}{4}A_{mn} \tilde{\gamma}_{mn}$, where $A_{mn} = \langle A\tilde {e}_{m}, \tilde{e}_{n} \rangle$. On $\cyf$ with $J$-compatible metric this gives $\mathcal{J}= \frac{1}{4}\Omega_{mn}\tilde{\gamma}_{mn}$, where $\Omega_{mn}$ are the components of the Kähler form.} 
On $\tilde{\mathrm{S}}{\otimes_{\mathbb{R}}\mathbb{C}}$ it can be diagonalised as
\begin{align}
    \mathcal{J} = \frac{1}{2}\sum_{a=1}^{4}\hat{\mathrm{s}}_{a} \ , \qquad \hat{\mathrm{s}}_{a}^2 = -1 \ ,
\end{align}
where the operators $\hat{\mathrm{s}}_{a}$ are mutually commuting elements of the Clifford algebra corresponding to rotations in four orthogonal complex planes in $T\cyf$, with eigenvalues $\pm i$. Restricting to the positive-chirality spinors $\tilde{\mathrm{S}}^{+}{\otimes_{\mathbb{R}}\mathbb{C}}$, the allowed eigenvalues of $\mathcal{J}$ on $\tilde{\mathrm{S}}^{+}{\otimes_{\mathbb{R}}\mathbb{C}}$ are $\{2i, 0, -2i\}$, with multiplicities $1,6,1$, respectively. The two complex-conjugate singlets span a two-dimensional real space of $SU(4)$-invariant parallel spinors on $\cyf$, which can be expressed as a pair of real, orthogonal spinors $\Psi_{1,2}$, satisfying
\begin{align}
    \mathcal{J}\Psi_{1} = 2\Psi_{2} \ , \quad \mathcal{J}\Psi_{2} = -2\Psi_{1} \ , \quad \tilde{\nabla}\Psi_{1,2} = 0 \ . \label{2.35}
\end{align}
Equivalently, the Kähler form on $\cyf$ is given in terms of the parallel spinors $\Psi_{1,2}$ by\footnote{In the conventions of Appendix \ref{A}, the normalisation coefficient can be checked explicitly using \eqref{2.35} and the eight-dimensional Fierz identity for commuting positive-chirality spinors
 $$\Psi'\Psi^{T} =\frac{1}{8} \big((\Psi^{T}\Psi')\id -\frac{1}{2!}(\Psi^{T}\tilde\gamma_{mn}\Psi')\tilde\gamma_{mn}+\frac{1}{2\cdot 4!}(\Psi^{T}\tilde\gamma_{mnpq}\Psi')\tilde\gamma_{mnpq}\big)P_{+} \ . $$}
\begin{align}
    \Omega = \frac{1}{2}\Omega_{mn}\, \tilde e^{m}\wedge \tilde{e}^{n} = \frac{1}{2}\langle\Psi_{2}, \tilde\gamma_{mn}\Psi_{1}\rangle \, \tilde{e}^{m}\wedge \tilde{e}^{n} \ . \label{2.36}
\end{align}
The two $SU(4)$-invariant parallel spinors 
$\Psi_{1,2}$ on $\cyf$, through Bär's correspondence, descend to two 
real Killing spinors on $\ses$. The Cayley four-form constructed from $\Psi_{1,2}$ is $SU(4)$-invariant and, with an appropriate normalisation, can be written in terms of the Kähler form $\Omega$ and the holomorphic $(4,0)$-form $\Upsilon$ on $\cyf$ as
\begin{align}
    \Phi_{1,2} = \frac{1}{2}\Omega \wedge \Omega \pm \mathrm{Re}\,\Upsilon \ .
\end{align}
The corresponding $G_2$-structures on $\ses$ can be conveniently expressed in terms of the canonical contact form $\eta$ and the holomorphic $(3,0)$-form $\alpha$ on the contact distribution with respect to the transverse complex structure $\phi$. We use conventions in which the characteristic vector field $\xi$ and the Kähler form $\Omega$ are given by
\begin{align}
    \xi = J(\zeta) \ , \qquad \Omega = \frac{1}{2}d(r^2\eta) \ .
\end{align}
The Sasaki-Einstein manifold $\mathrm{SE}_7$ is then endowed with a Sasakian structure $(\eta,\xi,\phi, ds^2_{\mathrm{SE}_7})$ satisfying
\begin{align}
    \phi^2 = -\id + \eta \otimes \xi \ , \qquad ds^2_{\mathrm{SE}_7} = \frac{1}{2} d\eta(\, \cdot\, , \phi \, \cdot\, ) + \eta \otimes \eta \ , \label{3.30c}
\end{align}
as well as
\begin{align}
    (\nabla_X \phi)(Y) = \eta(Y)X - \langle X,Y\rangle \xi \ , \quad \nabla_X \xi = \phi(X) \ , \quad \forall X,Y \in \Gamma(T\mathrm{SE}_7) \ , \label{3.31c}
\end{align}
where $\nabla$ is the Levi-Civita connection determined by the metric \eqref{3.30c}. The two $G_2$-structures $\varphi_{1,2}$, corresponding to the Killing spinors on $\ses$ obtained by the radial restriction $\psi_{1,2}=\Psi_{1,2}|_{r=1}$, satisfy
\begin{align}
    &\varphi_{1,2} = \iota_{\zeta}\Phi_{1,2}\big|_{r=1} = \frac{1}{2}\eta \wedge d\eta \pm \mathrm{Re}\,\alpha \ , \label{2.40} \\
    &\eta = \iota_{\zeta}\Omega \big|_{r=1} \ , \qquad \alpha = \iota_{\zeta}\Upsilon \big|_{r=1} \ . \label{2.40aa}
\end{align}

At this point, we note that among oriented associative M2-branes in $\ses$ there are those preserving the entire $S^1$-family of unit Killing spinors generated by linear combinations of $\psi_{1,2}$. Equivalently, this means that the same oriented three-cycle is associative
for every $G_{2}$-structure in the $S^{1}$-family
\begin{align}
    \varphi_{\Theta}
    =
    \frac{1}{2}\eta \wedge d\eta
    +
    \mathrm{Re}\big(e^{i\Theta}\alpha\big) \ . 
\end{align}
The volume form of these M2-branes $\Sigma$ satisfies
\begin{align}
    \vol(\Sigma) = \frac{1}{2}\eta \wedge d\eta \big|_{\Sigma} \ , \qquad \alpha\big|_{\Sigma} = 0 \ ,  \label{2.43}
\end{align}
and the corresponding cone $C(\Sigma)$ is a holomorphic cone in $\cyf$. Such M2-brane instantons are somewhat analogous to the M2-brane configurations describing BPS Wilson loops in $\mathcal{N}=2$ Chern-Simons-matter theories on $S^3$ \cite{Farquet:2013cwa, Farquet:2014bda, Boido:2023ojv, Gautason:2025bft}. Concretely, for backgrounds of Freund-Rubin type, M2-brane geometries dual to BPS Wilson loops are of the form $\mathrm{AdS}_{2}\times S^{1}$, where the hyperbolic space $\mathrm{AdS}_{2}$ is a totally geodesic subspace of $\ads$, while $S^{1}$ is identified with the M-theory circle. The $S^{1}$ geodesics preserving both internal spinors $\psi_{1,2}$ are then Reeb geodesics, calibrated by the contact form in the sense that $\vol(S^{1})=\eta|_{S^{1}}$.

The M2-brane worldvolumes satisfying \eqref{2.43} correspond to invariant
submanifolds $\Sigma \subset \ses$.\footnote{A similar generalised calibration condition recently appeared in \cite{BenettiGenolini:2026hmz} for supersymmetric probe D3-branes wrapping three-cycles $\Sigma\subset \mathrm{SE}_5$ and a circle in the external black-hole geometry, with the volume form on the internal three-cycle satisfying
$\vol(\Sigma)=\frac12\eta\wedge d\eta\big|_{\Sigma}$.} The tangent bundle of $\Sigma$ contains the Reeb
vector field and is preserved by the contact endomorphism $\phi$:
\begin{align}
    \xi \in \Gamma(T\Sigma) \ ,
    \qquad
    \phi(T\Sigma)\subset T\Sigma  \ . \label{2.44}
\end{align}
The invariant submanifold $\Sigma$ is minimal and inherits a Sasakian
structure from $(\eta,\xi,\phi,ds^2_{\ses})$ (see e.g. \cite{Blair2002} and references therein). Importantly, the
worldvolume theory contains the two real supercharges associated with the
preserved Killing spinors $\psi_{1,2}$, together with the $U(1)_r=S^{1}_{\xi}$ symmetry
generated by $\xi$. As we show in Section \ref{sec4}, this in turn allows one to reduce the computation
of the one-loop partition function to the equivariant indices of the transversely elliptic operators appearing in the quadratic action for fluctuations around the supermembrane embedding.

In contrast, the special Legendrian M2-brane worldvolumes $\Sigma_L$, arising as links of special Lagrangian cones
$C(\Sigma_L)\subset \cyf$, for a suitable choice of phase satisfy
\begin{align}
    \eta\big|_{\Sigma_L} = 0 \ , \qquad
    \varphi_{1,2}\big|_{\Sigma_L} = \pm \vol(\Sigma_L) \ ,
\end{align}
and, depending on the orientation, preserve precisely one of the Killing spinors $\psi_{1,2}$. For special Legendrian submanifolds, the contact endomorphism $\phi$ identifies $T\Sigma_L$ with a rank-three subbundle of the normal bundle, while the Reeb vector field provides the remaining normal direction. Thus the normal fluctuations decompose into a scalar mode along $\xi$ and a one-form on $\Sigma_L$, which simplifies the analysis of bosonic quadratic fluctuations \cite{Kawai2017HomAssoc}. However, since special Legendrian M2-branes preserve only one of the Killing spinors $\psi_{1,2}$ and are not invariant under the $U(1)_r$ action, they are not expected to contribute to the free energy computed using the localisation technique on the gauge-theory side, and will therefore play a less important role below.

\subsection{Example: M2-brane instantons in $S^7/\mathbb{Z}_k$} \label{sec2.3}

To illustrate the cone construction described above, we consider the example relevant for the duality between three-dimensional $\mathcal{N}=6$ ABJM theory \cite{Aharony:2008ug} and M-theory on $\ads\times S^7/\mathbb{Z}_{k}$. The Riemannian cone over $S^7/\mathbb{Z}_{k}$ is the orbifold $C(S^7/\mathbb{Z}_{k}) = \mathbb{R}^{8}/\mathbb{Z}_{k}$ with the flat metric. To consider the cone as a Calabi-Yau fourfold, we choose a complex structure on $\mathbb{R}^8$ such that $\mathbb{R}^8/\mathbb{Z}_{k}\simeq \mathbb{C}^4/\mathbb{Z}_{k}$ with $\mathbb{Z}_{k}$ acting on $\mathbb{C}^4$ coordinates as
\begin{align}\label{2.47f}
    \omega^{l}\cdot (z_{1},z_{2},z_{3},z_{4})
    =
    \big(e^{2\pi i l/k}z_{1},
        e^{2\pi i l/k}z_{2},
        e^{-2\pi i l/k}z_{3},
        e^{-2\pi i l/k}z_{4}\big) \ ,
    \qquad \omega^{l}\in \mathbb{Z}_{k} \ .
\end{align}
With this choice of $\mathbb{Z}_{k}$ action, the holomorphic four-form
\begin{align}
    \Upsilon = dz_{1}\wedge dz_{2}\wedge dz_{3}\wedge dz_{4} \ ,
\end{align}
is invariant under $\mathbb{Z}_{k}$ and descends to the quotient.

The link $S^7/\mathbb{Z}_{k}$ can be viewed as a toric Sasaki-Einstein manifold with a torus action $\mathbb{T}^4 = U(1)^4$. We denote by $U(1)_{r}\subset \mathbb{T}^4$ the Reeb action with weights $(1,1,1,1)$, and by $U(1)_{b}\subset \mathbb{T}^4$ the action with weights $(1,1,-1,-1)$, so that $\mathbb{Z}_{k}\subset U(1)_{b}$. The $\mathbb{Z}_{k}$ action is free on
$S^{7}$ and hence the quotient $S^{7}/\mathbb{Z}_{k}$ is smooth. Using the Hopf fibration $S^{7}\to \mathbb{P}^{3}$, the coordinates $z_{a}$ can be viewed as homogeneous coordinates on $\mathbb{P}^{3}$, and the $\mathbb{Z}_{k}\subset U(1)_{b}$
action induces the projective action
\begin{align}
    \omega^{l}\cdot [z_{1},z_{2},z_{3},z_{4}]
    =
    [e^{2\pi i l/k}z_{1},
     e^{2\pi i l/k}z_{2},
     e^{-2\pi i l/k}z_{3},
     e^{-2\pi i l/k}z_{4}] \ .
    \label{2.48}
\end{align}
The projective space $\mathbb{P}^{3}$ appearing here is the Kähler base of the Hopf fibration of $S^{7}$ associated with the Reeb circle $U(1)_{r}$. It should not
be confused with the type IIA base obtained by reduction along the M-theory circle $U(1)_{b}$. For $k>2$, the induced projective $\mathbb{Z}_{k}$ action has fixed loci on $\mathbb{P}^{3}$, which will play an important role below.

The simplest class of invariant M2-brane instantons is given by the toric edges of the moment polytope. Namely, the action of $\mathbb{T}^{4}=U(1)^{4}$ on $\mathbb{C}^{4}$, equipped with the Kähler form $\Omega$, defines the moment map
\begin{align}
    \mu:\mathbb{C}^4 \to \mathfrak{t}_{4}^{*}  \ , \quad \mu_{a}(z)=\frac{1}{2}|z_{a}|^2\geq0 \ , \quad a=1, \dots, 4 \ .
\end{align}
The image of the unit link $S^7$ under the moment map is the compact convex polytope $H(\xi)=\mu(S^7)$:
%, obtained by intersecting the moment cone with the characteristic hyperplane determined by the Reeb vector:
\begin{align}
    H(\xi) = \big\{y_{a}\geq 0 \, \big| \, \sum_{a=1}^{4}y_{a} = \frac{1}{2}\big \} \ .
\end{align}
The edges of $H(\xi)$ correspond to the holomorphic two-planes
\begin{align}
    L_{ab} = \{z_{a} = z_{b} = 0\} \ , \qquad a\neq b \ .
\end{align}
The two-planes $L_{ab}$ are setwise preserved by the $\mathbb{Z}_{k}$ action and the corresponding links are lens spaces of unit radius
\begin{align}
    \widehat\Sigma_{ab} =( L_{ab} \cap S^7)/\mathbb{Z}_{k} \simeq S^3/\mathbb{Z}_k \ , \qquad \mathrm{Vol}(\widehat\Sigma_{ab}) = \frac{2\pi^2}{k} \ .
\end{align}
Restoring the radius $\Le$ and the M2-brane tension
$T_{2}=1/(2\pi)^{2}\ell_{p}^{3}$, the classical action of any of these toric M2-branes is
\begin{align}
    T_{2}\mathrm{Vol}(\widehat\Sigma_{ab})
    =
    \frac{2\pi^{2}}{k}\frac{\Le^{3}}{(2\pi)^{2}\ell_{p}^{3}}
    =
    2\pi\sqrt{\frac{2N}{k}} \ ,
\end{align}
where we have used the AdS/CFT dictionary \cite{Aharony:2008ug}. This agrees with the classical M2-brane instanton action appearing in \cite{Park:2020hgt, Beccaria:2023ujc}.

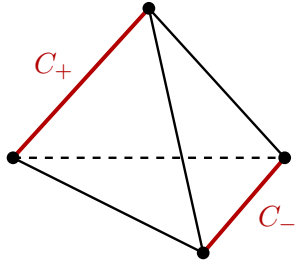
\begin{figure}[t]
    \centering
\begin{tikzpicture}[
    scale=0.9,
    vertex/.style={circle, fill=black, inner sep=1.7pt},
    edge/.style={black, line width=0.9pt},
    hiddenedge/.style={black, dashed, line width=0.9pt},
    rededge/.style={draw=red!70!black, line width=1.5pt}
]

% vertices of a projected tetrahedron
\coordinate (v1) at (0,2.2);
\coordinate (v2) at (-2,0);
\coordinate (v3) at (2,0);
\coordinate (v4) at (0.8,-1.4);

% visible black edges
\draw[edge] (v1) -- (v2);
\draw[edge] (v1) -- (v3);
\draw[edge] (v1) -- (v4);
\draw[edge] (v2) -- (v4);
\draw[edge] (v3) -- (v4);

% back edge dashed
\draw[hiddenedge] (v2) -- (v3);

% highlight two opposite edges in dark red
\draw[rededge] (v1) -- (v2);
\draw[rededge] (v3) -- (v4);

% fat vertices
\node[vertex] at (v1) {};
\node[vertex] at (v2) {};
\node[vertex] at (v3) {};
\node[vertex] at (v4) {};

% labels
\node[red!70!black] at (-1.4,1.35) {$C_{+}$};
\node[red!70!black] at (1.9,-0.9) {$C_{-}$};

\end{tikzpicture}

 \caption{The toric diagram of $\mathbb{C}^{4}$. The red edges correspond to the holomorphic two-planes $C_{+}\equiv L_{34}$ and $C_{-}\equiv L_{12}$.}
    \label{fig:toric-polytope-cpm}
\end{figure}

More precisely, although the volumes of the corresponding links are the same, different two-planes $L_{ab}$ may realise the lens space quotient through different induced actions of $\mathbb{Z}_{k}$ on $S^3$. In particular, among the two-planes $L_{ab}$, the planes $C_{+}\equiv L_{34}$ and $C_{-}\equiv L_{12}$ are distinguished by the fact that the $U(1)_{r}$-orbits coincide with the $U(1)_{b}$-orbits along them, with the same and opposite orientations respectively. In terms of the moment map, this implies that
\begin{align} \label{2.55}
    \langle \mu|_{C_{\pm}},\mathfrak{t}_{U(1)_{r}}\rangle
    =
    \pm
    \langle \mu|_{C_{\pm}},\mathfrak{t}_{U(1)_{b}}\rangle \ ,
\end{align}
where $\mathfrak{t}_{U(1)_{r,b}}\in \mathfrak{t}_{4}$ denote the Lie-algebra elements generating $U(1)_{r,b}$. Equivalently, the corresponding projective curves
\begin{align}
    B_{+} = \{z_{3}=z_4=0\} \subset \mathbb{P}^3 \ , \qquad  B_{-} = \{z_{1}=z_2=0\} \subset \mathbb{P}^3 \ ,
\end{align}
are fixed components of the projective $\mathbb{Z}_{k}$ action \eqref{2.48}. In contrast, for the remaining four edges of the polytope, the corresponding projective curves are not fixed components of the projective action, but contain isolated fixed points. These fixed points are the intersection points of the projective curve with the fixed loci $B_{\pm}$, and correspond to the two vertices of the toric polytope that are the endpoints of the edge (see Figure~\ref{fig:toric-polytope-cpm}).

We now pass from toric edges to more general invariant holomorphic cones. By the cone construction of Section \ref{sec22}, a holomorphic two-cone $C(\Sigma)\subset\mathbb C^4$ defines an invariant link
\begin{align}
\Sigma=C(\Sigma)\cap S^7  \ .
\end{align}
Since $\Sigma$ is required to be invariant under the $U(1)_r$ Reeb action, it determines a projective curve $B\subset\mathbb P^3$, which is the base of the circle fibration $\Sigma\to B$. With the unit-radius normalisation of $S^7$, the Reeb vector has unit norm and its orbits on the link $\Sigma$ are Hopf fibres of length $\ell_{\xi} = 2\pi$. We shall restrict to irreducible curves $B$, corresponding to single M2-brane instantons, and require $B$ to be preserved setwise by the projective $\mathbb Z_k$ action
\begin{align}
s\cdot B=B,\qquad \forall s\in\mathbb Z_k \ .
\end{align}
For such links the $\mathbb Z_k$ action on $\Sigma\subset S^7$ is free, so that $\widehat\Sigma=\Sigma/\mathbb Z_k$ is smooth, and
\begin{align}
\mathrm{Vol}(\widehat\Sigma)
=
\frac{1}{k}\mathrm{Vol}(\Sigma)
=
\frac{2\pi^2}{k}\deg B \  ,
\qquad
\deg B=\frac{1}{2\pi}\int_B d\eta \ .
\end{align}
An irreducible invariant curve $B$ either coincides with one of $B_\pm$, or intersects $B_\pm$ in a finite, possibly empty, set of points. These intersection points are fixed points of the induced projective $\mathbb Z_k$ action on $B$. The remaining four toric edges give the simplest examples of invariant curves with such finite intersections, while less trivial examples can be obtained by considering higher-degree curves.\footnote{For example, restricting to the $\mathbb Z_k$-invariant hyperplane $z_4=0$, one may consider the Brieskorn-type cone (see \cite{Brieskorn1966, Milnorsingular}) $C(\Sigma)=\{z_1^d+z_2^d+z_3^d=0,\ z_4=0\}\subset\mathbb C^4$, whose projective base is the degree $d$ Fermat curve. This cone is $\mathbb Z_k$-invariant provided $\omega^{2d}=1$, so the minimal positive degree in this family is $d=k/2$ for even $k$ and $d=k$ for odd $k$. The quotient link has volume $\mathrm{Vol}(\widehat\Sigma)=2\pi^2 d/k$, and for the minimal even $k$ case this gives $T_2\mathrm{Vol}(\widehat\Sigma)=\pi\sqrt{2kN}$. The curve has genus $g=\frac12(d-1)(d-2)$, does not intersect $B_-$,
and intersects $B_+$ in the $d$ fixed points
$(z_2/z_1)^d=-1$, $z_3=z_4=0$.}

\section{Quadratic fluctuations around invariant M2-branes}\label{sec3}

We now turn to the quadratic fluctuations around the BPS M2-brane saddles described in the previous section. For a general associative M2-brane preserving a single Killing spinor of a weak $G_{2}$ manifold, the quadratic bosonic and fermionic actions are given in Appendix \ref{B}. In the main text we specialise to invariant, or Sasakian, M2-brane three-cycles $\Sigma\subset\ses$ satisfying \eqref{2.43}, which preserve the two real internal Killing spinors associated with the Sasaki-Einstein structure.

For such invariant M2-branes, the quadratic fluctuations organise into two types of multiplets, corresponding respectively to fluctuations in the $\ads$ directions and to normal fluctuations of $\Sigma$ inside $\ses$. Below, we derive the relevant quadratic bosonic and fermionic actions in $\mathrm{AdS}_{4}\times\ses$. We first write the fluctuation operators in a unitary frame adapted to the decomposition $T\Sigma=\mathbb R\xi\oplus T_{H}\Sigma$, and then reformulate them in terms of horizontal Dolbeault-type operators associated with the two multiplets. In Section \ref{sec3.3}, we also present the on-shell worldvolume supersymmetry transformations that will be used in the subsequent one-loop analysis.

\subsection{Bosonic fluctuations}\label{sect3.1}

As in the previous section, we consider eleven-dimensional backgrounds of Freund-Rubin type and, for definiteness, take the non-compact factor to be global $\mathrm{AdS}_{4}$. Since the BPS M2-branes we consider are point-like in $\mathrm{AdS}_{4}$, the quadratic fluctuations do not couple to the background eleven-dimensional four-form flux. The eleven-dimensional background is the direct product $\mathrm{AdS}_{4}\times \mathrm{SE}_{7}$, and we take $\Sigma\subset \mathrm{SE}_{7}$. Consequently, the only non-trivial contributions to the extrinsic geometry arise from the embedding into $\mathrm{SE}_{7}$. The normal bundle connection and extrinsic curvature in the $\mathrm{AdS}_{4}$ directions are trivial, and the eleven-dimensional curvature decomposes as
\begin{align}
R_{11} = R_{\mathrm{AdS}_{4}}\oplus R_{\mathrm{SE}_{7}} \, .
\end{align}
It follows that the quadratic contribution from the four real bosonic fluctuations in $\mathrm{AdS}_{4}$ is
\begin{align}
    \frac{1}{2}\int d^3\sigma \, \sqrt{g} \, \langle x^{\hat{a}}, \Delta_{\Sigma}x^{\hat{a}}\rangle \ , \label{rw1} 
\end{align}
where $\Delta_{\Sigma}$ denotes the scalar Laplacian on $\Sigma$. These modes are the Goldstone bosons associated (in Euclidean signature) with the symmetry breaking $SO(4,1)\to SO(4)$.

The quadratic action for bosonic fluctuations normal to an invariant submanifold
$\Sigma \subset \ses$ follows from the general formula \eqref{bosonicmain}. The invariant submanifolds we consider are associative with respect to the
$S^{1}$-family of $G_{2}$-structures
\begin{align}
    \varphi_{\Theta}
    =
    \frac{1}{2}\eta \wedge d\eta
    +
    \mathrm{Re}\big(e^{i\Theta}\alpha\big) \ ,
\end{align}
where $\Theta$ is a constant phase. 
%As shown below, the action for normal fluctuations is independent of $\Theta$. 
The Dirac-type operator appearing in the quadratic action is defined by
\begin{align}
    \big\langle \mathrm{W},\slashed{\mathfrak{D}}\mathrm{V}\big\rangle
    =
    g^{\alpha\beta}\varphi_{\Theta}
    (\partial_{\alpha}, \nabla_{\beta}^{\perp}\mathrm{V}, \mathrm{W}) \ ,
    \qquad
    \mathrm{V},\mathrm{W}\in \Gamma(N\Sigma) \ , \label{3.2d}
\end{align}
and can be conveniently expressed in a unitary frame adapted to the invariant
submanifold $\Sigma$. 

The tangent bundle of a Sasaki-Einstein manifold admits a canonical decomposition
\begin{align}
    T\ses = \mathbb{R}\xi \oplus \ker \eta \ , 
\end{align}
where $\mathbb{R}\xi$ denotes the trivial line bundle spanned by the Reeb vector
field $\xi$. On the horizontal subbundle $\ker\eta$, it is convenient to introduce a local
unitary frame adapted to the transverse complex structure, such that
\begin{comment}\footnote{Explicitly, one can locally choose a real orthonormal frame $(\xi, e_{\mu}, \phi(e_{\mu}))$ on $\ses$. The corresponding unitary frame on $\ker\eta\otimes_{\mathbb{R}}\mathbb{C}$ is
$$
\mathrm{e}_{\mu}
=
\frac{1}{\sqrt{2}}\big(e_{\mu}-i\phi(e_{\mu}) \big),
\qquad
\mathrm{e}_{\bar\mu}
=
\frac{1}{\sqrt{2}}\big(e_{\mu}+i\phi(e_{\mu}) \big) \ .
$$}
\end{comment}
\begin{align}
 & \phi (\mathrm{e}_{\mu}) = i\mathrm{e}_{\mu} \ , \quad
   \phi(\mathrm{e}_{\bar\mu}) = -i\mathrm{e}_{\bar\mu} \ , \quad
   \mu =1,2,3 \ , \\ \label{unitfr}
 & \langle \mathrm{e}_{\mu} , \mathrm{e}_{\bar\nu}\rangle=\delta_{\mu\bar\nu} \ ,
   \quad
   \langle \xi , \mathrm{e}_{\mu}\rangle
   =
   \langle \xi , \mathrm{e}_{\bar\mu}\rangle
   =
   0 \ .
\end{align}
If the M2-brane wraps an invariant submanifold $\Sigma\subset \ses$, so that
$\xi\in \Gamma(T\Sigma)$ and $\phi(T\Sigma)\subset T\Sigma$, then
$\phi$ preserves both $T\Sigma$ and $N\Sigma$. It is therefore convenient to choose the unitary frame locally so that
\begin{align}
    &T\ses\big|_{\Sigma} = T\Sigma \oplus N\Sigma \ , \\
    &T\Sigma  = \mathbb{R}\xi \oplus \{\lambda\eun+\bar\lambda \bar\eun \, |\, \lambda \in \mathbb{C} \} \ , \quad N\Sigma = \{v^{a}\nn_{a}+\bar{v}^{\bar a}\nn_{\bar{a}} \, | \, v^{a}\in \mathbb{C}\} \ .  \label{unitfr1}
\end{align}
Writing 
\begin{align}
    T_H\Sigma=T\Sigma\cap\ker\eta \ ,
\end{align}
we denote by $T_H^{1,0}\Sigma$ and $T_H^{0,1}\Sigma$ the subbundles of
$T_H\Sigma\otimes_{\mathbb{R}}\mathbb C$ corresponding to the $+i$ and
$-i$ eigenvalues of $\phi$, respectively. Similarly, $N^{1,0}\Sigma$ and $N^{0,1}\Sigma$ denote the corresponding
subbundles of the complexified normal bundle $N\Sigma\otimes_{\mathbb{R}}\mathbb{C}$. For an invariant submanifold, the covariant derivative along $\xi$ preserves
the splitting $T\ses|_\Sigma=T\Sigma\oplus N\Sigma$. Therefore, for the Levi-Civita connection $\nabla$:
\begin{align}
    \nabla_\xi^{T}=\nabla_\xi
    \quad\text{on } \Gamma(T\Sigma) \ ,
    \qquad
    \nabla_\xi^{\perp}=\nabla_\xi
    \quad\text{on } \Gamma(N\Sigma) \ . \label{3.9b}
\end{align}
%This follows from $\nabla_{X}\xi=\phi(X)$, the fact that brackets of tangent vector fields are tangent, and the invariance property $\phi(T\Sigma)\subset T\Sigma$. 
It is also useful to express the covariant derivative along $\xi$ in terms
of the Lie derivative along the Reeb vector field. For any vector field $X$ on $\ses$, using $\nabla_X\xi=\phi(X)$, one finds
\begin{align}
    \mathcal{L}_{\xi}X
    =
    \nabla_{\xi}X-\nabla_X\xi
    =
    \nabla_{\xi}X-\phi(X) \ . \label{3.10aa}
\end{align}
In the adapted unitary frame, the Dirac-type operator \eqref{3.2d} can be
expressed in components as
\begin{align}
    \langle \nn_{a}, \slashed{\mathfrak{D}}\mathrm{V}\rangle = -i\nabla_{\xi}\bar{v}_{a}-\sqrt{2}\,e^{i\Theta}\epsilon_{ab}\nabla_{\bar \eun}^{\perp}v^{b} \ , \qquad \delta^{a\bar{b}}\langle \nn_{\bar b}, \slashed{\mathfrak{D}}\mathrm{V}\rangle = i\nabla_{\xi}v^{a}-\sqrt{2} \,e^{-i\Theta}\epsilon^{ab}\nabla_{\eun}^{\perp}\bar v_{b} \ ,  \label{3.13d}
\end{align}
where the indices are raised and lowered with $\delta_{a\bar b}$ and the orientation of the adapted unitary normal frame is chosen so that\footnote{The absolute value of the normalisation coefficient in
$\alpha(\eun,\nn_{a},\nn_{b})$ is fixed by the metric normalisation and can be checked either by
specialising to $\ses=S^{7}$, with cone $\cyf=\mathbb{C}^{4}$, or by expressing
$\alpha$ in terms of the Killing spinor and applying the Fierz identity, as
explained around \eqref{A.43b}.}
\begin{align}
\alpha(\eun,\nn_{a},\nn_{b})=2\sqrt{2}\,\epsilon_{ab} \ .
\end{align}
Thus, expressing \eqref{3.12} in terms of \eqref{3.13d} gives the quadratic action for bosonic fluctuations normal to $\Sigma$:
\begin{align}
     \frac{1}{2}&\int d^3\sigma \sqrt{g} \,
    \Big(
    |\slashed{\mathfrak{D}}\mathrm{V}+\mathrm{V}|^2
    -4\langle\mathrm{V}, \slashed{\mathfrak{D}}\mathrm{V}+\mathrm{V}\rangle
    \Big)  \nonumber \\
    =&\int d^3\sigma \sqrt{g} \, \big(\nabla_{\xi}\bar{v}_{a}\nabla_{\xi}v^{a}-i\bar{v}_{a}\nabla_{\xi}v^{a}+iv^{a}\nabla_{\xi}\bar{v}_{a}+2\nabla_{\eun}^{\perp}\bar{v}_{a}\nabla^{\perp}_{\bar{\eun}}v^{a}-3\bar{v}_{a}v^{a} \nonumber \\
    &\qquad \qquad \quad  +\sqrt{2}i\, e^{i\Theta}\epsilon_{ab}\nabla_{\bar \eun}^{\perp}v^{a}(\nabla_{\xi}v^{b}+iv^{b})-\sqrt{2}i\, e^{-i\Theta}\epsilon^{ab}\nabla_{ \eun}^{\perp}\bar{v}_{a}(\nabla_{\xi}\bar{v}_{b}-i\bar{v}_{b}) \big) \ . \label{3.15ab}
\end{align}

To write down the bosonic action in a frame-independent form, we treat $v=v^{a}\nn_{a}$ as a section of $N^{1,0}\Sigma$. It is convenient to introduce notation that will also be useful in the discussion of fermionic fluctuations and supersymmetry transformations below. We first define the Dolbeault-type operators built from the normal
connection:
\begin{align}
    &{\partial}_{N}: \Gamma(N^{0,1}\Sigma)
    \to
    \Omega^{1,0}_{H}(\Sigma , N^{0,1}\Sigma) \ ,
    \qquad
    {\partial}_{N}\bar{v}
    =
    {\mathrm{f}}\otimes \sqrt{2}\, \nabla^{\perp}_{ \eun}\bar{v}  \ , \\
    &\bar{\partial}_{N}: \Gamma(N^{1,0}\Sigma)
    \to
    \Omega^{0,1}_{H}(\Sigma , N^{1,0}\Sigma) \ ,
    \qquad
    \bar{\partial}_{N}v
    =
    \bar{\mathrm{f}}\otimes \sqrt{2}\, \nabla^{\perp}_{\bar \eun}v  \ , \label{3.17f}
\end{align}
where $\mathrm{f}, \bar{\mathrm{f}}$ denote the dual coframe, so that $\mathrm{f}(\eun)=1$ and $\bar{\mathrm{f}}(\bar\eun)=1$. We further define the $L^2$-pairings for the normal vectors and horizontal forms on $\Sigma$:
\begin{align}
    &(\bar{v} , v')
    =
    \int d^3\sigma \, \sqrt{g} \, \bar v_{a}v'^{a} \ ,
    \qquad
    v, v' \in \Gamma(N^{1,0}\Sigma) \ , \\
    &( \bar{\lambda}, \lambda')
    =
    \int d^3\sigma \, \sqrt{g} \,
    \bar{\lambda}_{\eun}\lambda'_{\bar\eun} \ ,
    \qquad
    \lambda = \lambda_{\bar\eun}\bar{\mathrm{f}}
    \in \Omega^{0,1}_{H}(\Sigma) \ ,
\end{align}
and similarly for the tensor-product extension of this pairing
%$\Omega^{1,0}_{H}(\Sigma,N^{0,1}\Sigma)\otimes \Omega^{0,1}_{H}(\Sigma,N^{1,0}\Sigma)$:
\begin{align}
    &(\bar{V}, V') = \int d^3\sigma \, \sqrt{g} \, (\bar{V}_{\eun})_{a}(V'_{\bar\eun})^{a} \ ,  \qquad\bar V = \mathrm{f} \otimes \nn_{\bar a} \,  (\bar V_{\eun})^{\bar a}  \ , \quad  V' = \bar{\mathrm{f}} \otimes \nn_{a} \, (V'_{\bar \eun})^{a} \ . \label{3.20c}
\end{align}
Using the above definitions and the identities derived in Appendix \ref{bosonicinv}, the $\Theta$-dependent terms in the action \eqref{3.15ab} drop out after integration by parts. Thus, for quadratic fluctuations around a closed invariant three-cycle  $\Sigma\subset\ses$ the action can be written in a frame-independent form as
\begin{align}
    &\frac{1}{2}\int d^3\sigma \, \sqrt{g} \,
    \langle \mathrm{V},
    (\slashed{\mathfrak{D}}-3)(\slashed{\mathfrak{D}}+1)
    \mathrm{V}\rangle = ( \bar v, -\mathcal{L}_{\xi}^2v-4i\mathcal{L}_{\xi}v
    +\bar{\partial}_{N}^{\dagger}\bar{\partial}_{N}v )  \ .
\end{align}

\subsection{Fermionic fluctuations}\label{sec3.2}  

The fermionic fluctuations around a classical membrane are described by the pullback of the eleven-dimensional Killing spinor derivative \eqref{2.9b} to the classical surface. In addition, due to the $\kappa$-symmetry of the BST action, one has to fix the fermionic gauge. We consider the ``static'' $\kappa$-symmetry gauge, so that the eleven-dimensional spinors $\theta$ describing fermionic fluctuations around M2-brane instantons satisfy
\begin{align}
    (1+\Gamma)\theta = 0 \ , \qquad \Gamma = \gamma_{\hat{5}}\otimes \gamma \ . \label{3.15f}
\end{align}
Since $\mathrm{AdS}_{4}$ is homogeneous and admits four independent Killing spinors, it is convenient to work in a basis adapted to them. For the membrane configurations of interest, the branes are placed at the centre $U_{0}$ of $\mathrm{AdS}_{4}$ where the Killing spinors can be chosen to be simultaneously chiral.

Once restricted to $U_{0}$, one may therefore use these Killing spinors to parametrise the fermionic fluctuations in the $\kappa$-symmetry gauge \eqref{3.15f}. Concretely, we consider
\begin{align}
    &\theta = \chi_{-}^{I}\otimes \Lambda_{I} + \chi_{+}^{I}\otimes \mathcal{V}_{I} \ , \quad I=1,2 \ , \\
    &\gamma_{\hat{5}}\chi_{\pm}^{I}=\pm \chi_{\pm}^{I} \ , \quad \gamma\Lambda_{I} = \Lambda_{I} \ , \quad \gamma\mathcal{V}_{I} = - \mathcal{V}_{I} \ , \label{3.17}
\end{align}
%where the seven-dimensional spinors $\lambda_{I}, \nu_{I}$ are taken to be Grassmann-odd. 
where $\chi^{I}_{\pm}$ are independent $\mathrm{AdS}_{4}$ Killing spinors (two for each chirality of $\gamma_{\hat{5}}$), normalised as
\begin{align}
    \langle \chi^{I}_{+}, \chi_{+}^{J}\rangle  = \epsilon ^{IJ} \ , \quad \langle \chi^{I}_{-}, \chi_{-}^{J}\rangle  = \epsilon ^{IJ} \ , \quad \langle \chi^{I}_{+}, \chi_{-}^{J}\rangle  = 0 \ .
\end{align}
Since $\chi_{\pm}^{I}$ are Killing spinors, the fermionic fluctuations are completely determined by the action of the pullback of the Killing spinor derivative on the seven-dimensional components $\Lambda_{I}$ and $\mathcal{V}_{I}$:
\begin{align} \label{3.19}
   \frac{1}{2} \int d^3\sigma \sqrt{g} \, \big( \langle \Lambda^{I}, i\slashed{D}\Lambda_{I}\rangle + \langle \mathcal{V}^{I}, i\slashed{D}\mathcal{V}_{I}\rangle\big) \ , 
\end{align}
where $\Lambda^{I}=\Lambda_{J}\epsilon^{JI}$, \ $\mathcal{V}^{I}=\mathcal{V}_{J}\epsilon^{JI}$, and $\slashed{D} = g^{\alpha\beta}e^{i}(\partial_{\alpha})\gamma_{i}D_{\beta}$. In the derivation of the fermionic quadratic actions  below, we suppress the doublet index and denote
$\Lambda_{I}$ and $\mathcal{V}_{I}$ simply by $\Lambda$ and $\mathcal{V}$,
respectively. The scalar product is then understood to include contraction
with $\epsilon^{IJ}$, namely
\begin{align}
    \langle \Lambda , \Lambda' \rangle
    =
    \epsilon^{IJ}\langle \Lambda_{I}, \Lambda'_{J}\rangle \ ,
\end{align}
and similarly for $\mathcal{V}$.

For the unitary frame $\eun_{\mu}=(\eun,\nn_a)$, adapted to the invariant
submanifold $\Sigma$, the corresponding Dirac matrices satisfy
\begin{align}
    &\{\gamma_\mu,\gamma_{\bar\nu}\}
    =
    2\delta_{\mu\bar \nu} \ , \qquad
    \{\gamma_\mu,\gamma_\nu\}
    =
    \{\gamma_{\bar\mu},\gamma_{\bar\nu}\}
    =
    0 \ , \qquad
    \{\gamma(\xi),\gamma_\mu\}
    =
    \{\gamma(\xi),\gamma_{\bar\mu}\}
    =
    0 \ ,
\end{align}
where $\gamma_\mu=\gamma(\eun_\mu)$ and
$\gamma_{\bar\mu}=\gamma(\eun_{\bar\mu})$. In this frame, and with the
orientation induced by \((\xi,\eun,\bar\eun)\), the internal part of the
\(\kappa\)-symmetry projector can be written as
\begin{align}
    \gamma
    =
    \gamma(\xi)\big(\gamma(\eun)\gamma(\bar\eun)-1\big) \ ,
    \qquad
    \gamma^2=1 \ .
\end{align}
To decompose the spinor bundle $\mathrm S$, we use a Fock-type frame adapted
to the transverse unitary frame. We first introduce the complex spinor
\begin{align}
    \psin
    =
    \frac{1}{\sqrt{2}}(\psi_1-i\psi_2) \ ,
    \qquad
    \gamma(\xi)\psin=-\psin \ ,
    \qquad
    \langle \bar\psi_{-},\psin\rangle=1 \ ,
\end{align}
where \(\psi_{1,2}\) are two real Killing spinors on \(\ses\) descending from
the parallel spinors \eqref{2.35} on \(\cyf\). With the above choice of
transverse complex frame, \(\psin\) is annihilated by the antiholomorphic
Clifford generators
\begin{align}
    \gamma_{\bar\mu}\psin=0 \ ,
\end{align}
and we collect further properties in Appendix \ref{appA3}. We also define
\begin{align}
    \mathrm e^\psi_\mu
    =
    \frac{i}{\sqrt{2}}\gamma_\mu\psin \ ,
    \qquad
    \langle \bar{\mathrm e}^{\psi}_{\mu},
    \mathrm e^\psi_\nu\rangle
    =
    \delta_{\mu\nu} \ ,
    \qquad
    \langle \bar\psi_{-},\mathrm e^\psi_\mu\rangle=0 \ ,
    \qquad
    \langle \mathrm e^\psi_\mu,\mathrm e^\psi_\nu\rangle=0 \ .
\end{align}
Together with their complex conjugates, these spinors form a local frame of the complexified spinor bundle. The constructed spinor frame can now be used to decompose the restriction of
the spinor bundle $\mathrm S$ to the M2-brane $\Sigma$ into the $\pm1$
eigenspaces of $\gamma$:
\begin{align}
    &\mathrm S\big|_{\Sigma}
    =
    \ker(\id-\gamma)\oplus \ker(\id+\gamma) \ , \\
    &\ker(\id-\gamma)
    =
    \mathrm{span}\{\psin,\, \bar\psi_{-}, \, \eun^\psi,\,\bar\eun^\psi\} \ ,
    \qquad
    \ker(\id+\gamma)
    =
    \mathrm{span}\{\nn_a^\psi,\,\bar\nn_a^\psi\} \ ,
\end{align}
where $a=1,2$, and both eigenspaces have %complex
rank four. Therefore the
fermionic fluctuations $\Lambda$ and $\mathcal V$ can be locally
decomposed as
\begin{align}
    \Lambda
    =
    a\psin+\lambda_{\bar \eun}\eun^\psi+\mathrm{c.c.} \ ,
    \qquad
    \mathcal V
    =
    \nu^a\nn_a^\psi+\mathrm{c.c.} \ .
\end{align}
Finally, to determine the quadratic action for the fermionic fluctuations, we
evaluate the pullback of the Killing-spinor derivative in the Fock-adapted
frame. The worldvolume
Dirac operator takes the form
\begin{align}
    \slashed{D}
    =
    \gamma(\xi)D_{\xi}
    +
    \gamma(\bar{\mathrm e})D_{\mathrm e}
    +
    \gamma(\mathrm e)D_{\bar{\mathrm e}} \ ,
\end{align}
where $D_{\mathrm e}$ and $D_{\bar{\mathrm e}}$ denote Killing-spinor
derivatives along the complexified tangent vectors $\eun$ and
$\bar\eun$, respectively. 

Explicitly, using the decomposition of the fermionic fluctuations in the
spinor frame introduced above, one finds\footnote{To obtain the action of
$\slashed D$ in components we use that $D\psi=0$ and that for the spinor
covariant derivative:
\mbox{$[\nabla,\gamma(X)]=\gamma(\nabla X)$}.}
\begin{align}
    \langle \Lambda, i\slashed{D}\Lambda\rangle
    &=
    -i\, \bar a\mathcal{L}_{\xi}a
    +i\, \bar\lambda_{\eun}(\nabla_{\xi}+i)\lambda_{\bar\eun}
    +\sqrt{2}\,\bar\lambda_{\eun}\nabla_{\bar{\mathrm e}}^{T}a
    -\sqrt{2}\,\bar a(\nabla_{\mathrm e}^{T}\lambda)_{\bar\eun}
    +\mathrm{c.c.} \ . \label{3.21}
\end{align}
Here $\nabla^T$ denotes the covariant derivative induced on $T\Sigma$,
$\mathcal{L}_{\xi}$ is the Lie derivative along the Reeb vector field on
$\Sigma$, and we have identified $a$ with a scalar on $\Sigma$ and $\lambda$
with a horizontal $(0,1)$-form,
\begin{align}
    \lambda=\lambda_{\bar\eun}\bar{\mathrm f} \ ,
    \qquad
    \bar{\mathrm f}(\bar{\mathrm e})=1 \ .
\end{align}
The expression \eqref{3.21} can be written in a frame-independent form by
introducing the horizontal Dolbeault operators
\begin{align}
    &\partial_H:\Omega_H^{0,0}(\Sigma)\to
    \Omega_H^{1,0}(\Sigma) \ ,
    \qquad
    \partial_H a=\sqrt{2}\, (\nabla^{T}_{\eun}a)\mathrm f  \ , \\
    &\bar\partial_H:\Omega_H^{0,0}(\Sigma)\to
    \Omega_H^{0,1}(\Sigma) \ ,
    \qquad
    \bar\partial_H a=\sqrt{2}\, (\nabla^{T}_{\bar\eun}a)\bar{\mathrm f} \ .
\end{align}
With respect to the $L^2$-pairings
\begin{align}
    (\bar a,a')
    =
    \int d^3\sigma\sqrt g\,\bar a a' \ ,
    \qquad
    ( \bar \lambda,\lambda')
    =
    \int d^3\sigma\sqrt g\,
    \bar\lambda_{\eun}\lambda'_{\bar\eun} \ ,
\end{align}
the formal adjoint of $\bar\partial_H$ is defined by\footnote{In other words,
$$
    (\bar{\lambda},\bar\partial_H a) = \int d^3\sigma\sqrt g\,
    \sqrt{2}\,\bar\lambda_{\eun}\nabla_{\bar\eun}^{T}a
    =
    -\int d^3\sigma\sqrt g\,
    \sqrt{2}\,
    \overline{(\nabla_{\eun}^{T}\lambda)_{\bar\eun}}\,a =(\, \overline{\bar\partial_H^\dagger\lambda}\, ,a)=(\partial_{H}^{\dagger}\bar{\lambda}, a) \ .
$$}
\begin{align}
    \bar\partial_H^\dagger:
    \Omega_H^{0,1}(\Sigma)\to \Omega_H^{0,0}(\Sigma) \ ,
    \qquad
    \bar\partial_H^\dagger\lambda
    =
    -\sqrt{2}\,(\nabla_{\eun}^{T}\lambda)_{\bar\eun} \ ,
\end{align}
and similarly, one defines $\partial_H^\dagger$ using the corresponding Hermitian
pairing on horizontal $(1,0)$-forms. The covariant derivative along the Reeb vector field can be expressed
in terms of the Lie derivative as
\begin{align}
    \mathcal L_\xi\lambda
    =
    \nabla_\xi\lambda-i\lambda \ ,
    \qquad
    \lambda\in\Omega_H^{0,1}(\Sigma) \ .
\end{align}
Therefore, the fermionic part \eqref{3.21} can be compactly written as
\begin{align}
    \frac{1}{2}\int d^3\sigma\sqrt g\,
    \langle \Lambda,i\slashed D\Lambda\rangle
    &=
    -i( \bar a,\mathcal L_\xi a)
    +i( \bar \lambda,\mathcal L_\xi\lambda+2i\lambda)
    +( \bar \lambda,\bar\partial_H a)
    +( \bar a,\bar\partial_H^\dagger\lambda) \ . \label{rw2}
\end{align}

For the fermionic fluctuations determined by $\mathcal{V}$, one obtains
\begin{align} \label{3.49}
    \frac{1}{2}
    \langle \mathcal V, i\slashed{D}\mathcal V\rangle
    =
    \frac{i}{2}\big(\bar\nu_{a}(\nabla_{\xi}\nu)^{a}
    -\nu^{a}(\nabla_{\xi}\bar\nu)_{a} \big)
    -3\bar\nu_{a}\nu^{a}
    -\frac{1}{\sqrt{2}}\epsilon_{ab}\nu^{a}(\nabla_{\bar \eun }^{\perp}\nu)^{b}
    -\frac{1}{\sqrt{2}}\epsilon^{ab}\bar \nu_{a}(\nabla_{\eun}^{\perp}\bar \nu)_{b} \ ,
\end{align}
where $\nabla^\perp$ denotes the connection induced on the normal bundle, and the fermionic field $\nu$ is regarded as a section of the holomorphic normal bundle
\begin{align}
    \nu=\nu^a \nn_a \ , \qquad
    \nu \in \Gamma(N^{1,0}\Sigma) \ .
\end{align}
We have also used that, in the chosen frame $\eun_{\mu}=(\eun,\nn_{a})$, the components of the transverse holomorphic three-form can be written as
\begin{align}
    \alpha(\eun, \nn_{a}, \nn_{b})
    =
    i\langle \psin , \gamma(\eun)\gamma(\nn_a)\gamma(\nn_b)\psin \rangle
    =
    2\sqrt{2} \, \epsilon_{ab} \ .
\end{align}
 Comparing \eqref{3.49} with the analogous expression for the normal bosonic fluctuations \eqref{3.13d}, and identifying $\nu$ with a section of the holomorphic normal bundle, with the constant phase $\Theta$ absorbed into the choice of spinorial frame, one finds
\begin{align}
    \frac{1}{2}\langle \mathcal{V}, i\slashed{D}\mathcal{V}\rangle = \frac{1}{2}\langle \mathcal{V},(\slashed{\mathfrak{D}}-3)\mathcal{V}\rangle \ .
\end{align}
This is a general property of fermionic fluctuations around associative M2-branes in weak $G_{2}$ manifolds, as discussed in Appendix \ref{B}. Using the same notation as in the bosonic case, the quadratic action for
the fermionic fluctuations determined by $\mathcal V$ takes the form
\begin{align}
    \frac{1}{2}\int d^3\sigma \, \sqrt{g} \, \langle \mathcal{V}, i\slashed{D}\mathcal{V}\rangle = i(\bar \nu , \mathcal{L}_{\xi}\nu +4i\nu) +\frac{1}{2}\big(\adj(\bar{\partial}_{N}\nu), \nu \big)+\frac{1}{2}\big(\bar{\nu}, \adjb (\partial_{N}\bar{\nu})\big) \ .
\end{align}
Here $\adj$ and $\adjb$ denote the adjunction-type maps
\begin{align}
    \adj:\Omega_H^{0,1}(\Sigma,N^{1,0}\Sigma)\to\Gamma(N^{0,1}\Sigma) \ ,
    \qquad
    \adjb:\Omega_H^{1,0}(\Sigma,N^{0,1}\Sigma)\to\Gamma(N^{1,0}\Sigma) \ ,
\end{align}
induced by the restriction of the transverse holomorphic three-form $\alpha$ to $\Sigma$, with their properties collected in Appendix \ref{bosonicinv}.

\subsection{Worldvolume supersymmetry} \label{sec3.3}

The worldvolume supersymmetry of the M2-brane is obtained by combining the preserved target-space supersymmetry with the residual $\kappa$-symmetry \cite{Kallosh:1997sw, Kallosh:1997ky}.\footnote{Worldvolume supersymmetry transformations for the supermembrane in flat space were also discussed recently in detail in \cite{Tseytlin:2025dae}.
} To interpret the worldvolume supersymmetry transformations as an equivariant $\mathcal{Q}$-complex, we choose a complex supercharge generated by the Killing spinor
\begin{align}
    \varepsilon\otimes \psi_{-} \ , \qquad
    \varepsilon = \chi_{+}^{I}\varepsilon_{I} \ ,
\end{align}
where $\varepsilon_I$ are fixed complex Grassmann-even parameters. The bosonic fluctuations $x^{\hat a}$ in $\ads$ and $v$ in $\ses$ belong to two distinct multiplets. As in \cite{Harvey:1999as}, we refer to the corresponding multiplets as the Rozansky-Witten (RW) and McLean (ML) multiplets.

To determine the supersymmetry transformations for the RW multiplet, it is convenient to twist the bosonic $\ads$ modes. The fluctuations $x^{\hat a}$ transform in the vector representation of
$\mathrm{Spin}(4)\cong SU(2)_{+}\times SU(2)_{-}$, and we use the preserved spinor to map $x^{\hat a}$ to an $SU(2)_{-}$ doublet. Let $\varepsilon^{c}$ denote the charge conjugate of $\varepsilon$, with the normalisation fixed by
\begin{align}
    &\varepsilon^{c} = C\bar\varepsilon \ , \qquad \langle \varepsilon^{c}, \varepsilon\rangle=\epsilon^{IJ}(\varepsilon^{c})_{I}\varepsilon_{J}   =1 \ . 
\end{align}
The vector $x^{\hat a}$ can then be encoded in two $SU(2)_{-}$ doublets of worldvolume scalars as
\begin{align}
    & x_{I} =\frac{1}{\sqrt{2}}\, x^{\hat{a}}\langle \chi_{-}^{I}, \gamma_{\hat{a}}\varepsilon\rangle \ , \quad \tilde{x}_{I} =\frac{1}{\sqrt{2}}\, x^{\hat{a}}\langle \chi_{-}^{I}, \gamma_{\hat{a}}\varepsilon^{c}\rangle  \ .
\end{align}
With this normalisation, the vector inner product is recovered as
\begin{align}
     \epsilon^{IJ}(\tilde{x}_{I}x_{J}'+\tilde{x}'_{I}x_{J})=\frac{1}{2}\, x_{\hat{a}}x_{\hat{b}}'\langle\varepsilon^{c}, (\gamma_{\hat{a}}\gamma_{\hat{b}}+\gamma_{\hat{b}}\gamma_{\hat{a}})\varepsilon \rangle = x_{\hat{a}}x'_{\hat{a}} \ .
\end{align}
In particular, the quadratic action for the $\ads$ fluctuations can be written as the $L^2$-pairing of $x_I$ and $\tilde{x}_I$:
\begin{align}
    \frac{1}{2}\int d^3\sigma \, \sqrt{g} \, x_{\hat{a}}\Delta_{\Sigma}x_{\hat{a}} = \int d^3\sigma \, \sqrt{g} \, \epsilon^{IJ}\tilde{x}_{I}\Delta_{\Sigma}x_{J} = (\tilde{x}^{I}, \Delta_{\Sigma}x_{I})=-(x^{I}, \Delta_{\Sigma}\tilde{x}_{I}) \ . \label{rw3}
\end{align}
Combining \eqref{rw3} and \eqref{rw2}, the quadratic action for the RW multiplet takes the form
\begin{align}
    S_{RW} = (\tilde{x}^{I}, \Delta_{\Sigma}x_{I}) -i(\bar a^{I},\mathcal L_\xi a_{I})
    +i(\bar \lambda^{I},\mathcal L_\xi\lambda_{I}+2i\lambda_{I})
    +(\bar \lambda^{I},\bar\partial_H a_{I})
    +(\bar a^{I},\bar\partial_H^\dagger\lambda_{I}) \ ,  \label{3.64}
\end{align}
where the $I,J$ indices have been restored for the fermions and are contracted with $\epsilon^{IJ}$. The scalar Laplacian on an invariant submanifold $\Sigma\subset \ses$, with volume form $\vol(\Sigma)=\frac{1}{2}\eta\wedge d\eta$, can be written as
\begin{align}
    \Delta_{\Sigma} &=-\mathcal{L}_{\xi}^2+\frac{1}{2}\big(\partial_{H}^{\dagger}\partial_{H}+ \bar{\partial}_{H}^{\dagger}\bar{\partial}_{H}\big) \nonumber \\
    &=-\mathcal{L}_{\xi}^2-2i\mathcal{L}_{\xi}+\bar{\partial}_{H}^{\dagger}\bar{\partial}_{H}=-\mathcal{L}_{\xi}^2+2i\mathcal{L}_{\xi}+{\partial}_{H}^{\dagger}{\partial}_{H} \ ,
\end{align}
where\footnote{Here we used metric compatibility on the Sasakian manifold $\Sigma$: $\langle \nabla_{\bar{\eun}}\eun, \xi\rangle = i , \ \langle \nabla_{\eun}\bar \eun, \xi\rangle = -i$. We also denote by $(\nabla^{T}_{\eun}\bar{\eun})^{H}$ the horizontal projection of $\nabla^{T}_{\eun}\bar{\eun}$.} 
\begin{align}
    &\bar{\partial}_{H}^{\dagger}\bar{\partial}_{H}= -2\nabla^{T}_{\eun}\nabla_{\bar{\eun}}^{T}+2\nabla^{T}_{(\nabla^{T}_{\eun}\bar\eun)^{H} } \ , \quad [\nabla_{\mathrm{e}}^{T}, \nabla_{\bar{\mathrm{e}}}^{T}]\big|_{\xi} = [\mathrm{e}, \bar{\mathrm{e}} ]\big|_{\xi} =-2i\xi \ ,
\end{align}
and similarly for $\partial_H, \partial_{H}^{\dagger}$. Furthermore, the operators $\bar{\partial}_{H}, \partial_{H}$ and their adjoints are invariant with respect to the Reeb action:
\begin{align}
    [\mathcal{L}_{\xi}, \bar{\partial}_{H}] = 0 \ , \quad [\mathcal{L}_{\xi}, \partial_{H}] = 0 \ .
\end{align}
Using these identities, and treating the bosonic fluctuations as Grassmann-even and the fermionic fluctuations as Grassmann-odd, the supersymmetry transformations generated by $Q_{\varepsilon}$ can be read off from \eqref{3.64} as
\begin{align}
    Q_{\varepsilon}\tilde{x}_{I} = \bar{a}_{I} \ , \quad Q_{\varepsilon} a_{I} = -i\mathcal{L}_{\xi}x_{I}+2 x_{I} \ , \quad Q_{\varepsilon}\lambda_{I} = \bar{\partial}_{H}x_{I} \ , 
\end{align}
with $Q_{\varepsilon}$ acting trivially on the remaining fields. Similarly,  for $\bar{Q}_{\varepsilon}$ one has
\begin{align}
    \bar{Q}_{\varepsilon}x_{I} = a_{I} \ , \quad \bar{Q}_{\varepsilon}\bar{a}_{I} = -i\mathcal{L}_{\xi}\tilde x_{I}-2\tilde x_{I} \ , \quad \bar{Q}_{\varepsilon}\bar{\lambda}_{I} = -\partial_{H}\tilde x_{I} \ .
\end{align}
Using the equations of motion for $\lambda$ and $\bar{\lambda}$, the supersymmetry algebra closes on-shell
\begin{align}
    &\{Q_{\varepsilon}, \bar{Q}_{\varepsilon}\} = - i\mathcal{L}_{\xi}+\mathcal{R}+[\mathrm{Eom}_{\lambda, \bar{\lambda}}] \ , \\
    &\mathrm{Eom}_{\lambda} =\bar{\partial}_{H}a_{I}+i\mathcal{L}_{\xi}\lambda_{I}-2\lambda_{I}   \ , \qquad \mathrm{Eom}_{\bar \lambda} = -\partial_{H}\bar{a}_{I}+i\mathcal{L}_{\xi}\bar \lambda_{I} +2\bar \lambda_{I} \ ,
\end{align}
where the R-charge operator $\mathcal R$ takes the values
\begin{align}
       \mathcal{R}[x,a,\lambda]=2 \ , 
    \qquad 
       \mathcal{R}[\tilde x,\bar a,\bar\lambda]=-2 \ .
\end{align}

For the ML multiplet, restoring the fermionic $I,J$ indices, the quadratic action reads
\begin{align}
    S_{ML} =
    &(\bar v, -\mathcal{L}_{\xi}^{2}v - 4i\mathcal{L}_{\xi}v
    + \bar{\partial}_{N}^{\dagger}\bar{\partial}_{N}v)
    + i(\bar{\nu}^{I}, \mathcal{L}_{\xi}\nu_{I} + 4i\nu_{I})+\frac{1}{2}\big(\adj(\bar{\partial}_{N}\nu^{I}), \nu_{I}\big)
    +\frac{1}{2}\big(\bar{\nu}^{I},
   \adjb(\partial_{N}\bar{\nu}_{I})\big) \ .
\end{align}
To write the $Q_{\varepsilon}$-transformations, it is convenient to reparametrise the fermionic fluctuations as
\begin{align}
    \nu_{I} = \varepsilon_{I}\, \mu +(\varepsilon^{c})_{I}\, \adjb(\tilde{\rho}) \ ,
    \qquad
    \bar{\nu}_{I} = -\varepsilon_{I}\, \adj(\rho)-(\varepsilon^{c})_{I}\,\tilde{\mu} \ ,
\end{align}
where
\begin{align}
\mu\in\Gamma(N^{1,0}\Sigma) \ ,\quad
\tilde\mu\in\Gamma(N^{0,1}\Sigma) \ ,\quad
\rho\in\Omega^{0,1}_{H}(\Sigma,N^{1,0}\Sigma) \ ,\quad
\tilde\rho\in\Omega^{1,0}_{H}(\Sigma,N^{0,1}\Sigma) \ .
\end{align} 
%$\rho$ is a fermionic fluctuation valued in $\Omega^{0,1}_{H}(\Sigma,N^{1,0}\Sigma)$. 
In these variables, the quadratic action becomes
\begin{align}
    S_{ML} =
    &(\bar v, -\mathcal{L}_{\xi}^{2}v - 4i\mathcal{L}_{\xi}v
    + \bar{\partial}_{N}^{\dagger}\bar{\partial}_{N}v)
    - i(\tilde{\mu}, \mathcal{L}_{\xi}\mu + 4i\mu)
    + i(\tilde{\rho}, \mathcal{L}_{\xi}\rho)+(\tilde{\rho}, \bar{\partial}_{N}\mu)
    +(\partial_{N}\tilde{\mu}, \rho) \ , \label{3.82}
\end{align}
where we have used \eqref{3.34}, together with the identities \eqref{3.27e} and \eqref{3.21d}, with the appropriate signs for Grassmann-odd variables. The corresponding $Q_{\varepsilon}$-transformations are
\begin{align}
    Q_{\varepsilon}\bar v = \tilde{\mu} \ , \qquad
    Q_{\varepsilon}\mu = -i\mathcal{L}_{\xi}v \ , \qquad
    Q_{\varepsilon}\rho = \bar{\partial}_{N}v \ ,
\end{align}
with $Q_{\varepsilon}$ acting trivially on the remaining fields. The conjugate supercharge acts by
\begin{align}
    \bar Q_{\varepsilon}v = \mu \ , \qquad
    \bar Q_{\varepsilon}\tilde{\mu} = -i\mathcal{L}_{\xi}\bar v \ , \qquad
    \bar Q_{\varepsilon}\tilde{\rho} = -\partial_{N}\bar v \  .
\end{align}
As for the RW multiplet, the algebra closes on-shell modulo the equations of motion for $\rho$ and $\tilde{\rho}$:
\begin{align}
    &\{Q_{\varepsilon},\bar Q_{\varepsilon}\}
    = -i\mathcal{L}_{\xi} + [\mathrm{Eom}_{\rho,\tilde{\rho}}] \  , \\
   & \mathrm{Eom}_{\rho}
    = \bar{\partial}_{N}\mu + i\mathcal{L}_{\xi}\rho  \ ,
    \qquad
    \mathrm{Eom}_{\tilde{\rho}}
    = -\partial_{N}\tilde{\mu} + i\mathcal{L}_{\xi}\tilde{\rho}  \ .
\end{align}
In contrast to the RW multiplet, all fields in the ML multiplet have zero R-charge.

\section{One-loop partition function for invariant M2-branes}\label{sec4}

\subsection{Off-shell supersymmetry and zero modes}\label{secc4.1}

The quadratic actions \eqref{3.64}, \eqref{3.82} for fluctuations around an invariant M2-brane $\Sigma\subset\ses$ are controlled by Dolbeault-type operators $\bar\partial_{H,N}$ together with the Lie derivative $\mathcal L_\xi$ along the Reeb vector field on $\Sigma$. Already at the level of the quadratic action, one can see that, away from the zero modes of $\bar\partial_{H,N}$ and $\bar\partial^{\dagger}_{H,N}$, bosonic and fermionic fluctuations are paired and their non-zero mode one-loop determinants cancel. It is nevertheless useful to reformulate this cancellation in cohomological language. The on-shell supersymmetry transformations constructed in Section \ref{sec3.3} suggest an extension to an off-shell $\mathcal Q$-complex whose square is the equivariant action generated by $\mathcal L_\xi$, possibly supplemented by an algebraic $U(1)_r$ charge. In the present case, this extension is achieved by introducing auxiliary bosonic fields, so that the algebra closes without using the linearised equations of motion. Once the worldvolume supersymmetry algebra has been closed off-shell, the remaining one-loop contribution can be expressed in terms of the $U(1)_r$-equivariant index of the corresponding transversely elliptic Dolbeault-type operator, in the sense of Section 4.4 of \cite{Pestun:2007rz}. The resulting off-shell supersymmetry transformations have a cohomological structure similar to the localisation complexes used for 3d CS theories on Seifert manifolds \cite{Kallen:2011ny} and for 5d SYM on contact/Sasakian manifolds \cite{Kallen:2012cs, Kallen:2012va}. 

For both RW and ML multiplets, this off-shell extension can be written, at the linearised level, in terms of fields
\begin{align}
    \Xr_{\mathrm{b}}, \Xr_{\mathrm{f}} \in \Gamma(\Sigma, E\otimes_{\mathbb{R}}\mathbb{C}) \ , \qquad \Hr_{\mathrm{b}} , \Hr_{\mathrm{f}} \in \Omega^{0, 1}_{H}(\Sigma, E\otimes_{\mathbb{R}}\mathbb{C}) \ ,
\end{align}
where $E$ denotes the relevant $U(1)_r$-equivariant vector bundle over $\Sigma$. If $t^E:E_{p\cdot t^{-1}}\to E_p$ denotes the lift of the $U(1)_{r}$ action to $E$, then sections $\mathrm{X}\in \Gamma(\Sigma, E)$ transform as
\begin{align}
    (t\cdot \mathrm{X})(p) = t^{E}\, \mathrm{X}(p \cdot t^{-1}) \ , \qquad t\in U(1)_{r} \ , \quad  p\in \Sigma \ .
\end{align}
The bosonic field $\Hr_{\mathrm{b}}$ is the auxiliary field required for the off-shell extension of the superalgebra. The supersymmetry transformations can be written in terms of the supercharge $\mathcal{Q} = Q_{\varepsilon}+\bar{Q}_{\varepsilon}$ as
\begin{align}\label{4.3s}
    &\mathcal{Q}\, \Xr_{\mathrm{b}} = \Xr_{\mathrm{f}} \ , \quad \mathcal{Q}\, \Xr_{\mathrm{f}} = R\, \mathrm{X}_{\mathrm{b}} \ , \quad \mathcal{Q}\, \mathrm{H}_{\mathrm{f}} = \bar{\partial}_{E}\Xr_{\mathrm{b}}+\Hr_{\mathrm{b}} \ , \quad\mathcal{Q}\, {\Hr}_{\mathrm{b}} = -\bar{\partial}_{E}{\Xr}_{\mathrm{f}}+R\, \Hr_{\mathrm{f}} \ ,
\end{align}
and similarly for the conjugate variables:
\begin{align}\label{4.4s}
    &\mathcal{Q}\, \bar\Xr_{\mathrm{b}} = \bar\Xr_{\mathrm{f}} \ , \quad \mathcal{Q}\,\bar \Xr_{\mathrm{f}} = R\, \bar{\mathrm{X}}_{\mathrm{b}} \ , \quad  \mathcal{Q}\, \bar{\mathrm{H}}_{\mathrm{f}} = -{\partial}_{E}\bar \Xr_{\mathrm{b}}+\bar \Hr_{\mathrm{b}} \ , \quad \mathcal{Q}\, \bar{\Hr}_{\mathrm{b}} = {\partial}_{E}\bar{\Xr}_{\mathrm{f}}+R\, \bar\Hr_{\mathrm{f}} \ .
\end{align}
With the auxiliary fields $\Hr_{\mathrm b}$ introduced, the superalgebra closes off-shell
\begin{align}
   \mathcal{Q}^2 = R \ .
\end{align}
Here $R$ denotes the generator of $\mathcal Q^2$ acting on the corresponding field and is defined as
\begin{align}
    R=-i\mathcal L_\xi+\mathcal R \ ,
\end{align}
where $\mathcal R$ is the algebraic part of the $U(1)_r$ action. The Dolbeault-type operator $\bar{\partial}_{E}$ appearing above is the first-order $U(1)_r$-invariant differential operator
\begin{align}
    \bar{\partial}_{E}: \Gamma(\Sigma, E) \to \Omega^{0,1}_H(\Sigma, E) \ , \qquad \bar{\partial}_{E} = \sqrt{2} \, \bar{\mathrm{f}}\otimes \nabla_{\bar\eun}^{E} \ , \qquad [R, \bar{\partial}_{E}] = 0 \ .
\end{align}
We shall assume that the Reeb flow on $\Sigma$ is regular and that the corresponding $U(1)_{r}$ action is free. In this case, after decomposing into $U(1)_r$ Fourier modes, the operator $\bar{\partial}_{E}$ on $\Sigma$ reduces to a family of elliptic operators on the orbit space $\Sigma/U(1)_{r}$, which, for Sasakian $\Sigma$, is a Riemann surface.\footnote{The principal symbol of the operator $\bar{\partial}_{E}$ is
\begin{align*}
    \sigma_{1}(\bar{\partial}_{E})(p, \beta) = \sqrt{2} \, \beta(\bar\eun) \, \bar{\mathrm{f}}\otimes \id_{E} \ , \qquad \beta \in T^{\ast}\Sigma  \ , \quad p\in \Sigma \ .
\end{align*}
The symbol is not invertible on the open set $\{(p,\beta)\mid \beta\neq 0\}$, since for $\beta=\eta$ it vanishes identically. 
%Thus $\bar\partial_E$ is not elliptic. For real covectors, the vanishing directions are the cotangent directions along the Reeb orbits, namely the covectors proportional to $\eta$. 
On the other hand, the transverse cotangent bundle to the $U(1)_r$ action is $T^{\ast}_{H}\Sigma=\{ \beta \in T^{\ast}\Sigma : \beta(\xi)=0 \}$.
For every non-zero real horizontal covector $\beta\in T^*_H\Sigma$, the factor
$\beta(\bar\eun)$ is non-zero. Hence the symbol is invertible on
$T^{\ast}_{H}\Sigma\setminus\{0\}$, and $\bar\partial_E$ is
$U(1)_r$-transversely elliptic.}

% The quasi-regular case should be treated similarly, with the quotient a Kähler orbifold, but it is not needed for the applications below.

Accordingly, for both RW and ML multiplets, the quadratic action can be written as a $\mathcal{Q}$-exact functional
\begin{align}\label{4.9q}
    S_{(2)} =  \mathcal{Q}V_{(2)}  \ , \qquad V_{(2)} = (\mathcal{Q}\bar{\Xr}_{\mathrm{b}}, (i\mathcal{L}_{\xi}-\mathrm{c})\Xr_{\mathrm{b}})+(\bar{\Hr}_{\mathrm{f}}, \mathcal{Q}\Hr_{\mathrm{f}})-(\bar\Hr_{\mathrm{f}}, \bar{\partial}_{E}\Xr_{\mathrm{b}})+(\partial_{E}\bar{\Xr}_{\mathrm{b}}, \Hr_{\mathrm{f}}) \ .
\end{align}
Explicitly,
\begin{align}\label{4.10q}
    S_{(2)} =& \ (R\bar{\Xr}_{\mathrm{b}}, (i\mathcal{L}_{\xi}-\mathrm{c})\mathrm{X}_{\mathrm{b}})+(\partial_{E}\bar{\Xr}_{\mathrm{b}}, \bar{\partial}_{E}\mathrm{X}_{\mathrm{b}})-(\bar{\Xr}_{\mathrm{f}},(i\mathcal{L}_{\xi}-\mathrm{c})\Xr_{\mathrm{f}}) \nonumber \\
    &-(\bar{\Hr}_{\mathrm{f}}, R\Hr_{\mathrm{f}})+(\bar{\Hr}_{\mathrm{f}}, \bar{\partial}_{E}\mathrm{X}_{\mathrm{f}}) +(\partial_{E}\bar{\Xr}_{\mathrm{f}}, \Hr_{\mathrm{f}})+(\bar{\Hr}_{\mathrm{b}}, \Hr_{\mathrm{b}}) \ .
\end{align}
The constant $\mathrm{c}$ denotes a shift in the kinetic operator of the corresponding multiplet. For the RW multiplet $\mathrm{c}=0$, while for the ML multiplet $\mathrm{c}=4$. The latter shift originates from the $U(1)_r$ charge $\mathcal{L}_\xi\alpha=4i\,\alpha$ of the transverse holomorphic three-form. 
%entering the adjunction-type map. 

The one-loop partition function for \eqref{4.10q} can be simplified. Indeed, due to the worldvolume supersymmetry and the fact that $[\mathcal{L}_{\xi},\bar{\partial}_{E}]=0$, the modes away from
$\ker\bar{\partial}_{E}$ and $\operatorname{coker}\bar{\partial}_{E}$
are paired between bosonic and fermionic degrees of freedom. The standard
argument \cite{Pestun:2007rz} then shows that the ratio of Gaussian
determinants depends only on the restriction of $R$ to these kernel and
cokernel spaces, giving
\begin{align}
    Z_{1}^{(E)} = \frac{\det_{\mathrm{coker} \, \bar{\partial}_{E}}R|_{\Hr_\mathrm{f}}}{\det_{\ker \bar{\partial}_{E}}R|_{\Xr_{\mathrm{b}}}} \ . \label{4.10}
\end{align}

There is, however, a subtlety in applying \eqref{4.10} in the present setup. In the standard localisation argument, the quadratic action $S_{(2)}$ is obtained by linearising the $\mathcal Q$-complex around the localisation locus of the $\mathcal Q$-exact deformation $\mathcal Q V$. The localising functional $V$ is usually chosen so that the bosonic part of $\mathcal Q V$ is positive semi-definite, with zero modes appearing only along the localisation locus.

By contrast, in the present setup, \eqref{4.10q} is obtained by first choosing a supersymmetric invariant M2-brane satisfying the generalised calibration condition
\begin{align}\label{4.12cal}
\vol (\Sigma) = \frac{1}{2}\eta \wedge d\eta\big|_{\Sigma} \ ,
\end{align}
and then linearising the BST action around this saddle. The linearised $\mathcal Q$-complex is then derived from the residual worldvolume supersymmetry of the quadratic action, as in Section \ref{sec3.3}, rather than obtained by linearising a full localisation complex. Thus, the constants $\mathrm c$ are not arbitrary deformation parameters, but are fixed by the embedding of the M2-brane in $\ads\times \ses$. For the given values of $\mathrm c$, the quadratic action \eqref{4.10q} may therefore develop additional zero modes which are not cohomological zero modes of the $\mathcal Q$-complex. For instance, such modes may arise from an accidental degeneracy of the quadratic operator when a Fourier mode satisfies $i\mathcal L_{\xi}=\mathrm c$, while not lying in the $\mathcal Q$-cohomology. Importantly, these non-cohomological zero modes occur in $\mathcal Q$-doublets, so the corresponding bosonic and fermionic degeneracies are equal. This suggests that such modes may be lifted by a regulator preserving the supersymmetry complex. In particular, shifting $\mathrm c$ in \eqref{4.9q} changes the quadratic action $S_{(2)}$, but does not change the supersymmetry transformations \eqref{4.3s}, \eqref{4.4s}. Thus, for the non-cohomological zero modes, one may replace $\mathrm c$ by $\mathrm c+\delta \mathrm c$ as a regulator, thereby lifting these paired non-cohomological zero modes and allowing the supersymmetric cancellation of paired modes to be performed. In the limit $\delta \mathrm c\to 0$, the index-type expression \eqref{4.10} is therefore not modified.\footnote{An alternative regularisation of zero modes in the type IIA superstring computation was used in \cite{Gautason:2023igo}, by considering a physical deformation of the background dual to mass-deformed ABJM theory. In contrast, in the present case $\delta \mathrm c$ is introduced only to lift the paired zero modes and does not modify the supersymmetry complex. Consequently, the resulting index-type one-loop expression is independent of $\delta \mathrm c$.}

There is, however, a second type of zero mode in \eqref{4.10}, corresponding to genuine BPS moduli. In what follows, BPS modes or BPS moduli will refer to modes preserving the particular worldvolume supercharge $\mathcal Q$ used in the above supersymmetry complex, rather than to modes associated with an arbitrary Killing spinor of the ambient background. After integrating out the bosonic auxiliary field in \eqref{4.3s}, the tangent space to the BPS locus is described by
\begin{align}\label{4.13su}
\mathcal{Q}\Xr_{\mathrm{f}} = R\Xr_{\mathrm{b}} = 0 \ , \qquad \mathcal{Q}\Hr_{\mathrm{f}} = \bar{\partial}_{E}\Xr_{\mathrm{b}} = 0 \ ,
\end{align}
which are precisely the bosonic zero modes appearing in \eqref{4.10}. Unlike the non-cohomological modes discussed above, these zero modes need not be paired. They correspond to infinitesimal deformations preserving the linearised BPS equations, and describe tangent vectors to the supersymmetric locus. Their presence signals that the M2-brane is not an isolated supersymmetric saddle at the linearised level, but rather may belong to a supersymmetric moduli space. The corresponding fermionic zero modes in the cokernel contribution to \eqref{4.10} come from
\begin{align}\label{4.14su}
R\Hr_{\mathrm{f}} = 0 \ , \qquad \bar{\partial}_{E}^{\dagger}\Hr_{\mathrm{f}} = 0 \ ,
\end{align}
and may be interpreted as fermionic modes associated with obstruction directions to integrating the infinitesimal BPS deformations to finite deformations of the instanton.\footnote{The distinction between the two types of zero modes also has a geometric interpretation. The equations \eqref{4.13su} select deformations preserving the linearised BPS, or generalised calibration, condition \eqref{4.12cal}, similar to the standard calibrated examples of \cite{McLean:1998}. At the same time, since in the present case the calibrating form is not closed, the space of infinitesimal minimal deformations can be larger than the space of infinitesimal calibrated deformations. For instance, in the case of compact associative submanifolds in a weak $G_2$ manifold, the space of infinitesimal minimal non-associative deformations can be non-empty \cite{Kawai2017HomAssoc, KawaiSecondOrder}, as is also clear from \eqref{3.12}, \eqref{bosonicmain} and \eqref{3.27}. There, $\ker(\slashed{\mathfrak{D}}+1)$ corresponds to associative deformations, which are also minimal, while $\ker(\slashed{\mathfrak{D}}-3)$ corresponds to minimal non-associative deformations.} %A similar structure appears in the topological A-model \cite{Witten:1988xj}, with the equivariant $\mathcal Q$-complex replaced by the BRST complex.

Thus, at a regular point of the BPS locus, the infinitesimal bosonic BPS deformations of the given multiplet are described by sections $\Xr_{\mathrm b}$ satisfying the linearised BPS equations \eqref{4.13su}.
Equivalently, assuming that these infinitesimal deformations are unobstructed, the tangent space to the corresponding BPS moduli space is
\begin{align}
T\mathcal{M}^{E}_{\mathrm{BPS}}
=
\ker R \cap \ker \bar{\partial}_{E} \ .
\end{align}
However, within the present semiclassical approach, the quadratic fluctuation complex by itself does not determine a canonical measure on $\mathcal{M}^{E}_{\mathrm{BPS}}$. In a standard localisation argument, this measure would be inherited from the full path-integral measure, together with the one-loop contribution from the normal directions to the localisation locus. Since such a full localisation construction is not available here, we do not attempt to derive the moduli-space measure from first principles.

Additionally, if $\ker R\cap\operatorname{coker}\bar{\partial}_{E}$ is non-trivial, the quadratic action contains fermionic zero modes associated with the obstruction directions \eqref{4.14su}. Unless these fermionic zero modes are saturated by additional fermionic vertex insertions, or by interaction terms beyond the quadratic approximation, the corresponding saddle gives a vanishing contribution to the partition function. More generally, one expects the effective dimension of the BPS moduli space to be given by
the index
\begin{align}
    \dim\left(\ker R\cap \ker\bar{\partial}_{E}\right)
    -
    \dim\left(\ker R\cap \operatorname{coker}\bar{\partial}_{E}\right) \ .
\end{align}
Therefore, instanton configurations for which this index is negative are typically not expected to give contributions to the partition function.

\subsection{Index formula for the one-loop partition function}

When $\mathcal{Q}$-cohomological zero modes are absent, or have been treated by introducing collective coordinates, the one-loop partition function \eqref{4.10} can be computed using the equivariant index of the corresponding transversely elliptic operator, as in \cite{Pestun:2007rz}. Although the kernel and cokernel of a $U(1)_r$-transversely elliptic operator need not be finite-dimensional, they decompose as $U(1)_r$-modules into irreducible representations with finite multiplicities:
\begin{align}
    \ker \bar{\partial}_{E} = \oplus_n \, m_{n}^{(0)}W_{n} \ , \qquad \mathrm{coker} \, \bar{\partial}_{E} = \oplus_{n} \, m_{n}^{(1)}W_{n} \ ,
\end{align}
where $\{W_{n}\}$ denotes a complete set of inequivalent irreducible representations of $U(1)_{r}$. Assuming that the Reeb orbits on $\Sigma$ have length $2\pi$, the fields can be expanded in Fourier modes of the Lie derivative $-i\mathcal{L}_{\xi}\big|_{W_n}=n$. The one-loop partition function \eqref{4.10} can then be formally expressed as
\begin{align}
     Z_{1}^{(E)} = \prod_{n\in \mathbb{Z}}\big(n+\mathcal{R}\big)^{-\chi_{n}(E)} \ , \qquad \chi_{n}(E) = m_{n}^{(0)}-m_{n}^{(1)} \ ,
\end{align}
where $\mathcal{R}=2$ for the RW multiplet and $\mathcal{R}=0$ for the ML multiplet. 

The equivariant index determines the coefficient
$\chi_n(E)=m_n^{(0)}-m_n^{(1)}$ of each $U(1)_r$ weight, which is the combination entering the one-loop partition function. The
$U(1)_r$-equivariant index can be expressed, in a distributional sense, in terms of the $U(1)_r$-characters:
\begin{align}
    &\ind_{U(1)_{r}}(t, \bar{\partial}_{E}) =\mathrm{Tr}(t, \ker {\bar{\partial}}_{E})-\mathrm{Tr}(t, \mathrm{coker}\, \bar{\partial}_{E}) = \sum_{n\in \mathbb{Z}}\chi_{n}(E)t^n \ , \qquad t\in U(1)_{r} \ .
\end{align}
In the present case, as the Reeb flow is assumed to be regular, instead of computing the full equivariant index
%\footnote{The equivariant index admits a cohomological expression in terms of the Berline-Vergne formula \cite{Berline1996, BerlineVergne1996Indice}  as an integral of a non-compactly supported equivariant form. In \cite{paradan2008indextransversallyellipticoperators}, alternative expression for the index was proposed in terms of the integral of a compactly supported one, but with generealised coefficients. In the present case, application of these formulas appears to be excessive.}
one can directly compute $\chi_{n}(E)$ for each Fourier mode \cite{Atiyah:1974obx} (see also \cite{Pestun:2016qko}).
Namely, the $n$-th
Fourier mode corresponds to the subspace of sections $\mathrm{X}\in\Gamma(\Sigma,E)$
transforming under the $U(1)_r$ action as
\begin{align}
    (t\cdot \mathrm{X})(p)
    =
    t^{E}\mathrm{X}(p\cdot t^{-1})
    =
    t^{n}\mathrm{X}(p) \ ,
    \qquad
    \mathrm{X}\in \Gamma(\Sigma, E) \ .
\end{align}
The multiplicity of each Fourier mode is determined by the space of $U(1)_r$-homomorphisms \mbox{$W_n\to\Gamma(\Sigma,E)$}. Since the Reeb flow is regular, $\Sigma$ can be viewed as a principal $U(1)_r$-bundle
\begin{align}
    \pi:\Sigma\to B=\Sigma/U(1)_r \ .
\end{align}
As $E$ is a $U(1)_r$-equivariant vector bundle and the $U(1)_r$ action on $\Sigma$ is free, $E$ descends to a vector bundle $\widetilde E$ over $B$, with $\pi^*\widetilde E\simeq E$.
The multiplicity space of the $n$-th Fourier mode can then be identified~as
\begin{align}
    \mathrm{Hom}_{U(1)_r}
    \bigl(W_n,\Gamma(\Sigma,E)\bigr)
    \simeq
    \Gamma(B,\widetilde E\otimes L_n) \ ,
\end{align}
where $L_n$ denotes the associated line bundle\footnote{Here $W_n$ has character $t^n$, so that $W_n^*$ has character $t^{-n}$. With the right $U(1)_r$ action on $\Sigma$, the associated bundle $L_n$ is obtained from $\Sigma\times W_n^*$ by the equivalence relation $(p\cdot t,w)\sim (p,t^{-n}w)$.}
\begin{align} \label{4.18}
    L_n=\Sigma\times_{U(1)_r}W_n^*  \ .
\end{align}
Therefore, the coefficients $\chi_n(E)$ are determined by the index of the elliptic operator
\begin{align}
    \chi_n(E)
    =
    \ind(\bar{\partial}_{\widetilde{\mathcal W}_n}) \  ,
\end{align}
where
\begin{align}
    \bar{\partial}_{\widetilde{\mathcal W}_n}:
    \Gamma(B,\widetilde{\mathcal W}_n)
    \to
    \Omega^{0,1}(B,\widetilde{\mathcal W}_n) \ ,
    \qquad
    \widetilde{\mathcal W}_n
    \equiv
    \widetilde E\otimes L_n \ .
\end{align}

In holographic applications, such as M-theory on $\ads\times S^{7}/\mathbb{Z}_{k}$,  %discussed in Section \ref{sec2.3}, 
the internal space may be a finite quotient $\widehat{\mathrm{SE}}_{7}=\ses/G$, where 
%$\ses$ is a simply connected Sasaki-Einstein manifold and 
$G$ is a finite group of isometries commuting with the $U(1)_r$ action. In general, the action of $G$ on $\ses$, or its restriction to the invariant link $\Sigma\subset \ses$, may have fixed loci, in which case $\widehat{\Sigma}=\Sigma/G$ should be understood as an orbifold link. We further assume that the relevant bundle $E\to\Sigma$ is $G$-equivariant, that the $G$ action on $E$ commutes with the $U(1)_r$ action, and that $\bar\partial_E$ is $G$-equivariant. The Fourier-mode decomposition described above can then be performed on the covering link $\Sigma$, and the projected fields on the quotient $\widehat\Sigma$ are obtained by taking the $G$-invariant part. Since $G$ is finite, the projection can be performed at the level of the $G$-equivariant index:
\begin{align}
    \ind_G(s,\bar{\partial}_{\widetilde{\mathcal W}_n})
    &=
    \mathrm{Tr}
    \bigl(s,\ker \bar{\partial}_{\widetilde{\mathcal W}_n}\bigr)
    -
    \mathrm{Tr}
    \bigl(s,\mathrm{coker}\,\bar{\partial}_{\widetilde{\mathcal W}_n}\bigr) \ ,
    \\
    \widehat{\chi}_n^{G}(E)
    &=
    \frac{1}{|G|}
    \sum_{s\in G}
    \ind_G(s,\bar{\partial}_{\widetilde{\mathcal W}_n}) \  . 
    \label{4.21}
\end{align}
Here $\widehat{\chi}_n^{G}(E)$ denotes the $G$-invariant contribution resulting from the orbifold projection. The individual terms appearing on the right-hand side of \eqref{4.21} can be computed using the Atiyah-Bott formula. For instance, for $s\neq 1$, if the fixed locus $B^s$ consists of isolated non-degenerate fixed points, one may use the equivariant Lefschetz formula
\begin{align}\label{4.23l}
\ind_{G}(s , \bar{\partial}_{\widetilde{\mathcal{W}}_{n}}) = \sum_{x_{0}\in B^{s}}\frac{\mathrm{Tr}(s^{\widetilde{{\mathcal{W}}}_{n}}_{x_{0}})}{\det_{T_{x_{0}}^{1, 0}B}(1-s_{x_{0}}^{-1})} \ .
\end{align}
For $s=1$, the fixed locus is all of $B$, and the contribution is the ordinary index, computed by the Riemann-Roch theorem.

Combining these contributions and allowing the Reeb orbits on the covering link $\Sigma$ to have length $\ell_{\xi}$, the non-zero mode part of the one-loop partition function for an invariant M2-brane in the quotient $\ses/G$ can be written as the Riemann $\zeta$-function regularised product
\begin{align}
    Z_{1}'
    =
    \prod_{n\in \mathbb{Z}}
    \Big(\frac{2\pi n}{\ell_{\xi}}+2\Big)^{-2\widehat{\chi}^{G}_{n}}
    \prod_{n\in \mathbb{Z}}
     \Big(\frac{2\pi n}{\ell_{\xi}}\Big)^{-\widehat\chi^{G}_{n}(N\Sigma)} \ ,
    \label{4.24}
\end{align}
where the prime indicates that the Fourier modes $2\pi n/\ell_\xi+2=0$ in the RW sector and $n=0$ in the ML sector, when present, correspond to zero weights of $R=\mathcal Q^2$ and should be omitted from the product. When these zero modes are absent, the non-zero mode partition function coincides with the ordinary one-loop partition function, $Z_1=Z_1'$. The additional factor of $2$ in the RW sector comes from the $SU(2)_{-}$ doublet structure, while
$\widehat\chi^{G}_{n}$ and $\widehat{\chi}_{n}^{G}(N\Sigma)$ are determined by \eqref{4.21} with $E$ taken to be the trivial bundle and the normal bundle, respectively.

\subsection{Examples: M2-brane instantons in $S^7/\mathbb{Z}_k$} \label{sec4.2}

To illustrate the formula \eqref{4.24}, we first apply it to the $\widehat\Sigma_{\pm}=\Sigma_{\pm}/\mathbb{Z}_k$ M2-brane instantons considered in Section \ref{sec2.3}, and then to their analogue in the $(p,q)$-model. We then consider the more general M2-brane instantons discussed in Section \ref{sec2.3}, corresponding to curves intersecting the distinguished components $B_{\pm}$. For these saddles, cohomological zero modes are generically present and must be taken into account. To apply \eqref{4.24}, we use that, for the covering invariant link $\Sigma\subset S^7$, the length of the Reeb circle fibre is $\ell_\xi=2\pi$, while the quotient by the finite group $G$ is implemented by restricting to the $G$-invariant modes on the covering link.

We recall that, for each Fourier mode, the unprojected coefficient $\chi_n(E)$ is the Euler characteristic of the holomorphic vector bundle $\widetilde{\mathcal{W}}_{n}$ over the Riemann surface $B$:
\begin{align}
\chi_{n}(E)
=
\chi(B,\widetilde{\mathcal{W}}_{n})
=
\deg \widetilde{\mathcal{W}}_{n}
-
\operatorname{rank}\widetilde{\mathcal{W}}_{n}(g-1) \ ,
\end{align}
where $g$ is the genus of $B$, and $\deg \widetilde{\mathcal{W}}_{n}$ and $\operatorname{rank}\widetilde{\mathcal{W}}_{n}$ denote the degree and rank of the holomorphic vector bundle $\widetilde{\mathcal{W}}_{n}$, respectively.\footnote{This is the Riemann-Roch theorem for holomorphic vector bundles over a compact Riemann surface (see, for example, the appendix of \cite{Wells:2008}).}

\subsection*{M2-brane instantons $\widehat\Sigma_{\pm}$ in $S^7/\mathbb{Z}_k$}

The simplest case corresponds to the links $\Sigma_{\pm}$ in $S^7$ for which the one-loop fluctuations were considered in \cite{Beccaria:2023ujc}, based on the type IIA superstring analysis of \cite{Gautason:2023igo}. In terms of the projective curves on the Kähler base of the $U(1)_r$ Hopf fibration $S^7\to \mathbb{P}^3$, these simply correspond to the projective lines
\begin{align}
    B_{\pm}\simeq \mathbb{P}^1 \subset \mathbb{P}^3 \ ,
\end{align}
where, as in Section \ref{sec2.3}, the curves $B_{\pm}$ are defined by
\begin{align}
    B_{+} = \{z_{3}=z_{4} = 0 \} \ , \qquad B_{-} = \{z_{1}=z_{2}=0\} \ .
\end{align}
On the corresponding links $\Sigma_{\pm}$, the orbits of $U(1)_r$ and $U(1)_b$ coincide, up to orientation. Hence the expansion in $U(1)_r$ Fourier modes is equivalent to the expansion in the M-theory circle modes. Since $\mathbb{Z}_{k}\subset U(1)_b$ acts on $\Sigma_{\pm}$ as a subgroup of the Reeb circle, the $\mathbb{Z}_k$ projection does not mix different $U(1)_r$ weights.

With the above conventions, the associated line bundle $L_n$ is simply $\mathcal{O}(n)$. By the Riemann-Roch theorem,
\begin{align} \label{4.27e}
    \chi(B_{\pm}, \mathcal{O}(n)) = n\deg B_{\pm}+1-g \ , 
\end{align}
where $\deg B_{\pm}=1$ and $g=0$ for $\mathbb{P}^1$. For $n\geq 0$, holomorphic sections of $\mathcal{O}(n)$ can be represented by degree $n$ homogeneous polynomials in homogeneous coordinates on $\mathbb{P}^1\subset\mathbb{P}^3$, while for $n=-1$ the index vanishes. For $n\leq -2$, one can use Serre duality, with canonical bundle $K_{\mathbb{P}^1}=\mathcal{O}(-2)$. Thus the $\mathbb{Z}_k$-equivariant index gives
\begin{align}
    \widehat{\chi}^{\mathbb{Z}_{k}}_n = n+1 \ , \qquad n\in k\mathbb{Z} \ ,
\end{align}
and vanishes otherwise. Importantly, for $k>2$, the $\mathbb{Z}_k$ projection removes the Fourier mode with weight $n=-2$:
\begin{align} 
\widehat{\chi}^{\mathbb{Z}_{k}}_{-2} = -\dim\big(\operatorname{coker}\bar{\partial}_{\mathcal{O}(-2)}\big)^{\mathbb{Z}_k} = 0 \ . 
\end{align}
Thus, after the $\mathbb{Z}_k$-projection, the RW cohomological zero modes discussed in Section \ref{secc4.1} are absent. For $k>2$, the RW contribution to the partition function is obtained by restricting the Fourier label to $n=km$:
\begin{align} \label{4.29e}
    Z_{1}^{\mathrm{RW}} = \prod_{m\in \mathbb{Z}}(km+2)^{-2(km+1)} \ .
\end{align}
Similarly, to compute the contribution from the ML multiplet, one can use that, by construction,
\begin{align}
    \widetilde{N\Sigma}_{\pm} \simeq N_{B_{\pm}/\mathbb{P}^3} \simeq \mathcal{O}(1)\oplus \mathcal{O}(1) \ , 
\end{align}
where the normal directions correspond to the $z_{3,4}$ directions for $B_{+}$ and to the $z_{1,2}$ directions for $B_{-}$. The Riemann-Roch theorem then gives
\begin{align}\label{4.31e}
    \chi_n(N\Sigma_{\pm}) = \chi\big(B_{\pm}, \mathcal{O}(n+1)\oplus \mathcal{O}(n+1)\big) = 2(n+2) \ .
\end{align}

For the ML multiplet, to impose the $\mathbb{Z}_k$ projection one should also account for the non-trivial transformation of the normal bundle:
\begin{align}
    (s\cdot \mathrm{X})(p) = s^{N\Sigma}\mathrm{X}(p\cdot s^{-1}) \ , \qquad s\in \mathbb{Z}_k \ .
\end{align}
For instance, for $B_{+}$, a section of the normal bundle corresponding to the $n$-th Fourier mode, with $n\geq -1$, can be represented by a pair of homogeneous polynomials
\begin{align}
    (\delta z_{3}, \delta z_4) = (f(z_{1}, z_2), g(z_1, z_2)) \ , \qquad f, g\in \mathbb{C}[z_1, z_2]_{n+1} \ .
\end{align}
Using the $\mathbb{Z}_k$ action \eqref{2.47f}, the $n$-th Fourier mode of $N_{B_{+}/\mathbb{P}^3}$ transforms under the generator $\omega$ of $\mathbb{Z}_k$ as
\begin{align}
    \omega \cdot (\delta z_3, \delta z_4) = \omega^{-n-2}(\delta z_3, \delta z_4) \ ,
\end{align}
and similarly for $N_{B_{-}/\mathbb{P}^3}$ with weight $\omega^{n+2}$. Thus, the $\mathbb{Z}_k$ projection, selecting $\mathbb{Z}_k$-invariant sections, gives
\begin{align}
    \widehat{\chi}_n^{\mathbb{Z}_k}(N\Sigma_{\pm}) = 2(n+2) \ , \qquad n+2\equiv 0 \bmod k \ ,
\end{align}
and vanishes otherwise. Crucially, as in the RW sector, the $n=0$ ML cohomological zero modes are projected out for $k>2$. Thus, for $k>2$, the contribution of the ML multiplet is
\begin{align} \label{4.36e}
    Z^{\mathrm{ML}}_1 = \prod_{m \in \mathbb{Z}} (km-2)^{-2km}  = \prod_{m\in \mathbb{Z}}(km+2)^{2km}\ .
\end{align}
Combining the RW and ML contributions \eqref{4.29e} and \eqref{4.36e}, and using Riemann $\zeta$-function regularisation, one obtains the total one-loop partition function
\begin{align}\label{4.37}
    Z_{1} = \prod_{m\in \mathbb{Z}}(km+2)^{-2} = \frac{1}{4\sin^2(\frac{2\pi}{k})} \ , 
\end{align}
which reproduces the result of \cite{Beccaria:2023ujc}. Since the $\widehat\Sigma_{\pm}$ M2-brane instantons have no BPS moduli after the $\mathbb{Z}_k$ projection, the resulting one-loop factor can be interpreted as an isolated-saddle contribution. In the present setup, the overall factor of $4$ may be attributed to the counting of the two saddles $\widehat\Sigma_{\pm}$ and their opposite orientations, equivalently the two choices of $\gamma_{\hat{5}}$ chirality in \eqref{2.13chiral}.\footnote{The analysis of $\frac{1}{2}$-BPS M2-brane instantons in $S^7/\mathbb Z_k$ was carried out in \cite{Park:2020hgt, Gautason:2025per}. An alternative counting prescription was proposed in \cite{Gautason:2025per}, where independent configurations, corresponding to internal cycles on $S^7$ preserving the same chirality of the AdS$_4$ Killing spinor, are obtained by quotienting the BPS family by the isometries of the background. In the present setup, however, the two distinguished M2-brane configurations $\Sigma_{\pm}$, related by $z_{1,2}\leftrightarrow z_{3,4}$, are selected by requiring the $U(1)_r$ and $U(1)_b$ orbits to coincide on them, up to orientation.}

\begin{comment}
\footnote{
We note that, if one formally keeps the degree $\deg B$ and genus $g$ arbitrary in \eqref{4.27e}, \eqref{4.31e}, and assumes that the $\mathbb{Z}_{k}$ projection acts in the same way as for $B_{\pm}$, the combined contribution becomes
\begin{align*}
    Z'_{1}
    =
    \prod_{m\in \mathbb{Z}}(km+2)^{-2+2g}
    =
    \frac{1}{\left[2\sin\left(\frac{2\pi}{k}\right)\right]^{2-2g}} \ .
\end{align*}
In this formal expression the dependence on $\deg B$ cancels between the RW and ML multiplets. Geometrically, however, the assumption that the $\mathbb{Z}_{k}$ projection acts in this simple way is special to the fixed components $B_{\pm}$. Thus, such a contribution is most naturally associated with cycles supported on these components, for example $B = d_{+}B_{+}+d_{-}B_{-}$, 
for which, since $B_{+}\cap B_{-}=\emptyset$, one has $\deg B=d_{+}+d_{-}$. It is tempting to relate these M2-brane contributions to the higher degree ``worldsheet'' instantons appearing in \eqref{ferminp} but this identification is not direct in the present setup. In particular, for general multiplicities $d_{\pm}$ the corresponding object is a reducible cycle rather than a single smooth irreducible M2-brane saddle.}
\end{comment}

\subsection*{M2-brane instantons $\widehat\Sigma_{\pm}$ in $(S^7/(\mathbb{Z}_{p}\times \mathbb{Z}_{q}))/\mathbb{Z}_k$}

To illustrate how the projection may act for a different choice of the finite group $G$, let us consider the case relevant for the $(p,q)$-model describing the low-energy effective theory of $N$ M2-branes probing the orbifold $(\mathbb{C}^4/(\mathbb{Z}_{p}\times \mathbb{Z}_{q}))/\mathbb{Z}_{k}$ \cite{Imamura:2008nn, Imamura:2008dt}, with M-theory dual on $\ads\times (S^7/(\mathbb{Z}_p \times \mathbb{Z}_{q}))/\mathbb{Z}_k$. In terms of the coordinates on $\mathbb{C}^4$, the orbifold action can be described as
\begin{align}
a\cdot (z_{1}, z_{2}, z_{3}, z_{4})& = (z_{1}, e^{2\pi i/p}z_{2}, z_{3}, e^{-2\pi i/p}z_{4}) \ , \quad b\cdot(z_{1}, z_{2}, z_{3}, z_{4}) =  (e^{2\pi i/q}z_{1}, z_{2}, e^{-2\pi i/q}z_{3}, z_{4}) \ , \nonumber \\
&\omega \cdot(z_{1}, z_{2}, z_{3}, z_{4}) =(e^{2\pi i/kq}z_{1}, e^{2\pi i/kp}z_{2}, e^{-2\pi i/kq}z_{3}, e^{-2\pi i/kp}z_{4}) \ ,
\end{align}
so that
\begin{align}
G = \langle a, b, \omega\rangle  \ , \quad  a^p=1 \ , \quad b^{q} = 1 \ , \quad \omega^k = a b \ .
\end{align}

The distinguished invariant projective lines, analogous to the fixed components in the $S^7/\mathbb{Z}_k$ case, are
\begin{align}
B_{+} = \{z_{3}=z_{4} = 0 \} \ , \qquad B_{-} = \{z_{1}=z_{2}=0\} \ ,
\end{align}
and are preserved setwise by the action of $G$. The corresponding links $\Sigma_{\pm}$ on the covering $S^7$ are the same as in the ABJM case, while the volume of the quotient links is
\begin{align}
\mathrm{Vol}(\widehat{\Sigma}) = \frac{1}{pqk}\mathrm{Vol}(\Sigma) = \frac{2\pi^2}{pqk} \ ,
\end{align}
where $pqk$ is the effective order of the quotient group acting on $\Sigma_{\pm}$. These M2-brane configurations were also proposed to be dual to the leading worldsheet-instanton appearing in the $(p,q)$-model in \cite{Hatsuda:2015lpa}.

%\footnote{For the symmetric case $p=q=r$, the corresponding leading contribution in the type IIA limit was also reproduced using a semiclassical one-loop analysis of the type IIA superstring in \cite{Gautason:2023igo}.}

The calculation of the one-loop partition function is then similar to the case of $S^7/\mathbb{Z}_k$, with a modification due to the finite-group quotient. Let us consider, for instance, the case of $\Sigma_{+}$. For the RW multiplet, the projection condition should be understood by requiring the homogeneous polynomials of degree $n\geq 0$ to be invariant under the action of the quotient group $G$. Imposing first the $\mathbb{Z}_{p}\times \mathbb{Z}_{q}$ invariance gives the allowed monomials
\begin{align}
z_1^{q n_1}z_2^{p n_2} \ , \quad n_1,n_2\in \mathbb{Z}_{\geq 0} \ ,
\end{align}
with $U(1)_r$ weight $n=qn_1+pn_2$. The remaining $\mathbb{Z}_k$ projection requires
\begin{align}
n_1+n_2\equiv 0 \bmod k \ .
\end{align}
The same calculation applies to $\Sigma_{-}$ with the opposite phases but the same multiplicities. To express the equivariant index as a number counting $G$-invariant sections, it is convenient to introduce the lattice
\begin{align}
\Lambda_{n}^{\epsilon_{1}, \epsilon_{2}} = \{(n_{1}, n_{2})\in \mathbb{Z}_{\geq\epsilon_{1}}\times\mathbb{Z}_{\geq \epsilon_{2}} \, | \, n=qn_{1}+pn_{2} , \ n_{1}+n_{2} \equiv 0 \bmod k \} \ .
\end{align}
With this notation, using Serre duality on the covering projective line and applying the projection there as in the $S^7/\mathbb{Z}_k$ case, the equivariant index controlling the RW multiplet is
\begin{align}
&\widehat{\chi}_{n}^{G} = \begin{cases}
|\Lambda_{n}^{0, 0}| \ , \quad n\geq0 \ , \\ \  0 \ , \quad n =-1 \ , \\
-|\Lambda_{-n}^{1, 1}|\  , \quad n\leq -2 \ .
\end{cases}
\end{align}
Importantly, as in the $S^7/\mathbb{Z}_k$ case, the projected RW cohomological zero modes are absent, provided $qk>2$ and $pk>2$. Similarly, for the ML multiplet we obtain
\begin{align}
&\widehat{\chi}_{n}^{G}(N\Sigma_{\pm}) = \begin{cases}
|\Lambda_{n+2}^{1, 0}|+|\Lambda^{0, 1}_{n+2}| \ , \quad n\geq-1 \ , \\ \  0 \ , \quad n =-2 \ , \\
-|\Lambda_{-(n+2)}^{1, 0}|-|\Lambda_{-(n+2)}^{0, 1}|\  , \quad n\leq -3 \ .
\end{cases} 
\end{align}
For the ML multiplet, the cohomological zero modes are also absent provided $qk>2$ and $pk>2$. Thus, after reindexing the ML contribution, the partition function in this case can be expressed as
\begin{align}
Z_{1} = \prod_{n\geq0}(n+2)^{-2|\Lambda_{n}^{0, 0}|}\prod_{n\geq 3}(n-2)^{2|\Lambda_{n}^{1,1}|}\prod_{n\geq 3}(n-2)^{-|\Lambda_{n}^{1, 0}|-|\Lambda_{n}^{0, 1}|}\prod_{n\geq 1}(n+2)^{|\Lambda_{n}^{1,0}|+|\Lambda_{n}^{0,1}|} \ .
\end{align}
Combining the RW and ML multiplets, and using the cancellations among the lattice points
\begin{align}
&\prod_{n\geq1}(n+2)^{|\Lambda_{n}^{1, 0}|+|\Lambda_{n}^{0, 1}|-2|\Lambda_{n}^{0, 0}|} = \prod_{m\geq 1}(kqm+2)^{-1}(kpm+2)^{-1} \ , \\
&\prod_{n\geq 3}(n-2)^{2|\Lambda_{n}^{1,1}|-|\Lambda_n^{1,0}|-|\Lambda_{n}^{0, 1}|} = \prod_{m\geq 1}(kqm-2)^{-1}(kpm-2)^{-1} \ ,
\end{align}
the one-loop partition function combines into
\begin{align}\label{4.56aaaa}
Z_{1} = \prod_{m\in \mathbb{Z}}(pkm+2)^{-1}\prod_{m\in \mathbb{Z}}(qkm+2)^{-1}
= \frac{1}{4\sin(\frac{2\pi}{pk})\sin(\frac{2\pi}{qk})} \ ,
\end{align}
where the last equality, as above, is understood in Riemann $\zeta$-function regularisation. The resulting one-loop partition function should be compared with the leading worldsheet-instanton contribution conjectured in \cite{Hatsuda:2015lpa}:
\begin{align}
d_{1}(p, q, k) = \frac{pq}{\sin(\frac{2\pi}{pk})\sin(\frac{2\pi}{qk})} \ . \label{4.57aaa}
\end{align}
%v3
This has the same non-trivial sine dependence as the one-loop partition function of a BPS M2-brane instanton \eqref{4.56aaaa}, while the remaining overall numerical factor $4pq$ can be separated into a factor $4$, attributed to the counting of BPS saddles, and a factor $pq=|\mathbb{Z}_{p}\times\mathbb{Z}_{q}|$, interpreted as an additional orbifold weight of the M2-brane instanton.\footnote{A useful check of this interpretation is provided by the symmetric case $p=q=r$. In the type IIA analysis of \cite{Gautason:2023igo}, the same additional factor $r^2$ is required to reproduce the matrix-model result. %and was argued to appear from the orbifold geometry of the string worldsheet $\mathbb{P}^{1}/\mathbb{Z}_{r}$.
At the same time, the normalisation of \eqref{4.56aaaa} is consistent with the type IIA limit $rk\gg1$. Indeed, the string partition function for a spherical worldsheet is expected to scale as $T/(2\pi g_s^2)$ \cite{Giombi:2020mhz,Gautason:2023igo,Beccaria:2023ujc}, where, for the orbifold background, $T=\sqrt{\lambda/2}$, $g_s^2=\pi(2\lambda)^{5/2}/(Nr)^2$, and $\lambda=N/k$. Thus, for $rk\gg 1$, \eqref{4.56aaaa} becomes $Z_{1}\approx (rk/4\pi)^2=T/(2\pi g_s^2)$, which reproduces the expected type IIA normalisation of an individual saddle, while the matrix-model coefficient \eqref{4.57aaa} contains, after separating the factor $4$ attributed to saddle counting, the additional factor $r^2$.}

\subsection*{M2-branes intersecting $B_{\pm}$ in $S^7/\mathbb{Z}_k$}

One can also consider M2-brane instantons in $S^7/\mathbb Z_k$ corresponding to $\mathbb{Z}_k$-invariant projective curves $B\subset \mathbb P^3$, distinct from $B_{\pm}$, and intersecting the components $B_{\pm}$ in finite, possibly empty, sets of points as discussed in Section \ref{sec2.3}. We assume that $B\cap B_{\pm}$ consists of smooth points of $B$, and that, for every element of $\mathbb Z_k$ acting projectively non-trivially on $\mathbb P^3$, each such point is a non-degenerate fixed point of the induced action on $B$. Let $N_{\pm}$ denote the number of intersection points $x_0^{\pm}\in B\cap B_{\pm}$.

Let us first analyse the contribution from the RW multiplet. Using that $\deg L_n=n\deg B$, the Riemann-Roch theorem gives
\begin{align}
\chi_n=\chi(B,L_n)
=
n\deg B+1-g \ .
\end{align}
In particular, before applying the $\mathbb Z_k$ projection, the index for the Fourier mode $n=-2$ is
\begin{align}
\chi_{-2}
=
1-g-2\deg B \ .
\end{align}
Thus, for every non-constant projective curve $B$, the unprojected RW index $\chi_{-2}$ is already negative and decreases further for higher-degree curves.

To consider the $\mathbb Z_k$-projection, one has to evaluate the equivariant fixed-point contributions. For $x_0^+\in B\cap B_+$, the projective action on $T\mathbb P^3|_{x_0^+}$ has weights
\begin{align}
T\mathbb P^3|_{x_0^+}:
\quad
\{1,\omega^{-2l},\omega^{-2l}\}\ .
\end{align}
For $\omega^{2l}\neq1$, the non-degeneracy assumption at the fixed point implies that the tangent line $T_{x_0^+}B$ has weight $\omega^{-2l}$. Hence the normal fibre has weights
\begin{align}
N_{B/\mathbb P^3}|_{x_0^+}:
\quad
\{1,\omega^{-2l}\}\ .
\end{align}
Similarly, for $\omega^{2l}\neq1$ and $x_0^-\in B\cap B_-$, one finds
\begin{align}
T_{x_0^-}B:
\quad
\{\omega^{2l}\}\ ,
\qquad
N_{B/\mathbb P^3}|_{x_0^-}:
\quad
\{1,\omega^{2l}\}\ .
\end{align}

The character of $L_n$ at the fixed point is determined by the lift of the $\mathbb Z_k$ action to the Hopf fibre. With the convention \eqref{4.18} and $s=\omega^{l}$, one obtains
\begin{align}
\mathrm{Tr}(s^{L_n}_{x_0^+})
=
\omega^{-ln}\ ,
\qquad
\mathrm{Tr}(s^{L_n}_{x_0^-})
=
\omega^{ln}\ .
\end{align}
For elements $s=\omega^l$ with $\omega^{2l}\neq1$ the projective action has isolated fixed points on $B$ and the equivariant Lefschetz formula \eqref{4.23l} gives
\begin{align}
\operatorname{ind}
(\omega^l,\bar\partial_{L_n})
=
N_+
\frac{\omega^{-ln}}{1-\omega^{2l}}
+
N_-
\frac{\omega^{ln}}{1-\omega^{-2l}}\ , \qquad \omega^{2l}\neq 1 \ .
\end{align}
To obtain the $\mathbb Z_k$-projected index \eqref{4.21}, it remains to treat the projectively trivial elements. The identity element $l=0$ contributes the ordinary Riemann-Roch index $\chi_n$. If $k$ is even, there is an additional element $l=k/2$, which acts trivially projectively on $\mathbb P^3$, but acts on $L_n$ by multiplication by $(-1)^n$. Thus, defining
\begin{align}
P_0(n)
=
\begin{cases}
1 \, , & k \ \mathrm{odd}\, ,\\
1+(-1)^n \, , & k \ \mathrm{even}\, ,
\end{cases}
\end{align}
one obtains
\begin{align}
\widehat\chi_n^{\mathbb Z_k}
=
\frac{1}{k}
\Big[
P_0(n)\chi_n
+
\sum_{\substack{1\le l\le k-1\\ \omega^{2l}\neq1}}
\left(
N_+
\frac{\omega^{-ln}}{1-\omega^{2l}}
+
N_-
\frac{\omega^{ln}}{1-\omega^{-2l}}
\right)
\Big] .
\end{align}
Specialising to $n=-2$, this gives the projected RW index
\begin{align}
\widehat{\chi}_{-2}^{\mathbb Z_k}
=
\begin{cases}
\frac{1}{k}
\left[
\chi_{-2}
-
\frac{k-1}{2}(N_++N_-)
\right] ,
& k \ \mathrm{odd}\ ,\\[3pt]
\frac{1}{k}
\left[
2\chi_{-2}
-
\frac{k-2}{2}(N_++N_-)
\right] ,
& k \ \mathrm{even}\ .
\end{cases}
\end{align}
For example, for the remaining four toric edges one has $g=0$, $\deg B=1$, and $N_+=N_-=1$. Hence $\chi_{-2}=-1$, and the projected RW index is $\widehat\chi_{-2}^{\mathbb{Z}_k}=-1$ for both odd and even $k$. More generally, since $B$ is not contained in either fixed component $B_{\pm}$, the number of intersection points is expected to be bounded by
\begin{align}
N_++N_-
\leq
2\deg B \ .
\end{align}
Under this assumption, one finds
\begin{align}
\chi_{-2}
\leq
\widehat\chi_{-2}^{\mathbb Z_k}
<0 \ .
\end{align}
This implies that such M2-brane configurations carry genuine fermionic cohomological zero modes for the RW multiplet. Therefore, the index governing the expected dimension of the BPS moduli space associated with the RW multiplet is negative.

One can also analyse the expected dimension of the BPS moduli space for the ML sector. The adjunction formula for a smooth projective curve $B\subset\mathbb P^3$ gives
\begin{align}
K_B
\simeq
K_{\mathbb P^3}|_B\otimes \det N_{B/\mathbb P^3}
\simeq
\mathcal O_B(-4)\otimes \det N_{B/\mathbb P^3}\ .
\end{align}
Taking degrees and using $\deg K_B=2g-2$, one obtains
\begin{align}
\deg N_{B/\mathbb P^3}
=
4\deg B+2g-2\ .
\end{align}
Consequently, since the cohomological modes in the ML multiplet appear in the $n=0$ Fourier mode,
\begin{align}
\chi(B,N_{B/\mathbb P^3})
=
\deg N_{B/\mathbb P^3}+2(1-g)
=
4\deg B \ .
\end{align}
Thus the unprojected index for the ML multiplet grows with the degree of the curve. 

More generic M2-brane instantons in $S^7/\mathbb{Z}_k$ therefore exhibit two related difficulties. The ML sector may contain bosonic BPS moduli, for which the present semiclassical analysis does not provide a canonical integration measure. More importantly, for the partition function without insertions, the RW sector contains fermionic $\mathcal{Q}$-cohomological zero modes. Unless these modes are saturated by additional insertions or by interaction terms beyond the quadratic approximation, the corresponding saddle is not expected to contribute to the partition function.

\section{Concluding remarks}\label{sec5}

In this paper, we have studied invariant BPS M2-brane instantons in Sasaki-Einstein seven-manifolds $\ses$, viewed as a special class of instantons in weak $G_{2}$ backgrounds. For such invariant M2-branes, which preserve both real internal Killing spinors of $\ses$, we derived the supersymmetric complex for quadratic fluctuations and expressed the one-loop partition function using the corresponding $U(1)_r$-equivariant indices. We then applied this formula to the basic M2-brane instantons in $S^{7}/\mathbb Z_{k}$ and in $(S^7/(\mathbb Z_p\times \mathbb Z_q))/\mathbb Z_k$, reproducing the characteristic sine dependence expected from the matrix-model results.

The simplification found for invariant M2-branes is a consequence of the preserved worldvolume supersymmetry of the quadratic fluctuation action. This supersymmetry allows one to decompose the fluctuation fields into $U(1)_r$ Fourier modes and organise them into equivariant $\mathcal Q$-complexes, with $\mathcal Q^2$ identified with the induced Reeb generator on the M2-brane worldvolume. For isolated BPS saddles, where the relevant zero modes are absent, the resulting index formula gives a well-defined Riemann $\zeta$-function regularised one-loop partition function. The simplest example is the $S^3/\mathbb Z_k\subset S^7/\mathbb Z_k$ instanton with $k>2$, for which the $\mathcal{Q}$-cohomological zero modes are removed by the $\mathbb Z_k$ quotient and the one-loop factor has an isolated-saddle interpretation. The cases $k=1,2$, where this zero-mode projection is modified, require a separate treatment.

%The simplest example is the $S^3/\mathbb Z_k\subset S^7/\mathbb Z_k$ instanton, for which the relevant moduli are absent from the zero $R$-weight sector after the quotient, and the one-loop factor has the expected isolated-saddle interpretation.

For more general invariant M2-brane instantons, however, the same index formula points to an important limitation of the single-saddle semiclassical analysis. The potentially problematic modes occur precisely in the zero $R=\mathcal Q^2$ weight part of the equivariant $\mathcal Q$-complex. In particular, the presence of RW multiplet fermionic zero modes in the obstruction space
\begin{align}
\ker R\cap\operatorname{coker}\bar\partial_{H}  \ ,
\end{align}
suggests that single M2-brane saddles corresponding to higher-degree links in $S^7/\mathbb Z_k$ are not expected to contribute to the purely bosonic partition function within this single-saddle approximation, unless these modes are saturated by additional insertions or interactions.\footnote{This limitation of the single-saddle approximation may also be relevant for understanding the D2-brane sectors in~\eqref{ferminp}, whose coefficients $f_{0,l}(k,\mu)$ depend on the chemical potential $\mu$ in the Fermi gas approach. For concreteness, in ABJM theory, in the type IIA limit the corresponding instanton action was identified with the classical action $\pi\sqrt{2kN}$ of a Euclidean D2-brane wrapping $\mathbb{RP}^{3}\subset\mathbb{CP}^{3}$ \cite{Drukker:2011zy} (see also \cite{Park:2020hgt,Beccaria:2023ujc} for its proposed M2-brane uplift; note, however, that for even $k>2$ the same value of the classical action is also obtained for higher-degree M2-brane links, such as the Brieskorn-type example discussed in footnote~6). From the matrix-model perspective, for even $k$, the single D2-instanton contribution and the $k/2$ worldsheet-instanton contribution have the same exponential weight $\exp(-2\mu)$ and, although their coefficients are separately divergent, their combined contribution is finite  \cite{Hatsuda:2012dt, Calvo:2012du, Hatsuda:2013gj, Hatsuda:2013oxa}. This suggests that a complete M2-brane description may require the combined treatment of such action-degenerate saddles and their zero modes, rather than their interpretation as isolated one-loop contributions.} %v3

This observation should not be interpreted as excluding the higher instanton corrections appearing in the double-series expansion of the modified grand potential \eqref{ferminp}. Rather, it raises a puzzle for their interpretation in terms of a single M2-brane semiclassical expansion. The present analysis is restricted to a single M2-brane wrapped on a BPS cycle in $\ses$. It should therefore be regarded as a semiclassical model for an individual membrane instanton saddle, rather than as a treatment of possible multi-M2-brane configurations. It would be interesting to understand whether the observed saddle-level equivariant $\mathcal Q$-complex can be promoted to a full equivariant localisation argument for the M2-brane path integral, in line with recent proposals for localising M2-branes \cite{Gautason:2025per, Gautason:2025plx, vanMuiden:2026nsp}. Such a formulation may clarify the role of obstruction zero modes and the range of validity of the single M2-brane semiclassical expansion.

Let us finally comment on extensions beyond the sphere-quotient examples considered explicitly in this paper, and on their possible holographic interpretation. The index formula \eqref{4.24} was derived for invariant M2-branes in Sasaki-Einstein seven-manifolds, and should therefore be directly applicable whenever an isolated BPS saddle can be identified. It is natural to ask whether such M2-brane instantons admit a counterpart in the localisation matrix models of the holographic dual Chern-Simons matter theories. In several $\mathcal N\geq 2$ examples, the large-$N$, fixed Chern-Simons level limit of the matrix model reproduces the expected volume dependence of the gravitational free energy \cite{Martelli:2011qj}. This includes the $\mathcal N=3$ theories dual to tri-Sasaki-Einstein spaces \cite{Herzog:2010hf}, as well as related $\mathcal N=2$ examples dual to $Q^{1,1,1}$ and $V^{5,2}$ \cite{Cheon:2011vi}. The corresponding non-perturbative corrections to the free energy are, however, much less understood in those cases. It would therefore be interesting to investigate whether the M2-brane instantons considered here, whose volumes are governed by the generalised calibration condition
\begin{align}
    \mathrm{vol}({\Sigma})
    =
    \frac{1}{2}\eta \wedge d\eta\big|_{\Sigma} \ ,
\end{align}
admit a direct matrix-model interpretation.

The simplest indication that such a relation may exist comes from the field-theoretic interpretation of the $\Sigma_{\pm}$ instantons in $S^7/\mathbb{Z}_k$. For the abelian ABJM theory with gauge group $U(1)_k\times U(1)_{-k}$, the moduli space is
$\mathbb C^4/\mathbb Z_k$ \cite{Aharony:2008ug}. With the conventions of Section \ref{sec2.3}, the matter fields $A_{1,2}$ and $B_{1,2}$ may be identified with the coordinates
\begin{align}
    z_{1,2}=A_{1,2}\ ,\qquad z_{3,4}=B_{1,2}\ .
\end{align}
The chiral ring is obtained by including monopole operators dressed by matter
fields, and, as a coordinate ring, is identified with the invariant ring
\begin{align}
\mathbb{C}[z_a]^{\mathbb Z_k}
=
\mathrm{Span}_{\mathbb{C}}
\left\{
z_1^{n_1}z_2^{n_2}z_3^{n_3}z_4^{n_4}
\,\middle|\,
n_1+n_2-n_3-n_4\equiv 0 \bmod{k}
\right\} .
\end{align}
In this description, the two elementary M2-branes $\Sigma_{+}$ and $\Sigma_{-}$
are associated with the two coordinate planes supported on $(A_1,A_2)$ and
$(B_1,B_2)$, respectively. The $\mathbb Z_k$ projection is implemented by the
baryonic $U(1)_b$ charge, and the two sectors are naturally related to
monopole-dressed operators of opposite magnetic charge.

This interpretation suggests a possible extension to more general $\mathcal N=3$, and possibly $\mathcal N=2$, Chern-Simons matter theories, although the connection with exponentially small corrections to the $S^3$ free energy is less direct. In a number of such examples, the asymptotic counting of chiral-ring operators, graded by R-charge and monopole charge, is related to the large-$N$ matrix-model eigenvalue distribution and to the volume of the dual Sasaki-Einstein manifold, as well as to the volumes of M5-brane cycles dual to baryonic-type operators
\cite{Gulotta:2011si,Gulotta:2011aa,Gulotta:2011vp,Kim:2012vza}. The M2-branes considered here correspond instead to three-dimensional Sasakian links of holomorphic two-cones in $\cyf$, the simplest examples of which can be identified with the edges of the toric polytope. It would therefore be interesting to understand whether single M2-brane saddles can be identified through suitably restricted asymptotic counting of chiral operators, using the character formula for the volume of Sasakian manifolds \cite{Martelli:2006yb}. A more conjectural question is whether such an identification can be used to relate the leading M2-brane saddles to non-perturbative corrections to the matrix-model free energy.

\section*{Acknowledgements}

I would like to thank M.~Beccaria, J.~P.~Gauntlett, F.~F.~Gautason, A.~Hanany, J.~van Muiden, A.~A.~Tseytlin and D.~Waldram for useful discussions. I am grateful to M.~Beccaria, J.~P.~Gauntlett, J.~van Muiden and A.~A.~Tseytlin for their comments on the draft, and especially to J.~van Muiden and A.~A.~Tseytlin for numerous discussions, encouragement and support throughout this work. This work was supported by the President's PhD Scholarship at Imperial College London.

\newpage

\appendix 
\section{Clifford algebra and Killing spinors} \label{A}

\subsection{Clifford algebra} \label{A.1}

We consider the eight-dimensional Clifford algebra $\mathrm{Cl}_{8}$ generated by real symmetric $16\times 16$ matrices
\begin{align}
    \{\tilde\gamma_m, \tilde\gamma_{n}\} = 2\delta_{mn} \ , \quad \tilde\gamma_{m}^{T} = \tilde\gamma_m \ , \quad m=0,1,\dots 7 \ ,
\end{align}
with the chirality operator defined as 
\begin{align}
    \tilde{\gamma}_9=\tilde{\gamma}_{0}\dots \tilde\gamma_{7} \ , \quad \tilde\gamma_9^2=\id \ , \quad \{\tilde\gamma_9,\tilde\gamma_m\}=0\ .
\end{align}
The eight-dimensional spinor module is taken to be real and, in the basis of real symmetric matrices, we consider the invariant symmetric spinor product
\begin{align} \label{A.17}
    \langle \Psi_{1}, \Psi_{2}\rangle   =  \Psi_{1}^{T}\Psi_{2} \ , \qquad \Psi_{1,2} \in \tilde{\mathrm{S}} \ .
\end{align}
%where the spinors $\Psi_{1,2}$ are taken to be Grassmann-even.

To construct a representation of the Clifford algebra $\mathrm{Cl}_{7}$ associated with a seven-dimensional vector space, we use the identification $\mathrm{Cl}_{7}\cong \mathrm{Cl}_{8}^{\mathrm{even}}$. Likewise, the spinor module $\mathrm{S}$ in seven dimensions may be realised as the restriction of a definite-chirality spinor module $\tilde{\mathrm S}$ in eight dimensions. Throughout the paper, whenever this identification is used, we choose $\mathrm{S}$ to be identified with positive-chirality spinors $\tilde{\mathrm{S}}^{+}$ in 8d. Concretely, we take the generators of $\mathrm{Cl}_{7}$ acting on $\mathrm{S}\cong \tilde{\mathrm{S}}^{+}$ to be
\begin{align}\label{A.18}
    &\gamma_{i} = i\tilde{\gamma}_{0}\tilde{\gamma}_{i}\big|_{\tilde{\mathrm{S}}^{+}}   \ ,  \qquad \{\gamma_{i},\gamma_{j}\} = 2\delta_{ij} \ , \qquad i =1, \dots, 7 \ ,  \\
    &\gamma_{i}^{T} = -\gamma_{i} \ , \qquad \gamma_{i}^{\ast} = -\gamma_{i} \ ,
\end{align}
where the matrices $\gamma_{i}$ are purely imaginary and antisymmetric. We note that the seven-dimensional spinor module admits a real structure inherited from $\tilde{\mathrm{S}}^{+}$, and the invariant form is obtained by restriction of \eqref{A.17}. With this convention, the orientation is fixed by
\begin{align}\label{A.20}
    -i \gamma_{1}\dots \gamma_{7} = \id \ , \qquad \gamma_{i_{1}i_{2}i_{3}i_{4}} =- \frac{i}{3!}\epsilon_{i_{1}i_{2}i_{3}i_{4}}{}^{ j_{1}j_{2}j_{3}}\gamma_{j_{1}j_{2}j_{3}} \ .
\end{align}
For $\psi_{1,2}\in \mathrm{S}$, the bilinear involving antisymmetrised Dirac matrices satisfies
\begin{align} \label{antisymferm}
    \psi_{1}^{T}\gamma_{i_{1}\dots i_{n}}\psi_{2} = (-1)^{n(n+1)/2}\, \psi_{2}^{T}\gamma_{i_{1}\dots i_{n}}\psi_{1} \ , \qquad \gamma_{i_{1}\dots i_{n}} = \gamma_{[i_{1}}\dots\gamma_{i_{n}]} \ .
\end{align}
We also include the useful Dirac matrix identities
\begin{align}
    &[\gamma_{ij}, \gamma^{l}] = - 4\delta_{[i}{}^{l}\gamma_{j]} \ , \quad \{\gamma_{ij}, \gamma_{l}\} = 2\gamma_{ijl} \ ,   \label{A.22a} \\
    &[\gamma_{ijk}, \gamma_{l}] = 2\gamma_{ijkl} \ , \quad \{\gamma_{ijk}, \gamma^{l}\} = 6\delta_{[i}{}^{l}\gamma_{jk]} \label{A.22} \ , \\
    &[\gamma_{ij}, \gamma^{kl}] = -8\delta_{[i}{}^{[k}\gamma_{j]}{}^{l]} \ , \quad \{\gamma_{ij}, \gamma^{kl}\} = 2\gamma_{ij}{}^{kl}-4\delta_{[ij]}{}^{kl} \ , \label{A.23a}
\end{align}
together with the seven-dimensional Fierz identity for commuting spinors
\begin{align}\label{A.23}
    \psi'\psi^{T} =\frac{1}{8} \big((\psi^{T}\psi')\id +(\psi^{T}\gamma_{i}\psi')\gamma_{i} -\frac{1}{2!}(\psi^{T}\gamma_{ij}\psi')\gamma_{ij}-\frac{1}{3!}(\psi^{T}\gamma_{ijk}\psi')\gamma_{ijk}\big) \ .
\end{align}
In particular, the identities \eqref{A.22}, \eqref{A.23} imply that for a unit $\psi^{T}\psi = 1$ seven-dimensional spinor
\begin{align} \label{A.24}
    \psi \psi^{T} = \frac{1}{8}\big(\id -\frac{1}{3!}(\psi^{T}\gamma_{ijk}\psi) \gamma_{ijk} \big) \ .
\end{align}

\subsection{Killing spinors and $G_2$-structure}\label{apB.1}

To obtain Killing spinors on seven-dimensional spin manifolds, we use Bär's correspondence \cite{Bar:1993}. Namely, Bär's correspondence establishes a one-to-one correspondence between Killing spinors on a compact Riemannian spin manifold $(M, g)$ and parallel spinors on the Riemannian cone $(C(M), \tilde{g})$. In the case of interest, we take $M = \ys$ to be a weak $G_2$ manifold with unit radius. The associated Riemannian cone has holonomy contained in $\mathrm{Spin}(7)$ and is equipped with the metric
\begin{align} \label{1.1}
    d\tilde{s}_{C(\ys)}^2  = dr^2 +r^2ds^2_{\ys} \ , \qquad r\in \mathbb{R}_{+} \ , 
\end{align}
where $ds^2$ denotes the metric on $\ys$. Let $\tilde{e}^{m} = (dr,  r e^{i})$, with $i=1, \dots, 7$, denote an  orthonormal coframe on the cone. The Levi-Civita connection 1-form $\tilde{\omega}$ on the cone $C(\ys)$ then satisfies
\begin{align}
    \tilde\omega^{ir} = \frac{1}{r}\tilde e^{i} = e^{i} \ , \qquad \tilde{\omega}^{ij} = \omega^{ij} \ , \label{212}
\end{align}
where $\omega^{ij}$ is the Levi-Civita connection 1-form on $\ys$. Using \eqref{212}, it follows that a parallel spinor $\Psi$ on the cone $C(\ys)$ obeys
\begin{align}
    \tilde{\nabla} \Psi = \nabla\Psi +\frac{1}{2r}\tilde{e}^{i}\tilde{\gamma}_{i}\tilde{\gamma}_{r}\Psi = 0   \ , 
\end{align}
where $\tilde{\nabla}$ and $\nabla$ denote the spin connection on $C(\ys)$ and $\ys$, respectively, and $\tilde\gamma_{r}\equiv \tilde{\gamma}_{0}$. We further restrict to $\Psi\in \tilde{\mathrm{S}}^{+}$, a real spinor of positive eight-dimensional chirality,
\begin{align}
    \tilde{\gamma}_{9} = \tilde{\gamma}_{r}\tilde\gamma_{1}\dots\tilde\gamma_{7} \ , \qquad \tilde{\gamma}_{9}\Psi = \Psi \ .
\end{align}
A real Killing spinor $\psi$ on $\ys$ is obtained by restricting $\Psi$ to the hypersurface $r=1$. With the conventions introduced in Appendix \ref{A.1}, the Killing spinor $\psi$ on $\ys$ is identified as
\begin{align}
    &\psi = \Psi\big|_{r=1} \ , \qquad \Psi \in \tilde{\mathrm{S}}^{+}\equiv\mathbf{8}_{s} \ , \label{A.33} \\
    &D\psi = \nabla\psi +\frac{i}{2}e^{i}\gamma_{i}\psi = 0 \ . \label{A.34a}
\end{align}
As in the main text, and using the Clifford algebra conventions specified above, we define the $G_{2}$-structure $\varphi$ as the three-form on $\ys$ with components
\begin{align}
    &\varphi_{ijk} = i\psi^{T}\gamma_{ijk}\psi \ , \quad (\ast \varphi)_{ijkl} = -\psi^{T}\gamma_{ijkl}\psi \ ,  \quad \psi^{T}\psi = 1 \ .
\end{align}
To derive the identities involving the $G_2$-structure, we use a local orthonormal frame $(\psi, e^{\psi}_{i})$ of the spinor bundle $\mathrm{S}$ determined by the Killing spinor
\begin{align}
    &e_{i}^{\psi} = -i\gamma_{i}\psi \ , \quad \langle e_{i}^{\psi}, e_{j}^{\psi}\rangle = \delta_{ij} \ , \quad \langle e_{i}^{\psi}, \psi \rangle  = 0 \ , \\
    & \forall \theta \in \Gamma(\mathrm{S}, \ys):\quad \theta = \theta_{0}\psi + \theta_{i} e_{i}^{\psi} \ .
\end{align}
Using orthogonal decomposition together with \eqref{antisymferm}, \eqref{A.22a} and \eqref{A.24}, we obtain
\begin{align}\label{B.7}
    &\gamma_{ij}\psi = c_{ijk}e_{k}^{\psi}  \ , \quad c_{ijk} = \langle e_{k}^{\psi}, \gamma_{ij}\psi \rangle = \frac{i}{2}\psi^{T}\{\gamma_{k}, \gamma_{ij}\}\psi = \varphi_{ijk} \ .
\end{align}
It then follows that
\begin{align}
    \langle \gamma_{i_{1}i_{2}}\psi, \gamma_{j_{1}j_{2}}\psi\rangle
    = \varphi_{i_{1}i_{2}i_{3}}\, \varphi_{j_{1}j_{2}j_{3}}\langle e_{i_{3}}^{\psi} , e_{j_{3}}^{\psi}\rangle
    = \varphi_{i_{1}i_{2}k}\, \varphi_{j_{1}j_{2}k} \ .
\end{align}
On the other hand, the left-hand side can be rewritten, using the definition of the spinorial inner product and \eqref{A.23a}, as
\begin{align}
    \langle \gamma_{i_{1}i_{2}}\psi, \gamma_{j_{1}j_{2}}\psi\rangle
    = -\psi^{T}\gamma_{i_{1}i_{2}}\gamma_{j_{1}j_{2}}\psi
    = -\psi^{T} \gamma_{i_{1}i_{2}j_{1}j_{2}}\psi +\delta_{i_{1}j_{1}}\delta_{i_{2}j_{2}}-\delta_{i_{1}j_{2}}\delta_{i_{2}j_{1}} \ .
\end{align}
We therefore obtain the standard $G_{2}$-structure identities:
\begin{align}
    &\varphi_{i_{1}i_{2}k} \, \varphi_{j_{1}j_{2}k}
    = \delta_{i_{1}j_{1}}\delta_{i_{2}j_{2}}-\delta_{i_{1}j_{2}}\delta_{i_{2}j_{1}}+(\ast\varphi)_{i_{1}i_{2}j_{1}j_{2}} \label{B.11} \ , \\
    & \varphi_{ikl}\varphi_{jkl} = 6\delta_{ij} \ .
\end{align}
Similarly, using orthogonal decomposition together with \eqref{antisymferm}, \eqref{A.22a}, and \eqref{A.24}, one finds the identity
\begin{align}
    i\gamma_{i}e_{j}^{\psi} = \delta_{ij}\psi +\varphi_{ijk}e_{k}^{\psi} \ . \label{B.13}
\end{align}

\subsection{Killing spinors and Sasakian structure}\label{appA3}

As discussed in Section \ref{sec22}, we consider a contact form $\eta$ on $\ses$ obtained from the reduction of the Kähler form on the $\cyf$-cone
\begin{align}
\eta = \iota_{\zeta}\Omega\big|_{r=1} \ , \qquad \Omega = \frac{1}{2}\langle\Psi_{2}, \tilde\gamma_{mn}\Psi_{1}\rangle \, \tilde{e}^{m}\wedge \tilde{e}^{n}
,
\end{align}
where $\zeta = r\partial_{r}$ and the parallel spinors $\Psi_{1,2}$ on the cone satisfy \eqref{2.35}. With the identification \eqref{A.33}, it follows that
\begin{align}
\eta =- i\langle \psi_{2}, \gamma_{i}\psi_{1}\rangle \, e^{i} = i\langle \psi_{1}, \gamma_{i}\psi_{2}\rangle \, e^{i} \ , \qquad \psi_{1,2} = \Psi_{1,2}\big|_{r=1} \ ,
\end{align}
where $e^{i}$ denote an orthonormal coframe on $\ses$. For the given contact structure $\eta$, the Reeb vector field $\xi$ is the unique vector field on $\ses$ satisfying $\eta(\xi) = 1$ and $\iota_{\xi}d\eta=0$. It then follows that
\begin{align}
\xi = \xi^{i}e_{i} \ , \quad \xi^{i} = -i\langle \psi_{2}, \gamma^{i}\psi_{1}\rangle \ , \quad \eta(\xi) = 1 \ ,
\end{align}
where the last identity can be checked with the help of \eqref{A.24}. Identity \eqref{A.24} also allows one to show that the Killing spinors $\psi_{1,2}$ are related to each other through the $U(1)_{r}$ action on the spinor module induced by the Reeb vector field $\xi$:
\begin{align} \label{A.31bb}
-i\gamma(\xi)\psi_{1} = \psi_{2} \ , \quad -i\gamma(\xi)\psi_{2} = -\psi_{1} \ ,
\end{align}
where $\gamma(\xi) = \xi^{i}\gamma_{i}$. This property descends from the $U(1)_r$ action \eqref{2.35} on $\Psi_{1,2}$ on $\cyf$. Using the fact that both $\psi_{1,2}$ satisfy the Killing spinor equation \eqref{A.34a}, it also follows that
\begin{align}
d\eta = 2\omega =\langle\psi_{2}, \gamma_{ij}\psi_{1}\rangle \, e^{i}\wedge e^{j} \ , \qquad \iota_{\xi}d\eta = 0 \ ,
\end{align}
where the closed 2-form $\omega$ defines the transverse Kähler form on the contact distribution $\ker \eta$. The components of the associated contact endomorphism $\phi$ can also be conveniently expressed in the orthonormal basis $e_{i}$ in terms of the Killing spinors:
\begin{align}
    &\nabla_{X}\xi = \phi(X) \ , \quad \forall X\in \Gamma(T\ses) \ , \qquad \phi^{i}{}_{j} = \langle e_{i},  \phi (e_{j})\rangle  \ , \\
    &\phi^{i}{}_{j} = -\langle \psi_{2}, \gamma^{i}{}_{j}\psi_{1}\rangle = \langle \psi_{1}, \gamma^{i}{}_j\psi_{2}\rangle \ , \quad \phi^{i}{}_{j}\phi^{j}{}_k = -\delta^{i}{}_{k}+\xi^{i}\xi_k \ , 
\end{align}
where the last identity follows from \eqref{B.7} and \eqref{B.11}. Similarly to the $U(1)_{r}$ action generated by the Reeb vector field, the action of the complex structure $\phi$ also naturally lifts to the spinor module. In particular, with the help of the Fierz identity, the action of $\phi$ on the Killing spinors $\psi_{1,2}$ becomes
\begin{align}
\gamma(\phi X)\psi_{1} = \gamma(X)\psi_{2}+i\langle \xi, X\rangle \psi_{1} \ , \quad \gamma(\phi X)\psi_{2} = -\gamma(X)\psi_{1}+i\langle \xi, X\rangle \psi_{2} \ , \quad \forall X\in T\ses \ , \label{3.6}
\end{align}
where $\gamma(\phi X) = \phi^{j}{}_{i}X^{i}\gamma_{j}$ and $\gamma(X)=X^{i}\gamma_{i}$ for $X = X^{i}e_{i}$. As in the main part of the paper, one can introduce a complex spinor $\psin$ as
\begin{align}
    \psin = \frac{1}{\sqrt{2}}(\psi_{1}-i\psi_{2}) \ .
\end{align}
Using \eqref{A.31bb} and \eqref{3.6}, it follows that
\begin{align}
    \gamma(\xi)\psin = -\psin \ , \qquad
    \gamma(\phi X)\psin = i \big(\gamma(X)+\langle \xi, X\rangle \big)\psin\ , \qquad \forall X\in \Gamma(T\ses) \ . \label{A.37b}
\end{align}
For the unitary frame \eqref{unitfr}, the identity
\eqref{A.37b} implies that $\psin$ is annihilated by the antiholomorphic Dirac matrices:
\begin{align}
    \gamma(\eun_{\bar \mu}) \psin = 0 \ , \quad \phi(\eun_{\bar \mu})= -i\eun_{\bar \mu} \ , \quad \eun_{\bar \mu}\in \ker \eta \ ,
\end{align}
where $\mu = 1,2,3$. The expressions for the transverse Kähler form $\omega$ and the transverse
holomorphic three-form $\alpha$ simplify in this frame. In the basis of the seven-dimensional Dirac matrices adapted to the unitary
frame\footnote{In the real basis $e_{i}$, the Dirac matrices $\gamma_i$ are chosen purely imaginary
and antisymmetric. In the complex frame this implies $\gamma_{\bar{\mu}} = \gamma_{\mu}^{\dagger} = -\bar{\gamma}_{\mu}$,
where $\bar{\cdot}$ denotes complex conjugation.}
\begin{align}
    \{\gamma_{\mu}, \gamma_{\bar\nu}\} = 2\delta_{\mu\nu} \ , \quad \{\gamma_{\mu}, \gamma_{\nu}\}= \{\gamma_{\bar\mu}, \gamma_{\bar\nu}\} = 0 \ , \quad \gamma_{\mu} \equiv \gamma(\eun_{\mu}), \ \gamma_{\bar\mu}\equiv \gamma(\eun_{\bar\mu}) \ .
\end{align}
The transverse Kähler form can be written as
\begin{align}
    \omega
    =
    -\frac{i}{2}\langle\bar\psi_{-}, \gamma_{ij}\psin \rangle \,
    e^{i}\wedge e^{j}
    =
    -i \langle \bar{\psi}_{-} ,  \gamma_{\mu\bar{\nu}}\psin \rangle \,
    \eun^{\mu}\wedge \eun^{\bar\nu}
    =
    i\delta_{\mu\nu}\, \eun^{\mu}\wedge \eun^{\bar{\nu}} \ ,
\end{align}
where $\eun^{\mu}, \eun^{\bar\nu}$ denote the dual coframe. The spinor
product \eqref{A.17} is extended to complex spinors
and is used throughout. When needed, we also consider the associated
Hermitian pairing
\begin{align}
    \langle \bar{\psi}, \psi'\rangle = \bar{\psi}^{T}\psi' \ .
\end{align}
Similarly, from \eqref{2.40} and \eqref{2.40aa}, the transverse holomorphic
three-form $\alpha$ is
\begin{align}
    \alpha
    =
    \frac{i}{3!}\langle \psin , \gamma_{ijk}\psin \rangle \,
    e^{i}\wedge e^{j}\wedge e^{k}
    =
    \frac{i}{3!}\langle \psin , \gamma_{\mu\nu\rho}\psin \rangle \,
    \eun^{\mu}\wedge \eun^{\nu}\wedge \eun^{\rho} \ ,
\end{align}
where the antiholomorphic components vanish because
$\gamma_{\bar\mu}\psin=0$. The Fierz identity \eqref{A.23}, applied to the complex null spinor
$\psi_{-}$, implies
\begin{align}
   \psi_{-}\psi_{-}^{T}
   =
   -\frac{1}{8\cdot 3!}
   \big(\psi_{-}^{T}\gamma_{ijk}\psi_{-}\big)\gamma_{ijk}
   =
   \frac{i}{8\cdot 3!}\alpha_{ijk}\gamma_{ijk} \ .
\end{align}
This also fixes the normalisation of $\alpha$:
\begin{align}
-\frac{1}{64\cdot (3!)^2}
\alpha_{i_{1}i_{2}i_{3}}\bar{\alpha}_{j_{1}j_{2}j_{3}}
\tr\big(\gamma_{i_{1}i_{2}i_{3}}\gamma_{j_{1}j_{2}j_{3}}\big)
=
\frac{1}{8\cdot 3!}
\alpha_{i_{1}i_{2}i_{3}}\bar{\alpha}^{i_{1}i_{2}i_{3}}
=
|\langle\bar{\psi}_{-}, \psi_{-}\rangle|^2
=
1 \ .
\end{align}
In particular, for the unitary frame considered above, this implies
\begin{align}
   |\alpha(\eun_{1}, \eun_{2}, \eun_{3})|
   =
   2\sqrt{2}  \ .\label{A.43b} 
\end{align}
By using the Killing spinor equation \eqref{A.34a}, it also follows that
\begin{align}
d\alpha = 4i \, \eta\wedge \alpha \ , \qquad \mathcal{L}_{\xi}\alpha = \iota_{\xi} d\alpha = 4i\, \alpha \ . \label{a.46}
\end{align}

\section{Associative M2-branes in weak $G_2$ manifolds} \label{B}

\subsection{Bosonic fluctuations in $\ys$}

Rather than directly applying the second-variation formula for a
minimal submanifold of $\ys$, written in terms of the Jacobi operator, we review
a McLean-type second-variation formula \cite{McLean:1998}, adapted to the case of
a nearly parallel $G_{2}$-structure. The deformation theory of generalised
calibrations was discussed in \cite{Gutowski:2002bx}, while explicit expressions
for associative submanifolds of weak $G_{2}$ manifolds were considered in
\cite{Kawai2017HomAssoc,KawaiSecondOrder}. Below, we review the relevant
second-variation formula using the properties of the Killing spinor $\psi$
preserved by the M2-brane.

To derive the second variation around a BPS surface, we consider a
one-parameter family of surfaces $\Sigma_{t}$ and use the expression
\eqref{2.23} for the volume functional:
\begin{align} \label{3.1}
    \mathrm{Vol}(\Sigma_t)
    &=
    \int d^3\sigma \,
    \sqrt{|\mathrm{A}_t|^2+\varphi^2(\partial_{1}^{t}, \partial_{2}^t, \partial_{3}^t)} \ , 
    \quad
    \mathrm{A}_t
    =
    \hodge \varphi(\, \cdot \, , \partial_{1}^{t}, \partial_{2}^{t}, \partial_{3}^{t}) \ .
\end{align}
For convenience, we also decompose the volume into its ``energy'' and
calibration parts:
\begin{align}
    &\mathcal{E}(\Sigma_t)
    =
    \int d^3\sigma \,
    \Big(
    \sqrt{|\mathrm{A}_t|^2+\varphi^2(\partial_{1}^{t}, \partial_{2}^t, \partial_{3}^t)}
    -\varphi(\partial_{1}^{t}, \partial_{2}^t, \partial_{3}^t)
    \Big) \ , \\
    &\mathrm{Vol}(\Sigma_t)
    =
    \mathcal{E}(\Sigma_t)+\int_{\Sigma_t}\varphi \ . \label{2.3}
\end{align}
In the case of a parallel $G_{2}$-structure, the second term in \eqref{2.3}
does not change under small variations, since $\varphi$ is closed. The second
variation in this case was obtained in \cite{McLean:1998} and used in the
context of M2-brane instantons in \cite{Harvey:1999as}.

We assume that the one-parameter family $\Sigma_t$ is generated by the flow of
a normal vector field $\mathrm{V}\in \Gamma(N\Sigma)$ such that
\begin{align}
    [\mathrm{V}, \partial_{\alpha}] = 0 \ , \qquad
    \nabla_{\mathrm{V}}\partial_{\alpha} = \nabla_{\alpha}\mathrm{V} \ ,
\end{align}
where $\nabla$ denotes the Levi-Civita connection on $\ys$. To
expand the action \eqref{3.1} about a BPS M2-brane $\Sigma$ up to quadratic
order in the normal directions, we first note that, for any normal vector
$\mathrm{V}\in N\Sigma$:
\begin{align} \label{3.5}
    \varphi (\partial_{\alpha}, \partial_{\beta}, \mathrm{V})\big |_{\Sigma} = 0 \ ,
    \qquad
    \varphi(\partial_{1}, \partial_{2}, \partial_{3})\big |_{\Sigma} = \sqrt{g} \ ,
\end{align}
which follows from \eqref{2.11}, \eqref{2.24a}, and $\gamma \psi = \psi$ on
$\Sigma$. We then arrive at
\begin{align}
    \delta\mathrm{A}_{i}
    =
    \nabla_{\mathrm{V}}\mathrm{A}_{i}
    =
    \sqrt{g}\,g^{\alpha\beta}
    \varphi(e_{i}, \nabla_{\alpha}^{\perp} \mathrm{V}, \partial_{\beta})
    -\sqrt{g}\,\mathrm{V}_{i} \ ,
\end{align}
where we have used \eqref{2.15a}, \eqref{2.24a}, and
\begin{align}
    \frac{1}{2}\epsilon^{\alpha\beta\gamma}
    (\hodge \varphi)(e_{i}, \nabla_{\alpha}\mathrm{V}, \partial_{\beta}, \partial_{\gamma})
    &=
    \frac{1}{2}\epsilon^{\alpha\beta\gamma}
    (\hodge \varphi)(e_{i}, \nabla^\perp_{\alpha}\mathrm{V}, \partial_{\beta}, \partial_{\gamma})=
    \sqrt{g}\,g^{\alpha\beta}
    \varphi(e_{i}, \nabla_{\alpha}^{\perp} \mathrm{V}, \partial_{\beta}) \ ,
\end{align}
with $\nabla^{\perp}$ denoting the connection on the normal bundle $N\Sigma$.
We also note that $\delta\mathrm{A}$ is a normal vector field on $\Sigma$, since
$\delta\mathrm{A}(\partial_{\alpha})=0$ by \eqref{3.5}. It is then
straightforward to see that the first variation of the ``energy'' part
vanishes, while the second variation gives
\begin{align}
    &\delta^2\mathcal{E}
    =
    \int d^3 \sigma \, \frac{1}{\sqrt{g}}|\delta\mathrm{A}|^2
    =
    \int d^3\sigma\sqrt{g}\,
    |\slashed{\mathfrak{D}}\mathrm{V}+\mathrm{V}|^2 \ , \\
    &\slashed{\mathfrak{D}}\mathrm{V}
    =
    g^{\alpha\beta}\varphi
    (\partial_{\alpha} , \nabla_{\beta}^{\perp}\mathrm{V}, \, \cdot \, )
     \ . \label{3.10}
\end{align}
In contrast to the case of a parallel $G_2$-structure, the calibration term is no longer
topological. Although its first variation vanishes on an associative surface,
its second variation is non-zero and can be computed similarly:
\begin{align}
    \delta^{2}\int_{\Sigma_t}\varphi
    &=
    4\int d^3\sigma\,
    \nabla_{\mathrm{V}}
    \big(\hodge\varphi(\mathrm{V}, \partial_{1}, \partial_{2}, \partial_{3})\big) =
    -4 \int d^3\sigma\sqrt{g} \,
    \langle \mathrm{V}, \slashed{\mathfrak{D}}\mathrm{V}+\mathrm{V}\rangle \ .
\end{align}
Thus, the resulting second-variation formula for the volume around a BPS
M2-brane becomes
\begin{align}\label{3.12}
    \delta^2 \mathrm{Vol}(\Sigma)
    =
    \int d^3\sigma \sqrt{g} \,
    \Big(
    |\slashed{\mathfrak{D}}\mathrm{V}+\mathrm{V}|^2
    -4\langle\mathrm{V}, \slashed{\mathfrak{D}}\mathrm{V}+\mathrm{V}\rangle
    \Big) \ . 
\end{align}
The operator
$(\slashed{\mathfrak{D}}+1):\Gamma(N\Sigma)\to \Gamma(N\Sigma)$ acts on the
normal bundle of the M2-brane and should be viewed as the mass-deformed
analogue of the Dirac operator considered in \cite{McLean:1998}. Moreover,
$\slashed{\mathfrak{D}}$ is self-adjoint on a closed M2-brane
\cite{Kawai2017HomAssoc}:
\begin{align}
    \int d^3\sigma \sqrt{g} \,
    \langle \mathrm{S}, \slashed{\mathfrak{D}}\mathrm{V}\rangle
    =
    \int d^{3}\sigma\sqrt{g} \,
    \langle \mathrm{V}, \slashed{\mathfrak{D}}\mathrm{S}\rangle \ ,
    \quad
    \forall\,  \mathrm{S}, \mathrm{V}\in \Gamma(N\Sigma) \ .
\end{align}
Therefore, the quadratic bosonic contribution resulting from \eqref{3.12} can
be written as
\begin{align}
    \delta^2\mathrm{Vol}(\Sigma)
    =
    \int d^3\sigma \, \sqrt{g} \,
    \langle \mathrm{V},
    (\slashed{\mathfrak{D}}-3)(\slashed{\mathfrak{D}}+1)
    \mathrm{V}\rangle \ . \label{bosonicmain}
\end{align}

\subsection{Fermionic fluctuations}

For a general eleven-dimensional background $\ads\times \ys$, with $\ys$ a weak $G_{2}$ manifold, the Fock-type spinor frame used in Section \ref{sec3.2} for invariant submanifolds of Sasaki-Einstein spaces is not available in general. That construction relied on the transverse complex structure and on the pair of Killing spinors defining the Sasaki-Einstein structure. For an associative M2-brane in a weak $G_{2}$ manifold, however, one can instead use the real spinor frame determined by the preserved Killing spinor $\psi$ on $\ys$, satisfying $\gamma\psi=\psi$ on $\Sigma$. In this frame the comparison with the fermionic operators appearing for associative cycles in manifolds with a parallel $G_{2}$-structure \cite{Harvey:1999as} becomes transparent, with the nearly parallel $G_{2}$-structure producing the constant mass shifts.

Concretely, to express the quadratic action \eqref{3.19} for fermions around a BPS M2-brane in $\ads \times \ys$, it is convenient to use the local orthonormal frame introduced in Appendix \ref{apB.1}. In particular, an arbitrary seven-dimensional spinor $\theta$ can be decomposed as
\begin{align}
    \theta = \theta_{0}\psi +\theta_{i}e^{\psi}_{i} \ , \qquad \theta_{0}=\langle \theta, \psi\rangle \ , \qquad \theta_{i} = \langle \theta, e_{i}^{\psi}\rangle \ ,
\end{align}
where the coefficients $\theta_{0}$ and $\theta_{i}$ are now taken to be Grassmann-odd to account for the fermionic statistics. It is then straightforward to determine the action of the seven-dimensional Killing spinor derivative $D$ in the chosen coordinates:
\begin{align}
    D\theta = d\theta_{0}\psi+(d\theta_{i}+\omega_{ij}\theta_{j}+\varphi_{ijk}e^{j}\theta_{k})e_{i}^{\psi}\ , 
\end{align}
where $D\psi = 0$ and \eqref{B.7} were used. The tangent bundle of $\ys$, restricted to $\Sigma$, admits a local decomposition
\begin{align}
    T\ys\big|_{\Sigma} = T\Sigma\oplus N\Sigma \ .
\end{align}
Accordingly, the generators of the Clifford algebra can be decomposed as
\begin{align}
    &\gamma_{\alpha} = \gamma(e_{\alpha})  \ , \quad \gamma_{\alpha}\gamma = \gamma \gamma_{\alpha} \ , \\ 
    &\gamma_{a} = \gamma(n_{a}) \ , \quad \gamma_{a}\gamma = -\gamma \gamma_{a} \ , 
\end{align}
where $e_{\alpha}$ ($\alpha=1,2,3$) and $n_{a}$ ($a=1, \dots , 4$) correspond to local orthonormal bases of $T\Sigma$ and $N\Sigma$, respectively. It then follows that the fermionic fluctuations in \eqref{3.17} can be decomposed as
\begin{align}
    &\Lambda_{I} = \rho_{I} \psi+\lambda_{\alpha I} e_{\alpha}^{\psi} \ , \qquad e_{\alpha}^{\psi} = -i \gamma_{\alpha}\psi \ , \\
    &\mathcal{V}_{I} = \nu_{aI}e_{a}^{\psi} \ , \qquad e_{a}^{\psi} = -i \gamma_{a}\psi \ .
\end{align}
The fermionic components $\rho_{I}$ and $\lambda_{\alpha I}$ correspond to a scalar and a one-form on the worldvolume $\Sigma$, so that the corresponding fluctuations can be written as
\begin{align}
    &\frac{1}{2}\int d^3\sigma \sqrt{g} \, \langle \Lambda^{I}, i\slashed{D}\Lambda_{I}\rangle = \frac{1}{2}\int d^3\sigma \sqrt{g} \, \big(\rho^{I}\nabla^{T}_{e_{\alpha}}\lambda_{\alpha I}-\lambda_{\alpha}^{I}\nabla^{T}_{e_{\alpha}}\rho_{I}+\varphi_{\alpha\beta\gamma}(\lambda_{\alpha}^{I}\nabla^{T}_{e_{\beta}}\lambda_{\gamma I})-\varphi_{\alpha\gamma\delta}\varphi_{\beta\gamma\delta}(\lambda_{\alpha}^{I}\lambda_{\beta I}) \big) \nonumber \\
    & = \frac{1}{2}\int_{\Sigma} \big(\lambda^{I}\wedge d\lambda_{I}-2\ast \lambda^{I}\wedge \lambda_{I}- \ast \rho^{I}\wedge d^{\dagger}\lambda_{I} -\ast \lambda^{I}\wedge d\rho_{I}\big) \ , \label{3.26}
\end{align}
where $\nabla^{T}$ is the covariant derivative on $T\Sigma$, $d^{\dagger} = (-1)^{3p}\ast d \, \ast$ when acting on a $p$-form in three dimensions, and we have used $\varphi_{\alpha\beta\gamma}|_{\Sigma} = \epsilon_{\alpha\beta\gamma}$, $\varphi_{\alpha\beta a}\big|_{\Sigma} = 0$, together with \eqref{B.13}. The fermionic fluctuations in \eqref{3.26} are governed by the operator
\begin{align}
    \widehat{L}_{-} = \begin{pmatrix}
        0 & -d^{\dagger} \\
        -d & \ast d-2
    \end{pmatrix} \ , 
\end{align}
which, up to a constant shift, is the same operator that appears at quadratic order in the Rozansky-Witten model \cite{Rozansky:1996bq}. Similarly, treating $\nu_{a I}$ as an element of $N\Sigma$ and using \eqref{B.11}, we obtain
\begin{align} \label{3.27}
    &\frac{1}{2}\int d^3\sigma \sqrt{g} \, \langle \mathcal{V}^{I}, i\slashed{D}\mathcal{V}_{I}\rangle = \frac{1}{2}\int d^3\sigma \sqrt{g} \, \big( \varphi_{a\alpha b} (\nu_{a}^{I}\nabla_{e_{\alpha}}^{\perp}\nu_{b I})- 3\delta_{ab}(\nu_{a}^{I}\nu_{b I})  \big) \nonumber\\
    &=\frac{1}{2}\int d^3\sigma \sqrt{g} \, \langle \mathcal{V}^{I}, (\slashed{\mathfrak{D}}-3)\mathcal{V}_{I}\rangle  \ ,
\end{align}
where $\slashed{\mathfrak{D}}$ is the same Dirac operator as in \eqref{3.10}. The fermionic operators in \eqref{3.26} and \eqref{3.27} are mass-deformed analogues of the fermionic
operators appearing in the Rozansky-Witten and McLean multiplets of \cite{Harvey:1999as}.

Note that the special case of an invariant link $\Sigma\subset \ses$ in the Sasaki-Einstein setting considered in Section \ref{sec3.2} is obtained from \eqref{3.26} and \eqref{3.27}. The fermionic part of the McLean multiplet follows directly from \eqref{3.27}. To recover the Rozansky-Witten sector \eqref{rw2} from \eqref{3.26}, one complexifies the real scalar and one-form components according to
\begin{align}
a_{I}
=
\frac{1}{\sqrt{2}}\big(\rho_{I}+i\lambda_{\xi I} \big) \ ,
\qquad
\lambda_{\bar{\eun}I}
=
-\frac{1}{\sqrt{2}}\big(\lambda_{1 I}+i\lambda_{2 I}\big) \ .
\end{align}
Here $\lambda_{\xi I}=\iota_{\xi}\lambda_{I}$ is the Reeb component of the one-form $\lambda_I$, while $\lambda_{1I},\lambda_{2I}$ are its horizontal components in a local orthonormal frame $(e_1,e_2)$ with $e_2=\phi(e_1)$ and $\eun=(e_1-i e_2)/\sqrt{2}$.

\section{The operators $\bar{\partial}_N, \partial_{N}$ and the maps $\adj, \adjb$}\label{bosonicinv}

In this appendix we collect the identities used in Section \ref{sec3} to rewrite the adapted-frame expression for the normal bosonic quadratic action in an invariant operator form, as well as the properties of the adjunction-type maps $\adj,\adjb$. We assume throughout that $\Sigma\subset\ses$ is a closed invariant three-cycle, and we use the notation introduced in Section \ref{sect3.1}, in particular $v\in\Gamma(N^{1,0}\Sigma)$ and the Dolbeault-type operators $\partial_N,\bar\partial_N$.

With the conventions for the Reeb vector field as in Section \ref{sect3.1}, the transverse holomorphic three-form transforms under the $U(1)_{r}$ action of the Reeb vector field as
\begin{align}
    \mathcal{L}_{\xi}\alpha
    =
    \iota_{\xi} d\alpha
    =
    4i\, \alpha \ , \qquad \alpha \in K_{\ker \eta}\simeq \Lambda^{3,0}(\ker \eta)^{\ast} \ . 
\end{align}
The restriction of the transverse holomorphic three-form $\alpha$ to $\Sigma$ provides the adjunction-type isomorphism
\begin{align}
    \Omega^{1,0}_{H}(\Sigma) \simeq K_{\ker\eta}\big|_{\Sigma}\otimes \det N^{1,0}\Sigma  \ . \label{3.23ab}
\end{align}
Since the normal bundle of the three-cycle  $\Sigma$ has complex rank two, and since the Hermitian metric identifies $(N^{1,0}\Sigma)^*\simeq N^{0,1}\Sigma$, the isomorphism \eqref{3.23ab} gives
\begin{align}
    \Omega^{1,0}_{H}(\Sigma, N^{0,1}\Sigma ) \simeq K_{\ker \eta }\big|_{\Sigma}\otimes N^{1,0}\Sigma  \ .
\end{align}
In practice, this allows one to introduce adjunction-type maps $\adj, \adjb$:
\begin{align}
    \adjb : \Omega^{1,0}_{H}(\Sigma, N^{0,1}\Sigma)\to \Gamma(N^{1,0}\Sigma) \ , \qquad \adj : \Omega^{0,1}_{H}(\Sigma, N^{1,0}\Sigma)\to \Gamma(N^{0,1}\Sigma)  \ .
\end{align}
For instance, in the unitary frame these act as 
\begin{align}
    \adjb(\bar V) =\frac{1}{2\sqrt{2}} \nn_{a} \, \delta^{a\bar{b}}\bar\alpha(\bar{\eun}, \bar{V}_{\eun},\nn_{\bar{b}}) =\nn_{ a}\, (-\epsilon^{ab}(\bar V_{\eun})_{b}) \ , \qquad \adjb^{-1} (v)  = \mathrm{f}\otimes \nn_{\bar a} \, (\delta^{\bar{a}b}\epsilon_{bc}v^c)  \ ,
\end{align}
and similarly for $\adj$:
\begin{align}
    \adj(V) =\frac{1}{2\sqrt{2}} \nn_{\bar a} \, \delta^{\bar{a} b}\alpha(\eun, V_{\bar \eun},\nn_{b}) =\nn_{ \bar a}\, (-\delta^{\bar{a} b}\epsilon_{bc}( V_{\bar \eun})^{c}) \ , \qquad \adj^{-1} (\bar v)  = \bar{\mathrm{f}}\otimes \nn_{ a} \, (\epsilon^{ab}\bar{v}_b)  \ .
\end{align}
With the $L^2$-pairings introduced above, the maps $\adj,\adjb$ give a way to express pairings involving the transverse holomorphic three-form:
\begin{align}
    &\big(\adj (V), v\big) = \big(\adjb^{-1}(v), V\big )  = \frac{1}{2\sqrt{2}}\int d^3\sigma \, \sqrt{g} \,
    \alpha(\eun, V_{\bar \eun},v) \ , \qquad \big(\adj (V), \adjb (\bar{V}')\big) = (\bar{V}', V) \ . \label{3.27e}
\end{align}
In addition, using the expression \eqref{3.17f} for $\bar{\partial}_{N}$, integration by parts on a closed
$\Sigma$ gives the symmetry relation
\begin{align}
   \big(\adj (\bar\partial_{N}v), v'\big) =\big(\adj(\bar{\partial}_{N}v'), v\big) \ . \label{3.21d}
\end{align}
Here we have used that the horizontal differential of the transverse holomorphic three-form $\alpha$ vanishes. In the adapted frame, this gives the total differential:
\begin{align}
    d(\eta\wedge \alpha(\eun, v, v')\mathrm{f})\big|_{\Sigma} = \big(\alpha(\eun, \nabla_{\bar \eun}^{\perp}v, v')+ \alpha(\eun, v, \nabla_{\bar \eun}^{\perp}v')\big)\, \eta  \wedge \mathrm{f}\wedge \bar{\mathrm{f}} \ .
\end{align}
Using the notation introduced above, and integrating by parts on the closed submanifold $\Sigma$, the quadratic action \eqref{3.15ab} can be written in operator form as
\begin{align}
    &\frac{1}{2}\int d^3\sigma \, \sqrt{g} \,
    \langle \mathrm{V},
    (\slashed{\mathfrak{D}}-3)(\slashed{\mathfrak{D}}+1)
    \mathrm{V}\rangle  \nonumber \\
    &=
    ( \bar v, -\mathcal{L}_{\xi}^2v-4i\mathcal{L}_{\xi}v
    +\bar{\partial}_{N}^{\dagger}\bar{\partial}_{N}v )
    +ie^{i\Theta}
    \big(\adj(\bar\partial_{N}v), \mathcal{L}_{\xi}v+2iv\big)
    -ie^{-i\Theta}
    \big( \mathcal{L}_{\xi}\bar{v}-2i\bar{v}, \adjb(\partial_{N}\bar{v})\big) \ ,
    \label{3.19d}
\end{align}
where $\bar{\partial}_{N}^{\dagger}$ denotes the Hermitian adjoint with respect to the $L^2$-pairing.

The induced Lie derivative along the Reeb vector field acts on
$v\in \Gamma(N^{1,0}\Sigma)$ and $\bar v\in \Gamma(N^{0,1}\Sigma)$ as
\begin{align}
    \mathcal{L}_{\xi}v
    =
    \nabla_{\xi}v-\phi(v)
    =
    \nabla_{\xi}v-iv \ ,
    \qquad
    \mathcal{L}_{\xi}\bar v
    =
    \nabla_{\xi}\bar v-\phi(\bar v)
    =
    \nabla_{\xi}\bar v+i\bar v \ .
\end{align}
We further note that the operators $\bar\partial_N$ and $\partial_N$ are invariant under the
Reeb action, in the sense that their commutators with the induced Lie derivative vanish. Namely, using the tensor-product
action of $\mathcal L_\xi$ on
$\Omega^{0,1}_{H}(\Sigma,N^{1,0}\Sigma)$, one finds\footnote{
For an invariant submanifold, the shape operator in the normal direction $\mathrm{V}\in \Gamma(N\Sigma)$, denoted by $\mathrm{A}_{\mathrm{V}}$, satisfies
$\mathrm{A}_{\mathrm{V}}\xi=0$, so the shape-operator contribution in the Ricci equation
vanishes (see e.g. \cite{Blair2002}). Using the interchange symmetry of the
Riemann tensor, the remaining $R_{\ses}$ contribution vanishes by
$$
R_{\ses}(\mathrm{W},\mathrm{V})\xi
=
\eta(\mathrm{V})\mathrm{W}
-
\eta(\mathrm{W})\mathrm{V}
=
0  \ ,
$$
since $N\Sigma\subset\ker\eta$.
}
\begin{align}
    [\mathcal L_\xi,\bar\partial_N]v
    =
    \sqrt{2}\, \bar{\mathrm{f}}\otimes R^\perp(\xi, \bar\eun )v
    =
    0  \ , \qquad v\in\Gamma(N^{1,0}\Sigma) \ ,
\end{align}
and similarly for $\partial_N$. Here $R^\perp$ is the curvature of the normal connection
\begin{align}
    R^\perp(X,Y)
    =
    \nabla^\perp_X\nabla^\perp_Y
    -
    \nabla^\perp_Y\nabla^\perp_X
    -
    \nabla^\perp_{[X,Y]} \ ,
    \qquad
    X,Y\in \Gamma(T\Sigma) \ .
\end{align}
By contrast, since $\alpha$ has Reeb charge $4i$ and $\xi$ is a Killing vector, the adjunction-type maps carry non-zero $U(1)_r$ charges
\begin{align}
    [\mathcal{L}_{\xi}, \adj] = 4i\, \adj \ , \qquad [\mathcal{L}_{\xi}, \adjb] = -4i \, \adjb \ . \label{3.34}
\end{align}
It now follows that the $\Theta$-dependent terms in \eqref{3.19d} are total Lie-derivative terms. Indeed, on a closed $\Sigma$:
\begin{align}
    0=\big( \mathcal{L}_{\xi}(\adj(\bar{\partial}_{N}v)), v\big)+\big(\adj(\bar{\partial}_{N}v), \mathcal{L}_{\xi}v\big) = 2\big(\adj(\bar\partial_{N}v), \mathcal{L}_{\xi}v\big )+4i\big(\adj(\bar{\partial}_{N}v), v\big) \ ,
\end{align}
where we have used Reeb invariance of $\bar{\partial}_{N}$, the charge of $\adj$, and the symmetry property \eqref{3.21d}. The conjugate identity gives the corresponding result for the term with $\adjb$. Thus, the action for quadratic fluctuations around a closed invariant three-cycle  $\Sigma\subset\ses$ takes the form
\begin{align}
    &\frac{1}{2}\int d^3\sigma \, \sqrt{g} \,
    \langle \mathrm{V},
    (\slashed{\mathfrak{D}}-3)(\slashed{\mathfrak{D}}+1)
    \mathrm{V}\rangle = ( \bar v, -\mathcal{L}_{\xi}^2v-4i\mathcal{L}_{\xi}v
    +\bar{\partial}_{N}^{\dagger}\bar{\partial}_{N}v )  \ .
\end{align}

\newpage
\small
\bibliographystyle{JHEP-v2.9}
\bibliography{M2-ext-bib}

@article{Farquet:2014bda,
    author = "Farquet, Daniel and Sparks, James",
    title = "{Wilson loops on three-manifolds and their M2-brane duals}",
    eprint = "1406.2493",
    archivePrefix = "arXiv",
    primaryClass = "hep-th",
    doi = "10.1007/JHEP12(2014)173",
    journal = "JHEP",
    volume = "12",
    pages = "173",
    year = "2014"
}

@book{Blair2002,
  author    = {Blair, David E.},
  title     = {Riemannian Geometry of Contact and Symplectic Manifolds},
  series    = {Progress in Mathematics},
  volume    = {203},
  publisher = {Birkh{\"a}user},
  address   = {Boston, MA},
  year      = {2002},
  isbn      = {0-8176-4261-7}
}

@article{BenettiGenolini:2026hmz,
    author = {Benetti Genolini, Pietro and Couzens, Christopher and L{\"u}scher, Alice},
    title = "{Probing black holes with equivariant localization}",
    eprint = "2604.26786",
    archivePrefix = "arXiv",
    primaryClass = "hep-th",
    month = "4",
    year = "2026"
}

@article{Farquet:2013cwa,
    author = "Farquet, Daniel and Sparks, James",
    title = "{Wilson loops and the geometry of matrix models in AdS$_4$/CFT$_3$}",
    eprint = "1304.0784",
    archivePrefix = "arXiv",
    primaryClass = "hep-th",
    doi = "10.1007/JHEP01(2014)083",
    journal = "JHEP",
    volume = "01",
    pages = "083",
    year = "2014"
}

@article{BenettiGenolini:2019jdz,
    author = "Benetti Genolini, Pietro and Perez Ipi{\~n}a, Juan Manuel and Sparks, James",
    title = "{Localization of the action in AdS/CFT}",
    eprint = "1906.11249",
    archivePrefix = "arXiv",
    primaryClass = "hep-th",
    doi = "10.1007/JHEP10(2019)252",
    journal = "JHEP",
    volume = "10",
    pages = "252",
    year = "2019"
}

@article{Calvo:2012du,
    author = "Calvo, Flavio and Marino, Marcos",
    title = "{Membrane instantons from a semiclassical TBA}",
    eprint = "1212.5118",
    archivePrefix = "arXiv",
    primaryClass = "hep-th",
    doi = "10.1007/JHEP05(2013)006",
    journal = "JHEP",
    volume = "05",
    pages = "006",
    year = "2013"
}

@article{Gulotta:2011aa,
    author = "Gulotta, Daniel R. and Herzog, Christopher P. and Pufu, Silviu S.",
    title = "{Operator Counting and Eigenvalue Distributions for 3D Supersymmetric Gauge Theories}",
    eprint = "1106.5484",
    archivePrefix = "arXiv",
    primaryClass = "hep-th",
    reportNumber = "PUPT-2384",
    doi = "10.1007/JHEP11(2011)149",
    journal = "JHEP",
    volume = "11",
    pages = "149",
    year = "2011"
}

@article{Gulotta:2011vp,
    author = "Gulotta, Daniel R. and Ang, J. P. and Herzog, Christopher P.",
    title = "{Matrix Models for Supersymmetric Chern-Simons Theories with an ADE Classification}",
    eprint = "1111.1744",
    archivePrefix = "arXiv",
    primaryClass = "hep-th",
    doi = "10.1007/JHEP01(2012)132",
    journal = "JHEP",
    volume = "01",
    pages = "132",
    year = "2012"
}

@article{Gulotta:2011si,
    author = "Gulotta, Daniel R. and Herzog, Christopher P. and Pufu, Silviu S.",
    title = "{From Necklace Quivers to the F-theorem, Operator Counting, and T(U(N))}",
    eprint = "1105.2817",
    archivePrefix = "arXiv",
    primaryClass = "hep-th",
    reportNumber = "PUPT-2374",
    doi = "10.1007/JHEP12(2011)077",
    journal = "JHEP",
    volume = "12",
    pages = "077",
    year = "2011"
}

@article{Kim:2012vza,
    author = "Kim, Hyojoong and Kim, Nakwoo",
    title = "{Operator Counting for N=2 Chern-Simons Gauge Theories with Chiral-like Matter Fields}",
    eprint = "1202.6637",
    archivePrefix = "arXiv",
    primaryClass = "hep-th",
    doi = "10.1007/JHEP05(2012)152",
    journal = "JHEP",
    volume = "05",
    pages = "152",
    year = "2012"
}

@article{Gautason:2025plx,
    author = "Gautason, Fridrik Freyr and van Muiden, Jesse",
    title = "{Ensembles in M-theory and holography}",
    eprint = "2505.21633",
    archivePrefix = "arXiv",
    primaryClass = "hep-th",
    doi = "10.1007/JHEP11(2025)078",
    journal = "JHEP",
    volume = "11",
    pages = "078",
    year = "2025"
}

@inproceedings{vanMuiden:2026nsp,
    author = "van Muiden, Jesse",
    title = "{Quantum M2-branes and Holography}",
    eprint = "2603.14544",
    archivePrefix = "arXiv",
    primaryClass = "hep-th",
    month = "3",
    year = "2026"
}

@article{Rozansky:1996bq,
    author = "Rozansky, L. and Witten, Edward",
    title = "{HyperKahler geometry and invariants of three manifolds}",
    eprint = "hep-th/9612216",
    archivePrefix = "arXiv",
    reportNumber = "IASSNS-HEP-96-128",
    doi = "10.1007/s000290050016",
    journal = "Selecta Math.",
    volume = "3",
    pages = "401--458",
    year = "1997"
}

@article{BenettiGenolini:2024lbj,
    author = {Benetti Genolini, Pietro and Gauntlett, Jerome P. and Jiao, Yusheng and L{\"u}scher, Alice and Sparks, James},
    title = "{Equivariant localization for D = 4 gauged supergravity}",
    eprint = "2412.07828",
    archivePrefix = "arXiv",
    primaryClass = "hep-th",
    doi = "10.1007/JHEP08(2025)211",
    journal = "JHEP",
    volume = "08",
    pages = "211",
    year = "2025"
}

@article{Kawai2017HomAssoc,
  author        = {Kotaro Kawai},
  title         = {Deformations of homogeneous associative submanifolds in nearly parallel {$G_2$}-manifolds},
  journal       = {Asian Journal of Mathematics},
  volume        = {21},
  number        = {3},
  pages         = {429--462},
  year          = {2017},
  eprint        = {1407.8046},
  archivePrefix = {arXiv},
  primaryClass  = {math.DG},
  doi           = {10.48550/arXiv.1407.8046}
}

@article{Sparks:2010sn,
    author = "Sparks, James",
    title = "{Sasaki-Einstein Manifolds}",
    eprint = "1004.2461",
    archivePrefix = "arXiv",
    primaryClass = "math.DG",
    doi = "10.4310/SDG.2011.v16.n1.a6",
    journal = "Surveys Diff. Geom.",
    volume = "16",
    pages = "265--324",
    year = "2011"
}

@article{Harvey:1982xk,
    author = "Harvey, R. and Lawson, Jr., H. B.",
    title = "{Calibrated geometries}",
    doi = "10.1007/BF02392726",
    journal = "Acta Math.",
    volume = "148",
    pages = "47",
    year = "1982"
}

@article{McLean:1998,
    author = "McLean, Robert C.",
    title = "{Deformations of calibrated submanifolds}",
    doi = "10.4310/CAG.1998.v6.n4.a4",
    journal = "Commun. Anal. Geom.",
    volume = "6",
    number = "4",
    pages = "705--747",
    year = "1998"
}

@article{Bar:1993,
  author    = {Christian B{\"a}r},
  title     = {Real Killing Spinors and Holonomy},
  journal   = {Communications in Mathematical Physics},
  volume    = {154},
  number    = {3},
  pages     = {509--521},
  year      = {1993},
  doi       = {10.1007/BF02102106},
}

@article{Hatsuda:2012dt,
    author = "Hatsuda, Yasuyuki and Moriyama, Sanefumi and Okuyama, Kazumi",
    title = "{Instanton Effects in ABJM Theory from Fermi Gas Approach}",
    eprint = "1211.1251",
    archivePrefix = "arXiv",
    primaryClass = "hep-th",
    reportNumber = "DESY-12-196, TIT-HEP-624",
    doi = "10.1007/JHEP01(2013)158",
    journal = "JHEP",
    volume = "01",
    pages = "158",
    year = "2013"
}

@article{friedrich1997nearly,
  title={On nearly parallel G2-structures},
  author={Friedrich, Th. and Kath, Ines and Moroianu, Andrei and Semmelmann, Uwe},
  journal={Journal of Geometry and Physics},
  volume={23},
  number={3-4},
  pages={259--286},
  year={1997},
  publisher={Elsevier}
}

@article{Gutowski:2002bx,
  author        = {Gutowski, J. and Ivanov, S. and Papadopoulos, G.},
  title         = {Deformations of generalized calibrations and compact non-Kahler manifolds with vanishing first Chern class},
  journal       = {Asian Journal of Mathematics},
  volume        = {7},
  pages         = {39--80},
  year          = {2003},
  eprint        = {math/0205012},
  archivePrefix = {arXiv},
  primaryClass  = {math.DG},
  doi           = {10.48550/arXiv.math/0205012}
}

@article{KawaiSecondOrder,
  author        = {Kawai, Kotaro},
  title         = {Second order deformations of associative submanifolds in nearly parallel {$G_2$}-manifolds},
  journal       = {The Quarterly Journal of Mathematics},
  volume        = {69},
  number        = {1},
  pages         = {241--270},
  year          = {2018},
  doi           = {10.1093/qmath/hax038},
  eprint        = {1610.07864},
  archivePrefix = {arXiv},
  primaryClass  = {math.DG}
}

@article{Pestun:2007rz,
    author = "Pestun, Vasily",
    title = "{Localization of gauge theory on a four-sphere and supersymmetric Wilson loops}",
    eprint = "0712.2824",
    archivePrefix = "arXiv",
    primaryClass = "hep-th",
    reportNumber = "ITEP-TH-41-07, PUTP-2248",
    doi = "10.1007/s00220-012-1485-0",
    journal = "Commun. Math. Phys.",
    volume = "313",
    pages = "71--129",
    year = "2012"
}

@article{Kallen:2012cs,
    author = {K{\"a}ll{\'e}n, Johan and Zabzine, Maxim},
    title = "{Twisted supersymmetric 5D Yang-Mills theory and contact geometry}",
    eprint = "1202.1956",
    archivePrefix = "arXiv",
    primaryClass = "hep-th",
    reportNumber = "UUITP-04-12",
    doi = "10.1007/JHEP05(2012)125",
    journal = "JHEP",
    volume = "05",
    pages = "125",
    year = "2012"
}

@article{Kallen:2011ny,
    author = "Kallen, Johan",
    title = "{Cohomological localization of Chern-Simons theory}",
    eprint = "1104.5353",
    archivePrefix = "arXiv",
    primaryClass = "hep-th",
    reportNumber = "UUITP-13-11",
    doi = "10.1007/JHEP08(2011)008",
    journal = "JHEP",
    volume = "08",
    pages = "008",
    year = "2011"
}

@article{Kallen:2012va,
    author = {K{\"a}ll{\'e}n, Johan and Qiu, Jian and Zabzine, Maxim},
    title = "{The perturbative partition function of supersymmetric 5D Yang-Mills theory with matter on the five-sphere}",
    eprint = "1206.6008",
    archivePrefix = "arXiv",
    primaryClass = "hep-th",
    reportNumber = "UUITP-17-12",
    doi = "10.1007/JHEP08(2012)157",
    journal = "JHEP",
    volume = "08",
    pages = "157",
    year = "2012"
}

@article{Pestun:2016qko,
    author = "Pestun, Vasily",
    title = "{Review of localization in geometry}",
    eprint = "1608.02954",
    archivePrefix = "arXiv",
    primaryClass = "hep-th",
    doi = "10.1088/1751-8121/aa6161",
    journal = "J. Phys. A",
    volume = "50",
    number = "44",
    pages = "443002",
    year = "2017"
}

@book{Wells:2008,
  author    = {Wells, Raymond O.},
  title     = {Differential Analysis on Complex Manifolds},
  edition   = {3},
  series    = {Graduate Texts in Mathematics},
  volume    = {65},
  publisher = {Springer},
  address   = {New York, NY},
  year      = {2008},
  doi       = {10.1007/978-0-387-73892-5},
  isbn      = {978-0-387-73892-5}
}

@article{Martelli:2011qj,
    author = "Martelli, Dario and Sparks, James",
    title = "{The large N limit of quiver matrix models and Sasaki-Einstein manifolds}",
    eprint = "1102.5289",
    archivePrefix = "arXiv",
    primaryClass = "hep-th",
    doi = "10.1103/PhysRevD.84.046008",
    journal = "Phys. Rev. D",
    volume = "84",
    pages = "046008",
    year = "2011"
}

@article{Martelli:2006yb,
    author = "Martelli, Dario and Sparks, James and Yau, Shing-Tung",
    title = "{Sasaki-Einstein manifolds and volume minimisation}",
    eprint = "hep-th/0603021",
    archivePrefix = "arXiv",
    reportNumber = "CERN-PH-TH-2006-039, HUTP-06-A0002",
    doi = "10.1007/s00220-008-0479-4",
    journal = "Commun. Math. Phys.",
    volume = "280",
    pages = "611--673",
    year = "2008"
}

@article{Brieskorn1966,
  author  = {Brieskorn, Egbert},
  title   = {Beispiele zur Differentialtopologie von Singularit{\"a}ten},
  journal = {Inventiones Mathematicae},
  year    = {1966},
  volume  = {2},
  number  = {1},
  pages   = {1--14},
  doi     = {10.1007/BF01403388},
  url     = {https://doi.org/10.1007/BF01403388}
}

@article{Drukker:2011zy,
    author = "Drukker, Nadav and Marino, Marcos and Putrov, Pavel",
    title = "{Nonperturbative aspects of ABJM theory}",
    eprint = "1103.4844",
    archivePrefix = "arXiv",
    primaryClass = "hep-th",
    reportNumber = "IMPERIAL-TP-2011-ND-01",
    doi = "10.1007/JHEP11(2011)141",
    journal = "JHEP",
    volume = "11",
    pages = "141",
    year = "2011"
}

@article{Bergshoeff:1987qx,
    author = "Bergshoeff, E. and Sezgin, E. and Townsend, P. K.",
    title = "{Properties of the Eleven-Dimensional Super Membrane Theory}",
    reportNumber = "IC-87-255",
    doi = "10.1016/0003-4916(88)90050-4",
    journal = "Annals Phys.",
    volume = "185",
    pages = "330",
    year = "1988"
}

@book{Milnorsingular,
 ISBN = {9780691080659},
 URL = {http://www.jstor.org/stable/j.ctt1bd6kvv},
 author = {John Milnor},
 publisher = {Princeton University Press},
 title = {Singular Points of Complex Hypersurfaces. (AM-61)},
 urldate = {2026-06-01},
 year = {1968}
}

@article{Gautason:2023igo,
    author = "Gautason, Fridrik Freyr and Puletti, Valentina Giangreco M. and van Muiden, Jesse",
    title = "{Quantized strings and instantons in holography}",
    eprint = "2304.12340",
    archivePrefix = "arXiv",
    primaryClass = "hep-th",
    doi = "10.1007/JHEP08(2023)218",
    journal = "JHEP",
    volume = "08",
    pages = "218",
    year = "2023"
}

@book{Atiyah:1974obx,
    author = "Atiyah, M. F.",
    title = "{Elliptic Operators and Compact Groups}",
    doi = "10.1007/BFb0057821",
    isbn = "978-3-540-37811-2, 978-3-540-37811-2",
    publisher = "Springer-Verlag",
    address = "Berlin, Germany",
    volume = "401",
    year = "1974"
}

@article{Becker:1996ay,
    author = "Becker, Katrin and Becker, Melanie and Morrison, David R. and Ooguri, Hirosi and Oz, Yaron and Yin, Zheng",
    title = "{Supersymmetric cycles in exceptional holonomy manifolds and Calabi-Yau 4 folds}",
    eprint = "hep-th/9608116",
    archivePrefix = "arXiv",
    reportNumber = "DUKE-TH-96-124, LBL-39156, LBNL-39156, UCB-PTH-96-33, NSF-ITP-96-65, WIS-96-34-PH",
    doi = "10.1016/S0550-3213(96)00491-9",
    journal = "Nucl. Phys. B",
    volume = "480",
    pages = "225--238",
    year = "1996"
}

@article{Boido:2023ojv,
    author = {Boido, Andrea and L{\"u}scher, Alice and Sparks, James},
    title = "{Matrix models from black hole geometries}",
    eprint = "2312.11640",
    archivePrefix = "arXiv",
    primaryClass = "hep-th",
    doi = "10.1007/JHEP05(2024)226",
    journal = "JHEP",
    volume = "05",
    pages = "226",
    year = "2024"
}

@article{Cagnazzo:2009zh,
    author = "Cagnazzo, Alessandra and Sorokin, Dmitri and Wulff, Linus",
    title = "{String instanton in AdS(4) x CP**3}",
    eprint = "0911.5228",
    archivePrefix = "arXiv",
    primaryClass = "hep-th",
    doi = "10.1007/JHEP05(2010)009",
    journal = "JHEP",
    volume = "05",
    pages = "009",
    year = "2010"
}

@article{Lotay2012Associative,
  author    = {Jason D. Lotay},
  title     = {Associative submanifolds of the 7-sphere},
  journal   = {Proceedings of the London Mathematical Society},
  volume    = {105},
  number    = {6},
  pages     = {1183--1214},
  year      = {2012},
  doi       = {10.1112/plms/pds029},
  url       = {https://doi.org/10.1112/plms/pds029},
  issn      = {0024-6115}
}

@article{Harvey:1999as,
    author = "Harvey, Jeffrey A. and Moore, Gregory W.",
    title = "{Superpotentials and membrane instantons}",
    eprint = "hep-th/9907026",
    archivePrefix = "arXiv",
    reportNumber = "EFI-99-22A, YCTP-P15-99, IASSNS-99-57, EFI-99-22",
    month = "7",
    year = "1999"
}

@article{Gautason:2025per,
    author = "Gautason, Fridrik Freyr and van Muiden, Jesse",
    title = "{Localization of the M2-Brane}",
    eprint = "2503.16597",
    archivePrefix = "arXiv",
    primaryClass = "hep-th",
    doi = "10.1103/67bh-xd42",
    journal = "Phys. Rev. Lett.",
    volume = "135",
    number = "10",
    pages = "101601",
    year = "2025"
}

@article{Becker:1995kb,
    author = "Becker, Katrin and Becker, Melanie and Strominger, Andrew",
    title = "{Five-branes, membranes and nonperturbative string theory}",
    eprint = "hep-th/9507158",
    archivePrefix = "arXiv",
    reportNumber = "NSF-ITP-95-62",
    doi = "10.1016/0550-3213(95)00487-1",
    journal = "Nucl. Phys. B",
    volume = "456",
    pages = "130--152",
    year = "1995"
}

@article{Giombi:2020mhz,
    author = "Giombi, Simone and Tseytlin, Arkady A.",
    title = "{Strong coupling expansion of circular Wilson loops and string theories in AdS$_5 \times {\rm S}^5$ and AdS$_4 \times {\rm CP}^3$}",
    eprint = "2007.08512",
    archivePrefix = "arXiv",
    primaryClass = "hep-th",
    reportNumber = "Imperial-TP-AT-2020-05",
    doi = "10.1007/JHEP10(2020)130",
    journal = "JHEP",
    volume = "10",
    pages = "130",
    year = "2020"
}

@article{Kallosh:1997ky,
    author = "Kallosh, Renata",
    title = "{World volume supersymmetry}",
    eprint = "hep-th/9709069",
    archivePrefix = "arXiv",
    reportNumber = "SU-ITP-97-36",
    doi = "10.1103/PhysRevD.57.R3214",
    journal = "Phys. Rev. D",
    volume = "57",
    pages = "3214--3218",
    year = "1998"
}

@article{Kapustin:2009kz,
    author = "Kapustin, Anton and Willett, Brian and Yaakov, Itamar",
    title = "{Exact Results for Wilson Loops in Superconformal Chern-Simons Theories with Matter}",
    eprint = "0909.4559",
    archivePrefix = "arXiv",
    primaryClass = "hep-th",
    reportNumber = "CALT-68-2750",
    doi = "10.1007/JHEP03(2010)089",
    journal = "JHEP",
    volume = "03",
    pages = "089",
    year = "2010"
}

@article{Kallosh:1997sw,
    author = "Kallosh, R.",
    editor = "Bais, F. A. and Bergshoeff, E. A. and de Wit, B. and Dijkgraaf, R. and Schellekens, A. N. and Verlinde, Erik P. and Verlinde, Herman L.",
    title = "{Quantization of p-branes, D-p-branes and M-branes}",
    eprint = "hep-th/9709202",
    archivePrefix = "arXiv",
    reportNumber = "SU-ITP-97-39",
    doi = "10.1016/S0920-5632(98)00153-4",
    journal = "Nucl. Phys. B Proc. Suppl.",
    volume = "68",
    pages = "197--205",
    year = "1998"
}

@article{Aharony:2008ug,
    author = "Aharony, Ofer and Bergman, Oren and Jafferis, Daniel Louis and Maldacena, Juan",
    title = "{${\mathcal{N}}\!=6$ Superconformal Chern-Simons-Matter Theories, M2-Branes and Their Gravity Duals}",
    eprint = "0806.1218",
    archivePrefix = "arXiv",
    primaryClass = "hep-th",
    reportNumber = "WIS-12-08-JUN-DPP",
    doi = "10.1088/1126-6708/2008/10/091",
    journal = "JHEP",
    volume = "10",
    pages = "091",
    year = "2008"
}

@article{Bergshoeff:1987cm,
    author = "Bergshoeff, E. and Sezgin, E. and Townsend, P. K.",
    title = "{Supermembranes and Eleven-Dimensional Supergravity}",
    reportNumber = "IC-87-10",
    doi = "10.1201/9781482268737-10",
    journal = "Phys. Lett. B",
    volume = "189",
    pages = "75--78",
    year = "1987"
}

@article{Drukker:2010nc,
    author = "Drukker, Nadav and Marino, Marcos and Putrov, Pavel",
    title = "{From weak to strong coupling in ABJM theory}",
    eprint = "1007.3837",
    archivePrefix = "arXiv",
    primaryClass = "hep-th",
    reportNumber = "HU-EP-10-39",
    doi = "10.1007/s00220-011-1253-6",
    journal = "Commun. Math. Phys.",
    volume = "306",
    pages = "511--563",
    year = "2011"
}

@article{Herzog:2010hf,
    author = "Herzog, Christopher P. and Klebanov, Igor R. and Pufu, Silviu S. and Tesileanu, Tiberiu",
    title = "{Multi-Matrix Models and Tri-Sasaki Einstein Spaces}",
    eprint = "1011.5487",
    archivePrefix = "arXiv",
    primaryClass = "hep-th",
    reportNumber = "PUPT-2359",
    doi = "10.1103/PhysRevD.83.046001",
    journal = "Phys. Rev. D",
    volume = "83",
    pages = "046001",
    year = "2011"
}

@article{Marino:2011eh,
    author = "Marino, Marcos and Putrov, Pavel",
    title = "{ABJM theory as a Fermi gas}",
    eprint = "1110.4066",
    archivePrefix = "arXiv",
    primaryClass = "hep-th",
    doi = "10.1088/1742-5468/2012/03/P03001",
    journal = "J. Stat. Mech.",
    volume = "1203",
    pages = "P03001",
    year = "2012"
}

@article{Hatsuda:2013gj,
    author = "Hatsuda, Yasuyuki and Moriyama, Sanefumi and Okuyama, Kazumi",
    title = "{Instanton Bound States in ABJM Theory}",
    eprint = "1301.5184",
    archivePrefix = "arXiv",
    primaryClass = "hep-th",
    reportNumber = "DESY-13-010, TIT-HEP-626",
    doi = "10.1007/JHEP05(2013)054",
    journal = "JHEP",
    volume = "05",
    pages = "054",
    year = "2013"
}

@article{Hatsuda:2013oxa,
    author = "Hatsuda, Yasuyuki and Marino, Marcos and Moriyama, Sanefumi and Okuyama, Kazumi",
    title = "{Non-perturbative effects and the refined topological string}",
    eprint = "1306.1734",
    archivePrefix = "arXiv",
    primaryClass = "hep-th",
    reportNumber = "DESY-13-096, TIT-HEP-627",
    doi = "10.1007/JHEP09(2014)168",
    journal = "JHEP",
    volume = "09",
    pages = "168",
    year = "2014"
}

@article{Park:2020hgt,
    author = "Park, Jaemo and Shin, Hyeonjoon",
    title = "{1/2-BPS membrane instantons in AdS$_4 \times$S$^7 / \mathbf{Z}_k$}",
    eprint = "2008.11380",
    archivePrefix = "arXiv",
    primaryClass = "hep-th",
    doi = "10.1103/PhysRevD.102.126021",
    journal = "Phys. Rev. D",
    volume = "102",
    number = "12",
    pages = "126021",
    year = "2020"
}

@article{Beccaria:2023ujc,
    author = "Beccaria, Matteo and Giombi, Simone and Tseytlin, Arkady A.",
    title = "{Instanton Contributions to the ABJM Free Energy from Quantum M2 Branes}",
    eprint = "2307.14112",
    archivePrefix = "arXiv",
    primaryClass = "hep-th",
    reportNumber = "PUPT-2645, Imperial-TP-AT-2023-04",
    doi = "10.1007/JHEP10(2023)029",
    journal = "JHEP",
    volume = "10",
    pages = "029",
    year = "2023"
}

@article{Tseytlin:2025dae,
    author = "Tseytlin, Arkady A. and Wang, Zihan",
    title = "{On world-volume supersymmetry of supermembrane action in~static gauge}",
    eprint = "2512.04948",
    archivePrefix = "arXiv",
    primaryClass = "hep-th",
    reportNumber = "Imperial-TP{\textendash}2025-AT-02",
    doi = "10.1098/rspa.2025.1058",
    journal = "Proc. Roy. Soc. Lond. A",
    volume = "482",
    number = "2334",
    pages = "20251058",
    year = "2026"
}

@article{Gautason:2025bft,
    author = "Gautason, Fridrik Freyr and Nix, Alexia",
    title = "{Universal holographic Wilson loops in 3d SCFTs}",
    eprint = "2511.04596",
    archivePrefix = "arXiv",
    primaryClass = "hep-th",
    month = "11",
    year = "2025"
}

@article{Imamura:2008nn,
    author = "Imamura, Yosuke and Kimura, Keisuke",
    title = "{On the moduli space of elliptic Maxwell-Chern-Simons theories}",
    eprint = "0806.3727",
    archivePrefix = "arXiv",
    primaryClass = "hep-th",
    reportNumber = "UT-08-20",
    doi = "10.1143/PTP.120.509",
    journal = "Prog. Theor. Phys.",
    volume = "120",
    pages = "509--523",
    year = "2008"
}

@article{Imamura:2008dt,
    author = "Imamura, Yosuke and Kimura, Keisuke",
    title = "{N=4 Chern-Simons theories with auxiliary vector multiplets}",
    eprint = "0807.2144",
    archivePrefix = "arXiv",
    primaryClass = "hep-th",
    reportNumber = "UT-08-23",
    doi = "10.1088/1126-6708/2008/10/040",
    journal = "JHEP",
    volume = "10",
    pages = "040",
    year = "2008"
}

@article{Hatsuda:2015lpa,
    author = "Hatsuda, Yasuyuki and Honda, Masazumi and Okuyama, Kazumi",
    title = "{Large N non-perturbative effects in $\mathcal{N}=4$ superconformal Chern-Simons theories}",
    eprint = "1505.07120",
    archivePrefix = "arXiv",
    primaryClass = "hep-th",
    reportNumber = "DESY-15-078, HRI-ST-1505",
    doi = "10.1007/JHEP09(2015)046",
    journal = "JHEP",
    volume = "09",
    pages = "046",
    year = "2015"
}

@article{Freund:1980xh,
    author = "Freund, Peter G. O. and Rubin, Mark A.",
    editor = "Salam, A. and Sezgin, E.",
    title = "{Dynamics of Dimensional Reduction}",
    reportNumber = "EFI 80/35-CHICAGO",
    doi = "10.1016/0370-2693(80)90590-0",
    journal = "Phys. Lett. B",
    volume = "97",
    pages = "233--235",
    year = "1980"
}

@article{Cheon:2011vi,
    author = "Cheon, Sangmo and Kim, Hyojoong and Kim, Nakwoo",
    title = "{Calculating the partition function of N=2 Gauge theories on $S^3$ and AdS/CFT correspondence}",
    eprint = "1102.5565",
    archivePrefix = "arXiv",
    primaryClass = "hep-th",
    doi = "10.1007/JHEP05(2011)134",
    journal = "JHEP",
    volume = "05",
    pages = "134",
    year = "2011"
}
\end{document}